\documentclass[final]{york-thesis}
\usepackage[latin1]{inputenc}
\usepackage{multirow}
\usepackage{color}
\usepackage[english]{babel}
\usepackage{subfigure}
\usepackage{bbm}
\usepackage{graphicx}
\usepackage{amsmath}
\usepackage{amsfonts}
\usepackage{amssymb}
\usepackage{here}
\usepackage[ps2pdf]{hyperref}
\hypersetup{colorlinks=true,linkcolor=blue,citecolor=red,pdfpagemode=UseNone}
\newcommand{\promille}{%
  \relax\ifmmode\promillezeichen
        \else\leavevmode\(\mathsurround=0pt\promillezeichen\)\fi}
\newcommand{\promillezeichen}{%
  \kern-.05em%
  \raise.5ex\hbox{\the\scriptfont0 0}%
  \kern-.15em/\kern-.15em%
  \lower.25ex\hbox{\the\scriptfont0 00}}
\def\gsim{\mathrel{\rlap{\lower4pt\hbox{\hskip1pt$\sim$}} \raise1pt\hbox{$>$}}} 
\def\lsim{\mathrel{\rlap{\lower4pt\hbox{\hskip1pt$\sim$}} \raise1pt\hbox{$<$}}} 

\definecolor{dark-magenta}{rgb}{.5,0,.5}
\setboolean{masters}{false}
\setboolean{hasfigures}{true}
\setboolean{hastables}{true}

\title{Calculation of supercritical Dirac resonances in heavy-ion collisions}
\author{Edward Ackad}
\department{Department of Physics and Astronomy}

\committeememberslist{
\begin{enumerate}
\item Marko Horbatsch (Supervisor)
\item A. Kumarakrishnan (Chair)
\item Jurij Darewych (Program Member)
\item Roman Koniuk (Program Member)
\item Douglas E. Smylie (Outside Member)
\item Richard Hall (External)
\item Randy Lewis (Dean's Representative)
\end{enumerate}}

\abstractfile{abstract.tex}
\dedicationfile{dedication.tex}
\acknowledgementsfile{Acknow.tex}

\begin{document}

\makefrontmatter

\chapter{Introduction}
In nature, two types of particles are found and are distinguished by their multi-particle statistical behavior. Bosons are particles obeying Bose-Einstein statistics, and allow any number of particles to be in the same quantum state. Fermions obey Fermi-Dirac statistics. Fermions must all be in a unique quantum state and this is responsible for the structure observed in atoms, nuclei and in baryons. The fundamental building blocks in nature are fermions, namely the quarks (the neutron, proton and other baryonic matter are composed of quarks and anti-quarks) and electrons (including muons and tauons). The fundamental bosons appear as mediators of interactions (photons and gluons), but composite bosons, made up of an even number of fermions, also exist (e.g. the deuterium nucleus, $^2H$). While atomic physics is concerned with modeling the behavior of all atomic systems, the electron plays a central role, since it is stable, abundant and the lightest charged fermion making it very accessible experimentally.

The electron has an intrinsic spin of $\frac{1}{2}$ i.e., a half-integer value as do all fermions. Bosons possess integer spin (e.g. photons have an intrinsic spin of 1). A notable characteristic of particles is that different particles with the same spin behave in the same manner by obeying the same quantum mechanical equations of motion. For instance, the muon is also a spin-$\frac{1}{2}$ particle with the same electric charge as the electron, but it is more massive (around 207 times the mass of the electron) and has a finite lifetime (primary decay channel: $\mu^- \rightarrow e^- \bar{\nu_e}\nu_{\mu}$). Prior to decaying, the muon will behave as an electron with a heavier mass. Therefore, any model of the electron will also describe other fundamental (point) particles with spin-$\frac{1}{2}$, once the constants such as the mass or sign of the charge are adjusted. This includes the anti-particles of the fermions. The positron is the anti-particle of the electron. It has the same mass and spin, but the opposite charge.

In order to describe the behavior of the electron it is necessary to specify its equation of motion. Initially the electron was modeled as a classical charged point particle with quantized orbital angular momentum. This was hypothesized by Bohr to explain the level structure of the hydrogen atom which consists of an electron orbiting a proton. The Bohr model of the atom was relatively successful and explained the Balmer series (spectral lines of hydrogen), but fell short in many areas, e.g. when applied to the spectrum of helium or other many-electron atoms. The Bohr-Sommerfeld model, incorporated elliptical orbits, and addressed the radiative decay problem, but could not resolve the helium problem. It was replaced by the model of an electron as a wave obeying the Schr\"odinger equation. Pauli augmented this theory by including the intrinsic spin of the electron, \textit{ad hoc}, in what is now called the Schr\"odinger-Pauli equation. This is a useful model of the electron that describes most the quantum phenomena dealing with the electron. 

When describing the electron as a wave, each orbital has room for two electrons of opposite spin due to the spin differentiating the quantum state. Since the electron is a fermion, once an orbital is occupied by a pair of electrons no other electron can be in that state. This forces other electrons into higher states, and causes higher orbitals to be filled giving structure to the atomic system. The blocking of an orbital, due to being filled with an electron, is called Pauli-blocking.

For all the success of the Schr\"odinger-Pauli equation, it was not the final model of the electron. This equation accounts for the fact that the electron is a fermion but lacks the effects of special relativity. This equation is, therefore, only valid at low energies ($E\ll$ m$_{\rm e}$c$^2$ where m$_{\rm e}$ is the mass of the electron and c is the speed of light in vacuum). A model which takes into account special relativity must be valid for kinetic energies that are comparable to, or larger than, the rest energy of the electron. With such sizable kinetic energy it is possible to create an electron-positron pair ($E \approx$ 2m$_{\rm e}$c$^2$). A relativistic theory of the electron, which would to fulfill these requirements, is needed.

\section{Relativistic quantum mechanics}
To satisfy special relativity, the total energy of a free particle must be equal to $\sqrt{{p^2\rm c}^2+{\rm m}_{\rm e}^2{\rm c}^4}$. It is also necessary to have the same-order derivatives for the coordinate and time dependence. Dirac demanded that they be of first order to have simple time evolution. This resulted in a multi-component wave equation, with four components being the lowest number able to satisfy Poincar\'e invariance. This was the start of relativistic quantum mechanics (RQM) for fermions. RQM describes the electron at very high and very low energies. Using hole theory, RQM also accounts for the possibility of electron-positron pair creation. To accomplish this, the tools of quantum field theory (QFT) are used. 

In QFT, the classical field degrees of freedom are quantized. This is done by expressing the classical field in normal modes and describing each mode by a harmonic oscillator equation. Canonical quantization is then used to promote the expansion coefficients to creation and annihilation operators. The first complete quantum field theory was created, by Dirac, in 1927 and described the electromagnetic field. By combining the Dirac equation for the electron with the Maxwell equations for the photon and quantizing them, by imposing anticommutation and commutation relations respectively on the field operators, a robust part of the model of charged fermions is obtained. This is called quantum electrodynamics (QED). In QED the photon acts as the force carrier for the electric charge, communicating the interaction between the different particles. 

RQM predicts that an electron-positron pair can be created both dynamically, through excitation by a time-varying field, or statically though very strong potentials. While the former is well understood and verified by experiment, the static creation of pairs has yet to be demonstrated experimentally. This method of pair creation is intimately tied to the fundamentals of QED. Thus, it presents a good test of the theory in the presence of very strong potentials.

In RQM, the spin-$\frac{1}{2}$ particles are described by the Dirac equation. This is done in the same way as non-relativistic spin-0 (hypothetical) particles are described by the Schr\"odinger equation. The Dirac equation has been around for almost 80 years and remains a formidable challenge to solve in all but the most trivial cases. Time-dependent problems are especially difficult. It is only with the use of modern computational resources and methods that situations of interest can be solved \cite{krekoranegtrans,krekoraklien}. These problems are very challenging, both theoretically and computationally. Their interpretation is also challenging, since they describe multi-particle situations, e.g., particle-antiparticle pair creation.

\subsection{Supercritical pair creation}
One of the predictions of the relativistic theory of the electron is that for fields of sufficient strength, so-called supercritical fields, the ground state will have its energy inside the negative-energy continuum. This is due to a feature of relativistic theories of particles: a lower continuum. The spectrum for a non-relativistic particle that vanishes asymptotically, such as the attractive Coulomb interaction contains both continuum states and bound states. The bound states start with the most negative energy (the rest energy is not included), namely the ground state and become denser as the energy rises (Rydberg series). At the ionization energy (the energy which the particle needs in order to become unbound) and above, there is a continuum of states the particle can occupy. These states, in the case of hydrogen, represent free electrons scattering from a proton. In Dirac theory this structure is also present, but there also exists a lower continuum below the bound states, starting at $E=-$m$_{\rm e}$c$^2$ and extending to $E\rightarrow-\infty$. As the attractive potential becomes stronger, the system's ground state energy is lower, which causes no dramatic change in the non-relativistic theories. In Dirac theory, as the potential increases the ground state is lowered to eventually reside in the lower continuum (which is usually called the negative-energy continuum). This causes the ground state to change from a bound state to a resonant state, called a supercritical resonance state. 

Resonances are formed, in general, when a bound state is embedded into a continuum of scattering states. Often this is due to the mixture of two potentials, e.g., when the Coulomb potential of an atom is perturbed by a unform electric field. In one direction, this results in the bound states having their eigenenergies among the continuum of scattering states. Thus, resonances are created (cf. section~\ref{resonancessec} for more details).

Supercritical resonances have all of the features of a usual atomic resonance with the primary difference being in the interpretation. The supercritical resonance is a pair creation resonance and not just the familiar scattering resonance from quantum mechanics. Supercritical resonances are characterized by two parameters in the usual way: the energy position and the lifetime. Since the lifetime is inversely proportional to the energy width, these parameters completely describe the shape of the resonance as it appears in the energy density of states (obtained from the spectrum) or in the scattering problem (viewed as scattering of the negative-energy continuum states). The resonance's shape is called a Breit-Wigner distribution which has the form
\begin{equation}\label{bweqn}
F(E) =\frac{\Gamma/2}{(E_{\mathrm{res}}-E)^2+\Gamma^2/4},
\end{equation}
where $E_{\mathrm{res}}$ and $\Gamma$ are the energy position and width respectively. It is, therefore, useful to determine the resonance parameters in order to facilitate this experimental search.

Two direct methods of obtaining the resonance parameters using the mapped Fourier grid method (cf. appendix \ref{appendix1} for details of the mapped Fourier grid method) are demonstrated in chapter~\ref{tdscstates}. These extended methods are general enough to be applicable to other resonance phenomena in quantum mechanics.

\section{Analytic continuation methods to determine resonance parameters}
There are a few well-established methods for obtaining the resonance parameters using analytic continuation of the Hamiltonian (which means making the Hamiltonian complex by an artificial transformation). These methods have been employed, compared, extended and improved in this thesis by using the supercritical resonance as a challenging example. This has led to very accurate determinations of the supercritical resonance parameters and shows general trends in the behavior of supercritical resonances. The improvements in the methods are equally applicable to normal scattering resonances in quantum mechanics. The resulting increase in performance is demonstrated along with the fact that several competing methods are shown to agree to very high precision.

The method of complex scaling (CS) \cite{moiseyevcs} and its extension, smooth exterior scaling (SES) \cite{RissSESCAP,SES2cap}, are compared to the method of adding a polynomial complex absorbing potential (CAP)  \cite{santragf,sahoo,masuir2r4}, by determining the supercritical Dirac resonance parameters. In this thesis, the CAP method is extended to the relativistic Dirac equation for the first time. All these methods use a parameter to characterize the transformation of the Hamiltonian from real to complex and then vary that parameter. Stability in the resonance parameters is then used to find the best approximation to the true resonance parameters; hence they are called stabilization methods. A comparison of several stabilization methods is carried out in chapter~\ref{CSCAP}.

A further extension of these analytic continuation methods was motivated by R. Lefebvre \textit{et al.} in Ref.~\cite{lebrevePade} where a Pad\'e extrapolation was employed to increase the accuracy of the CAP method. In chapter~\ref{padechp}, this idea is extended to SES. Convergent results are obtained by both methods when extrapolation is applied. The stability and parameter independence of the results are demonstrated. The extrapolation also increases the accuracy of the analytic continuation methods, allowing for the exploration of effects beyond the first-order approximation to the two-center potential, which is used to model the potential of two highly charged ions. These investigations provide detailed knowledge of the behavior of supercritical resonance states, including how the mapped Fourier grid method handles supercritical resonance states. Understanding the behavior of the mapped Fourier grid method with a supercritical resonance state is needed for time-dependent collisions with supercritical intermediate states.

In this thesis, the mapped Fourier grid method, a pseudospectral method, was chosen due to its suitability for the non-perturbative strong-field problems of interest. By extending the mapped Fourier grid method to the Dirac equation (cf. appendix~\ref{appendix1} and Ref.~\cite{me1}) we then have a suitable and efficient way to solve the Dirac equation numerically. The method builds a matrix representation of the Hamiltonian and obtains the eigenvalues and eigenvectors by diagonalization. The representation is carried out in coordinate-space and the method maps the coordinates allowing for an efficient coverage of phase-space \cite{Kosloff96}. It has been demonstrated to yield very accurate results for the hydrogenic problem, which is used as a test case. It is well suited for solving time-dependent problems due to the fact that the diagonalization yields a complete set of states. Expanding into this set of states, therefore, preserves unitarity. A key feature of the method, which is exploited in time-dependent problems (cf. chapter~\ref{tdscchp}), is that the complete set of states contains the supercritical resonance. Thus, the resonance does not need to be added or constructed, as was done in prior calculations in the literature \cite{pra10324,PhysRevA.37.1449}. 

\section{Searching for supercritical resonances in collisions}
The collision of two heavy, fully-ionized atoms becomes supercritical when the atoms are a few tens of femtometers ($10^{-15}$m) apart. This is the only known route to supercritical resonance states that is accessible experimentally, making it the focus of work seeking to demonstrate their existence. By treating each of the two nuclei in a collision as a homogeneously charged sphere separated by a distance $R(t)$, a suitable model of the time-dependent electrostatic potential in the collision is obtained. The collision allows access to a supercritical potential, and therefore supercritical resonance states, by solving the time-dependent Dirac equation with this potential. Magnetic and retardation effects can be ignored, due to the slow-down of the nuclear motion near the Coulomb barrier.

To solve the time-dependent Dirac equation, the propagator, which operates on an initial state and advances it according to the time-dependent hamiltonian, is approximated. The approximation assumes the system will not change over a short time interval, $\Delta t$, allowing for the use of the propagator with time-independent states. By solving the time-independent Dirac equation at a particular time, amounting to a fixed internuclear separation, $R(t)$, and then using the approximate propagator to move forward by $\Delta t$, the time-dependent Dirac equation for the collision is solved as a series of quasi-stationary steps.

To propagate the state forward the eigenvalues and associated eigenvectors of the time-independent hamiltonian are needed. Also, in order to preserve unitarity, a complete set of states is required. As shown in Ref.~\cite{me1}, the mapped Fourier grid method can satisfy these conditions efficiently. Thus, by using the mapped Fourier grid method to solve the Dirac eigenvalue problem at successive internuclear separations (which amounts to subsequent times) and by using these eigenstates in the approximate propagator to advance from $t$ to $t+\Delta t$, it is possible to obtain the time evolution, not only for a single initial state, but for an entire set of independent initial states.

\section{Previous searches}\label{prevsearch}
To demonstrate the existence of supercritical resonances, some signature of them must be sought. This work began in the 1970's and has continued to this day owing to the difficulty of both the theoretical predictions and the required experimental conditions.

In 1981 the Frankfurt group published calculations of the resonance parameters and dynamical collision calculations. The group looked for evidence of supercritical resonance states in the available data from the Gesellschaft f\"ur Schwerionenforschung (GSI) ~\cite{pra10324}. The supercritical resonance calculations were performed in the center-of-mass frame for extended nuclei using the monopole approximation to the two-center potential. The group obtained results for the supercritical resonance parameters (to within a percent accuracy), and constructed the supercritical resonance state wavefunction for use in the collision calculation. This was achieved by solving the Dirac equation for a potential, which matched the two-center potential up to some small value, $r_c$, and then remained constant afterwards, with a value of $V(r)=V(r_c)$. The choice of $r_c$ was guided by the knowledge that the transition from bound-type to continuum-type character for the supercritical resonance state, took place around $E_{\rm res} -V(r) \pm {\rm m}_{\rm e}{\rm c}^2=0$ (cf. section~\ref{resex}). This resulted in a wavefunction that could be used to calculate the required matrix elements. Subsequently the resonance state was used together with an adiabatic quasi-molecular basis, in order to solve the coupled differential equations for the expansion coefficients and calculate the expected positron spectrum (and total positron creation probability). They found that although total positron production increases dramatically with the combined nuclear charge of different systems, the increase is smooth, making detection of the effects of the supercritical resonance decay difficult. The calculations were performed for different Fermi levels (the lowest level filled with electrons) to represent a collision in which a single $K-$shell vacancy was created prior to closest approach. 

The GSI accelerator at the time was not able to provide fully stripped heavy nuclei and a solid, neutral, target was used to provide the collision partners. As expected, this damped the signal, since the electrons would be Pauli-blocked from being created below the Fermi level. 

\begin{figure}[H]\centering
\includegraphics[angle=0,scale=0.32]{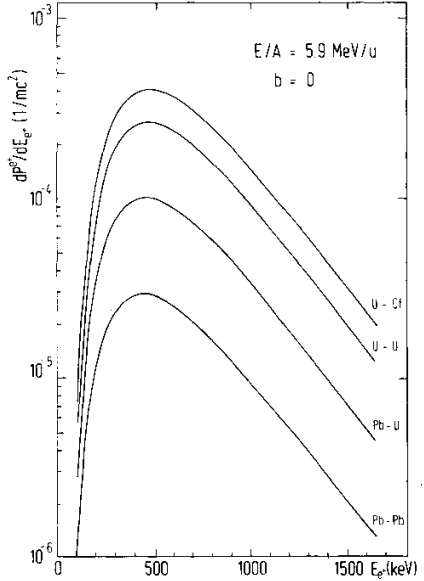}
\caption[Closed coupling calculation results from Ref.~\cite{pra10324}]{\spacing{1} \label{pra24103} Positron spectrum from Reinhardt \textit{et al.} \cite{pra10324} for collision systems with states up to 3S$\sigma$ and 4P$_{1/2}\sigma$ occupied at a center-of-mass energy of 700MeV using a closed-coupling calculation. The results show no qualitative difference for supercritical collisions (U-U, U-Cf) as compared to subcritical collisions (Pb-U and Pb-Pb systems).  }
\end{figure}

Figure.~\ref{pra24103} shows the results obtained for both supercritical collisions and collisions where even at closest approach the potential is not supercritical, i.e. a subcritical collision.
Each curve represents the calculated positron spectrum for different collision systems from uranium-californium to lead-lead with states up to 3S$\sigma$ and 4P$_{1/2}\sigma$ filled initially. The lead-lead collision does not have a supercritical potential even at closest approach ($Z_{\rm united}\approx 164$), but has the same qualitative shape as the uranium-californium spectrum ($Z_{\rm united}\approx 190$) indicating that the decay of the supercritical resonance is not resulting in marked features. The authors compared their results to some unpublished data from GSI obtaining excellent agreement except for the case of uranium-uranium (experimental uranium-californium data were not available).

\begin{figure}[ht]\centering
\includegraphics[angle=0,width=8cm]{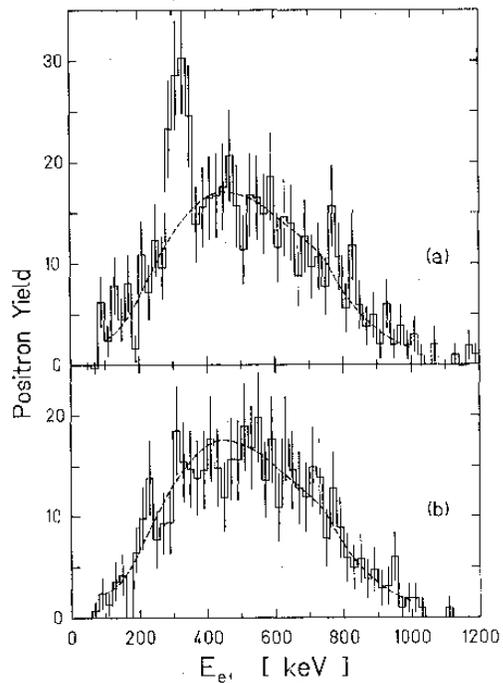}
\caption[Experimental positron spectrum from Ref.~\cite{PhysRevLett.51.2261}]{\spacing{1} \label{schweppe} Experimental positron spectrum from Schweppe \textit{et al.} \cite{PhysRevLett.51.2261} for a uranium-californium collision system at 6.0MeV/u. The top plot is for preferentially backward elastic scattering $100^{\circ} < \theta_{c.m.} < 130^{\circ}$, while the lower plot is for favored forward elastic scattering $50^{\circ} < \theta_{c.m.} < 80^{\circ}$ for the ion (uranium). The dashed lines show the theoretical expectation for dynamical positrons including the nuclear background. Of interest is the peak in the top plot which is narrower than 80keV.  }
\end{figure}

In 1983, the EPOS collaboration at GSI published their findings in Ref.~\cite{PhysRevLett.51.2261} on a sharp peak ($<$80keV) from uranium-californium collisions at 6.05MeV/u. A similar peak was found in uranium-uranium collisions in unpublished work, but there were still questions as to whether the peaks had a common origin. Figure~\ref{schweppe} shows the collaboration's data and the peak in (a) is reported with a confidence level of $<$0.1\%.
The lines with error bars on both plots represent the positron spectrum, while the dashed lines show the theoretically expected spectrum from dynamical positrons and the nuclear background. The top plot is for preferentially backward elastic scattering $100^{\circ} < \theta_{c.m.} < 130^{\circ}$ of the uranium ion, while on the lower plot it is for favored forward elastic scattering $50^{\circ} < \theta_{c.m.} < 80^{\circ}$. The authors initially thought this peak was from the spontaneous decay of the supercritical resonance, since it was near the expected energy of the supercritical resonance at closest approach. They argued that the possibility that the positrons were from a single internal pair conversion of nuclear transitions did not seem likely, since no emission line could account for the intensity of the peak. Explaining the peak as the decay of a supercritical resonance was a problem because to obtain such a narrow peak the nuclei would have to stick for a long time ($\gsim10^{-20}$s). In another analysis of a very similar narrow peak in uranium-thorium collisions \cite{PhysRevLett.56.444}, the hypothesis of very long nuclear sticking, was found to be very unlikely. Instead, led by the fact that a similar peak was found in thorium-thorium and thorium-curium collisions, the authors examined the earlier hypothesis \cite{PhysRevLett.54.1761} that the peak was due to the decay of a previously unknown neutral particle.

\begin{figure}[ht]\centering
\includegraphics[angle=0,width=8cm]{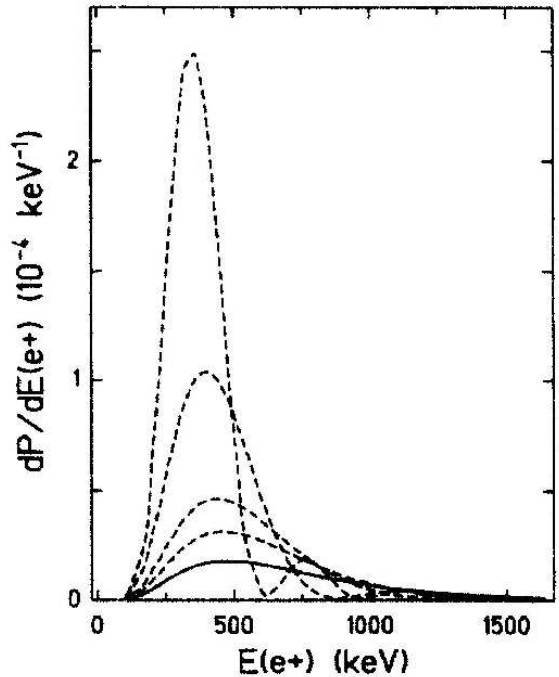}
\caption[Closed coupling calculation results with nuclear sticking from Ref.~\cite{PhysRevA.37.1449}]{\spacing{1} \label{mullerfig}Positron spectrum from M\"uller \textit{et al.} \cite{PhysRevA.37.1449} for bare uranium-uranium collisions at a center-of-mass energy of 740MeV. Dashed curves are for sticking times of 10,5,2,1 zepto-seconds in order from the tallest to the smallest peak. The solid line is for the collision without sticking. }
\end{figure}

The theory group revisited the problem in Ref.~\cite{PhysRevA.37.1449} with the added detailed effects of nuclear sticking in an attempt to explain the narrow peak which showed up in the results at the EPOS and ORANGE experimental groups at GSI. The cause of the peaks are still under investigation \cite{EurPJA.14.191} but it is now clear that the peak is not due to supercritical resonance decay, since they were also found in subcritical collisions such as thorium-tantalum \cite{PhysLettB.245.153}. M\"uller \textit{et al.} were able to calculate the expected positron spectra for collisions in which the nuclei remain almost static at closest approach for up to 10 zepto-seconds ($10^{-20}$s). Nuclear sticking times of a few zepto-seconds are experimentally obtainable in deep inelastic collisions as demonstrated in Ref.~\cite{PhysRevLett.50.1838} and Ref.~\cite{PhysRevC.34.562}. The sticking has a strong influence on the positron spectra, and it was shown that the peak of the positron spectrum will center closer to the resonance peak providing a clearer signal of the decay of a supercritical resonance shown in Fig.~\ref{mullerfig}.

These results also demonstrate that the sticking causes interference between the dynamical pair production, due to the motion of the highly charged nuclei, and the decay of the resonance. This is clearly visible from the second interference peak at around 800keV for the largest peak in the figure, which corresponds to a sticking time of 10 zepto-seconds. Calculations corresponding to a sticking time of 5 zepto-seconds (2$^{\rm nd}$ curve from top) also exhibits a second peak although it is at around 1100keV. The other dashed curves correspond to nuclear sticking times of 2 and 1 zepto-seconds and still show a small shifting of each peak towards the resonance energy. This prior work remains the basis for comparison to the work presented here in chapter~\ref{tdscchp}, and is the most recent (prior) work on the subject known to the author.

Work continues on these collisions, but it is mainly focused on explaining the sharp peaks and examining the possibility of longer nuclear sticking times; nevertheless the search for supercritical states continues. The new GSI-FAIR experiments will perform bare uranium-uranium collisions removing the Pauli-blocking of the previous experiments. The head-on collisions in these experiments represent the best hope for detecting supercritical resonances in the near future.

\chapter{Theoretical background}\label{physics}
Using his equation, P.A.M. Dirac correctly predicted the existence of the positron more than 70 years ago and the equation has remained a focus of study ever since. The equation was intended to describe electrons relativistically, combining quantum mechanics with special relativity, but it ended up having a much broader scope. As a wave equation, the Dirac equation describes spin-$\frac{1}{2}$ particles (fermions) and antiparticles in accord with relativistic kinematics. As a field equation, the Dirac equation describes spin-$\frac{1}{2}$ fields. When coupled to quantized electromagnetism, it forms the cornerstone of quantum electrodynamics (QED). QED is the most successful physical theory accounting for phenomena such as the electron's anomalous magnetic moment and the energy level shifts of hydrogen (Lamb shift).

Solutions to the Dirac equation are four-component wavefunctions, called spinors (or more precisely bispinors or Dirac spinors). In canonical Dirac-Pauli representation, the Dirac equation for an electron (or positron) with mass m$_{\rm e}$ is
\begin{equation}\label{desimp1}
\left(\vec{\mbox{\boldmath$\alpha$}}\cdot \vec{{\bf p}}+\mbox{$\hat{\beta}$m$_{\rm e}$c$^2$}\right)\Psi=i\hbar\frac{\partial \Psi}{\partial t}
\end{equation}
where the $\alpha$ matrices are commonly defined as
\begin{equation}
\alpha_i=\left(\begin{array}{cc} 0 & \sigma_i\\ \sigma_i & 0 \end{array} \right).
\end{equation}
Here $\sigma_i$ are the $2\times2$ Pauli matrices and $\hat{\beta}=diag\{1,1,-1,-1\}$ with all other entries being $0$. The solutions to Eq.~\ref{desimp1} contain both positive- and negative-energy four-component wavefunctions with both spin states ($\frac{1}{2}$ and $-\frac{1}{2}$). The existence of negative-energy solutions are a core feature of the Dirac equation. They hint that the interpretation of the Dirac equation as a single-particle theory is problematic. The interpretational difficulty posed by the negative-energy states is explored in the next section using hole theory.

\section{Hole theory}\label{holetheory}
\begin{figure}[!ht]\centering
\includegraphics[angle=0,scale=0.4]{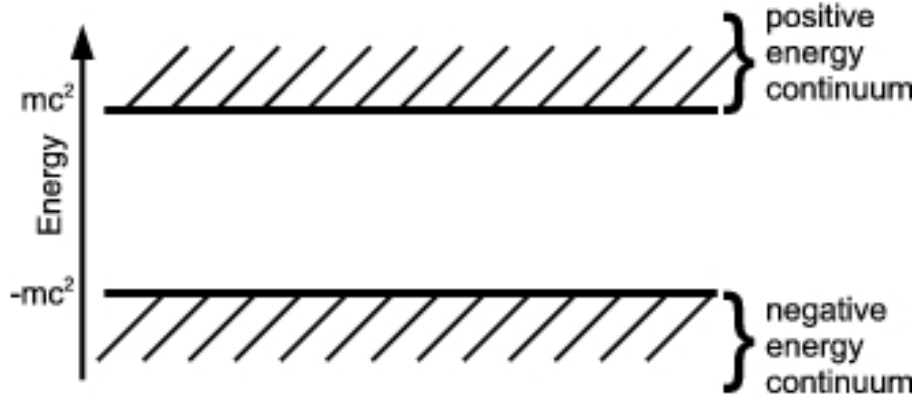}
\caption[Illustration of the free Dirac spectrum]{\spacing{1} \label{DEspec} Illustration of the free Dirac equation spectrum. The upper continuum is similar to what is obtained in the free Schr\"odinger equation (except that the rest energy appears explicitly) but the lower continuum is only found in relativistic theories. The lower continuum is also called the Dirac sea and in hole theory is assumed to be completely filled with electrons preventing the decay of positive-energy particles into these states. The gap between the two continua is 2m$_{\rm e}$c$^2$, the minimum energy required to create a free electron-positron pair. }
\end{figure}

The interpretational difficulties caused by the negative-energy solutions of the Dirac equation are due to the expectation that the relativistic theory of the electron should resemble the non-relativistic theory (the Schr\"odinger equation) and yield a probabilistic single-particle interpretation. The relativistic theory differs since it is valid at high energies allowing particles to be created and destroyed. Therefore, any description of the relativistic electron containing these phenomena will be a multi-particle theory. Figure.~\ref{DEspec} illustrates the spectrum of the potential-free Dirac equation. The negative-energy states form a spectrum similar to the positive-energy continuum, but covering down to $E\rightarrow-\infty$.
The negative-energy states pose the following problem: Positive-energy states could decay to negative-energy states continuously, thus emitting an arbitrary amount of energy \cite{greiner}. This is not what is observed. Dirac proposed that the negative-energy states were all filled with electrons (the highest filled energy level is called the Fermi level). The Pauli principle then ensures that no positive-energy state could decay into a (filled) negative-energy state. This restored stability against radiative decay at the expense of losing the single-particle interpretation. Now all solutions to the Dirac equation contain information about the filled negative-energy states (the Dirac sea). They are inherently multi-particle solutions. In the low-energy regime the multi-particle component is small and justifiably ignored. 

\begin{figure}[!ht]\centering
\includegraphics[angle=0,scale=0.3]{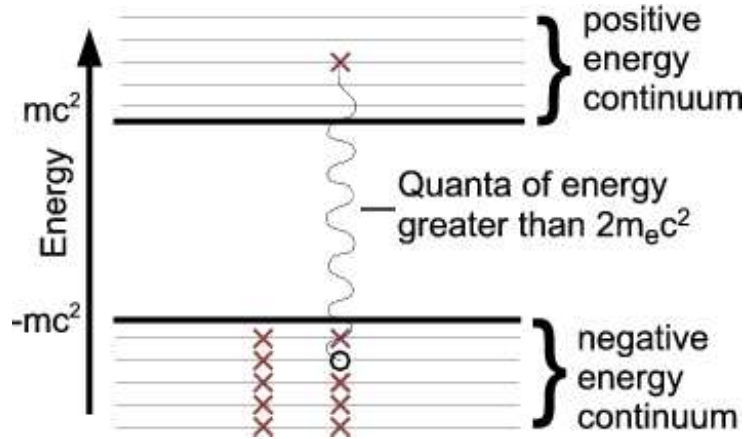}
\caption[Illustration of pair creation]{\spacing{1} \label{DEcreate} Illustration of pair creation in hole-theory. A negative-energy electron receives sufficient energy to jump to a positive-energy state. The electron is then free and leaves behind a hole in the Dirac sea. This hole represents a missing negative charge and acts as a positively charged particle. The hole is the antiparticle of the electron since the electron can then fall back into the hole and fill it leaving the spectrum with no positive-energy electrons.}
\end{figure}

The Dirac sea allows the theory to incorporate the phenomenon of particle creation within so-called hole-theory. It takes into account the creation of particle-hole pairs (which can also be interpreted as particle-antiparticle pairs). In Fig.~\ref{DEcreate}, a negative-energy electron is shown being excited into a state of positive energy. This can be interpreted as the creation of an electron and a positively charged particle. 
The absence of a negative charge in the Dirac sea will appear as a positively charged particle, a hole. The excitation process of a negative-energy electron ($e_{E<0}^-$) to a positive-energy state by a photon ($\gamma$) is written schematically \cite{sakurai} as
\begin{equation}
e_{E<0}^- + \gamma \rightarrow e_{E>0}^-,
\end{equation}
and is reinterpreted in hole-theory as
\begin{equation}
\gamma \rightarrow e_{E>0}^- + e_{E>0}^+. 
\end{equation}
The reverse process, annihilation, where a positive-energy electron falls into an empty negative-energy state filling a hole (vacancy) is written schematically as
\begin{equation}
e_{E>0}^- \rightarrow  e_{E<0}^- + 2\gamma,
\end{equation}
which is reinterpreted in hole-theory as
\begin{equation}
e_{E>0}^- +e_{E>0}^+\rightarrow   2\gamma.
\end{equation}
The positively charged particle with positive-energy, $e_{E>0}^+$, is therefore to be interpreted as a positron since it annihilates the electron. To do this the hole must be viewed as a particle, which means that its dynamical quantities are those of a hole with positive-energy. For instance, the spin of the positron must be the opposite of the missing negative-energy electron, since a missing spin-up electron would behave as a spin-down positron and vice versa. The momentum must also change sign. The absence of momentum $\bf p$ in the Dirac sea will appear as the presence of a particle of momentum $-{\bf p}$. In this way, hole theory allows for the Dirac equation to describe physical situations in which the number of particles is not conserved making it much more useful than a single-particle theory. Dirac predicted the existence of the positron using his equation, despite initially not being taken too seriously as demonstrated by this quote from W. Pauli's \textit{Handbuch} article in Ref.~\cite{sakurai}:
\begin{quotation}
Recently Dirac attempted the explanation, already discussed by Oppenheimer, of identifying the holes with anti-electrons, particles of charge +$|e|$ and the electron mass. Likewise, in addition to protons there must be antiprotons. The experimental absence of such particles is then traced back to a special initial state in which only one of the two kinds of particles is present. We see that this already appears to be unsatisfactory because the laws of nature in this theory with respect to electrons and antielectrons are exactly symmetrical. Thus $\gamma$-ray photons (at least two in order to satisfy the laws of conservation of energy and momentum) must be able to transform, by themselves, into an electron and an antielectron. We do not believe, therefore, that this explanation can be seriously considered.
\end{quotation}
This article ended up appearing in print after C. D. Anderson had already verified the existence of the positron after analyzing particle tracks and looking for a particle with the electron's mass and opposite charge \cite{PhysRev.43.491}.

\section{The Dirac equation with a potential}
To discuss more than free particles a potential should be added to the Dirac equation and there are some interesting consequences with respect to the negative-energy states. The relativistic covariance of the Dirac equation has the feature that a scalar potential will affect the negative- and positive-energy states identically. For details of the behavior of the Dirac equation with a scalar potential the reader is referred to Ref.~\cite{greiner}.

A vector potential of relevance to the present work, is the electrostatic Coulomb potential. The Coulomb interaction is the time-component of the four-vector potential in the Coulomb gauge. It is written in natural units ($\hbar=$ m$_{\rm e}=$ c $=1$) as
\begin{equation}
V(r)=-\frac{Z\alpha}{r},
\end{equation}
where $Z$ is the charge and $\alpha$ is the fine-structure constant. The Dirac equation with an attractive Coulomb potential, the relativistic hydrogen atom, represents a major success in the spectroscopy of atomic hydrogen. It accounts for the corrections due to $L \cdot S$ coupling, i.e., orbital angular momentum coupling to the intrinsic spin angular momentum of the electron. This is due to the spin being a fundamental piece of the Dirac equation. As an aside, we mention that high-precision spectroscopy revealed an inadequacy of the equation as early as 1947: the Dirac equation predicts degeneracies in energy levels (e.g. 2$S_{\frac{1}{2}}$ vs 2$P_{\frac{1}{2}}$) which are only approximate in nature. The 2$P_{\frac{1}{2}}$ state can decay radiatively to the ground state by one-photon emission, and thus becomes a resonance in quantum electrodynamics. In the Dirac theory all excited states are infinity long lived (exact energy eigenstates).

\begin{figure}[!ht]\centering
\includegraphics[angle=0,scale=0.71]{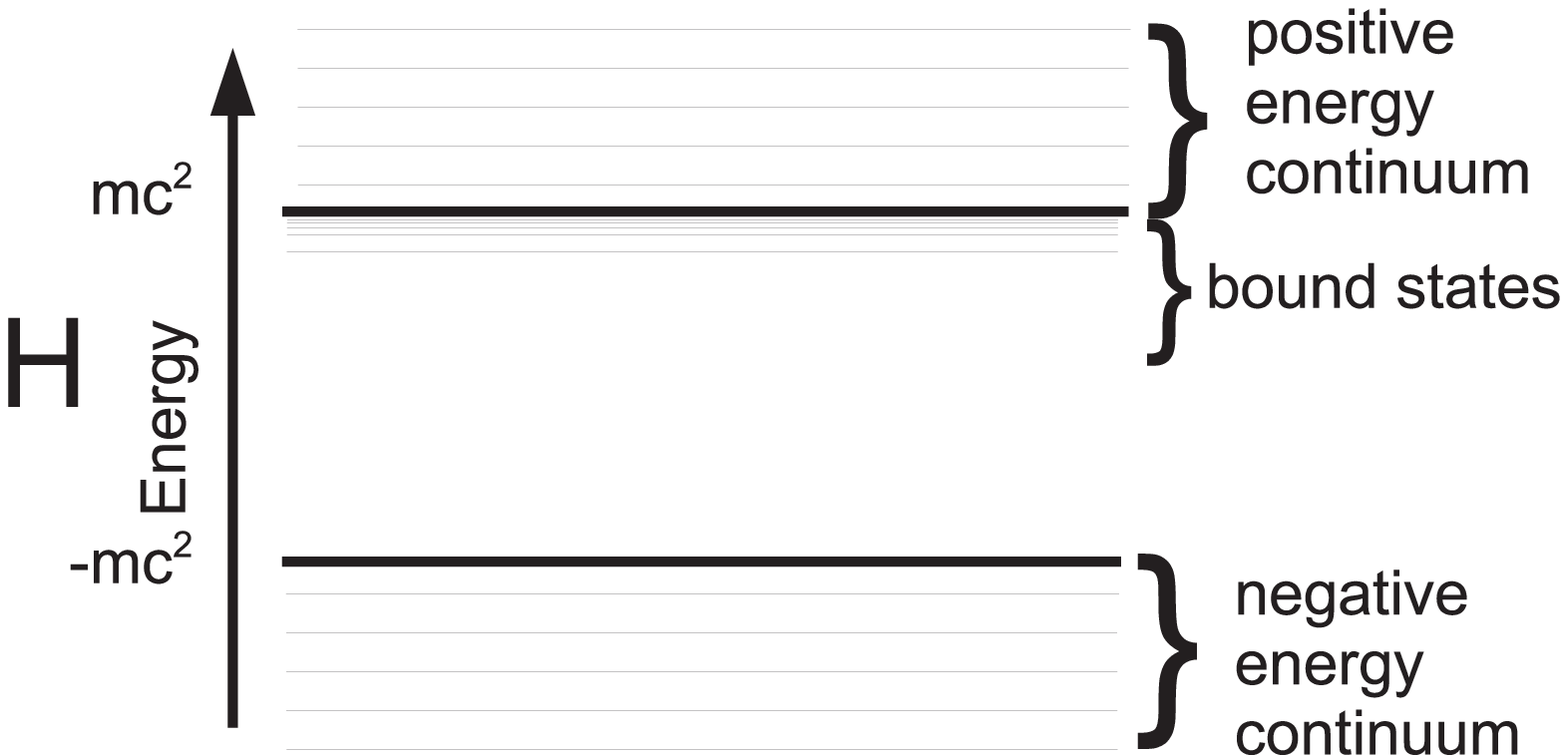}
\includegraphics[angle=0,scale=0.71]{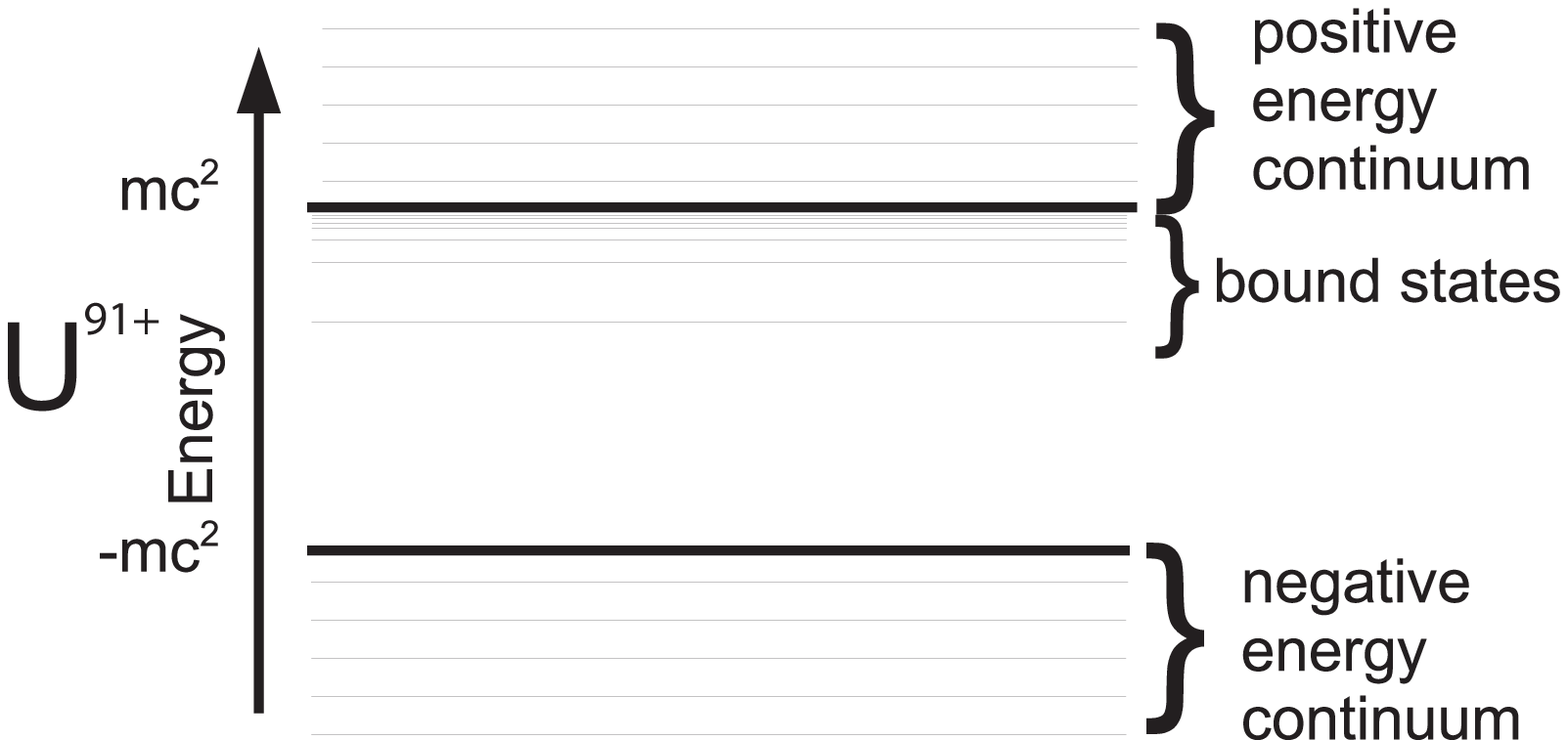}
\caption[Hydrogen and hydrogen-like uranium spectrums]{\spacing{1} \label{DEH} Illustration of the hydrogen spectrum for hydrogen (left) and hydrogen-like uranium (right) with neither to scale. The lowest bound state, the ground state, is lower for hydrogen-like uranium than for hydrogen. Thus it takes more energy to ionize the ground state electron bound to a uranium nucleus than to a proton. }
\end{figure}

A key difference between scalar and vector potentials is that a vector potential acts differently for positive- and negative-energy states. The relativistic hydrogen atom has the same continuum spectrum as the potential-free Dirac equation with added states in the gap just below the positive continuum shown in Fig.~\ref{DEH}. 
These states are called bound states since they are localized in coordinate-space. Also depicted in Fig.~\ref{DEH} is the spectrum of hydrogen-like uranium (U$^{91+}$), although not to scale. The state closest to the negative-energy continuum is the ground state since it is the lowest positive-energy state. It is called the 1S state in atomic systems (where the digit stands for the principle quantum number ranging from 1 to infinity and the letter stands for the orbital angular momentum with S representing no orbital angular momentum or a spherically symmetric state \cite{liboff}). As the strength of the central binding potential (such as the Coulomb potential of a nucleus) becomes stronger the ground state, as well as the other bound states, will sink deeper into the gap. This explains why it takes more energy to excite the ground-state electron to the continuum in hydrogen-like uranium (compared to hydrogen). In the Schr\"odinger theory this energy scales like $Z^2$.

Solving the Dirac equation for the Coulomb potential, as found in Ref.~\cite{sakurai} or \cite{greiner}, one obtains the Sommerfeld formula for the energy of the bound states,
\begin{equation}\label{sommerfeld}
E=\frac{1}{\sqrt{1+\frac{(Z \alpha)^2}{(n_r+\sqrt{(j+\frac{1}{2})^2-(Z \alpha)^2})^2} }},
\end{equation}
where $Z$ is the charge of the nucleus, $n_r$ is the relativistic principal quantum number ranging from 0 to infinity (where the principle quantum numbers are related by $n=n_r+(j+\frac{1}{2})$), $\alpha$ is the fine-structure constant and $j$ is the total angular momentum quantum number and is equal to $\frac{1}{2}$ for S-states. As $Z$ increases, the denominator in Eq.~\ref{sommerfeld} increases. A state with the same $n_r$ and $j$ will therefore have a lower energy for higher $Z$. A lower energy eigenvalue means the electron will be more tightly bound to the nucleus. Figure~\ref{DEH} shows this difference for $Z=1$ (hydrogen) and $Z=92$ (hydrogen-like uranium).

While uranium-238 is already one of the heaviest stable nuclei, the following question may be posed: What are the consequences of having a (hypothetical) nucleus with higher charge $Z$? A pure Coulomb potential originates from a point source and the Dirac equation for this case does not yield solutions for a charge larger than $Z>137$ (for $Z\lsim 137$ one obtains a ground-state energy very close to the center of the gap, i.e. $E_{\rm 1S}\approx 0$).

The pure Coulomb potential is unrealistic for heavy nuclei. A more accurate model of a nucleus involves a finite charge distribution (such as a homogeneously charged sphere) which allows solutions for any $Z$ value. For very large $Z$ ($Z \gsim 170$) the ground state energy resides in the negative-energy continuum leading to interesting effects discussed in the next section. Unfortunately, there is no evidence for the existence of such very heavy nucleus ($Z>170$) in nature. The effects are mimicked in fully-ionized atomic collisions when two nuclei are nearly touching. We, therefore, will use the model of a single heavy nuclei as a simple explanatory tool, but look for the effects in the collisions of heavy, fully-ionized atoms. 

\subsection{Supercritical ground states}\label{scgs}
When considering the difference between the two spectra in Fig.~\ref{DEH} and examining the Sommerfeld formula (cf. Eq.~\ref{sommerfeld}), one can see that increasing the charge of the nucleus will result in deeper bound states. The ground state is always the most affected and decreases fastest in energy. Eventually, as the charge of the nucleus is increased beyond about 170 (depending on the model of the nucleus used), the ground-state energy will reside in the negative-energy continuum. An electron bound in such a state would require a photon with sufficient energy to create an electron-positron pair ($E>$ 2m$_{\rm e}$c$^2$) to ionize the electron. If this state was empty it would be possible for one of the negative-energy electrons, with $E \approx E_{\rm 1S}$, to fill the ground state. This is because the empty ground state introduces a hole (which is initially ``bound" to the nucleus) into the negative-energy continuum, since unlike the negative-energy states, this ground state is still a partially localized state. Any electron which occupies it will remain very tightly bound since it is Pauli-blocked from tunneling out to the negative energy states. Once a negative-energy electron fills the ground state the hole will then be in the negative continuum. This hole will escape from the nucleus since it acts as a positively charged particle and is repelled by the highly charged positive nucleus. This situation can be reinterpreted as pair production, since the previously fully-ionized nucleus now has an electron in the ground state and a free positron has escaped (this is commonly referred to as bound-free pair production, since the electron is bound and the positron is free in the final state). The hole (positron) will have an energy very close to that of the ground state since it is most likely that the negative-energy electron which filled the ground state has the same energy. 

It is also possible for the potential to be even stronger causing the ground state to become more deeply embedded in the negative-energy continuum. This increased potential will also lower the energy of the other bound states. The 2$P_{\frac{1}{2}}$, the first excited state, becomes supercritical at $Z\gsim 180$ \cite{rafelski}. This will result in other states, besides the ground state, becoming supercritical. These states will behave as already described and, therefore, although the description given is for a supercritical ground state they equally apply to any bound state which becomes supercritical.

The spontaneous pair production due to the potential of such heavy nuclei is often referred to as charged-vacuum decay \cite{rafelski}. This is because the vacuum itself is unstable and decays by spontaneous pair production. This is reminiscent of the Klein-paradox (where a step potential of $V_0>2$m$_{\rm e}$c$^2$ shows reflection and transition of an impinging relativistic wavepacket) and a feature of the multi-particle nature of the Dirac equation with very strong potentials \cite{krekoraklien}. While such large potentials, so called supercritical potentials, are not available from single nuclei, colliding nuclei can come into close contact and for a short time create a potential large enough for charged-vacuum decay to take place. It may also soon become possible for an intense laser pulse to add to the potential of a nucleus which would also allow for charged-vacuum decay.

\section{Resonances and supercritical resonances}\label{resex}
\begin{figure}[!ht]\centering
\includegraphics[angle=270,scale=0.28]{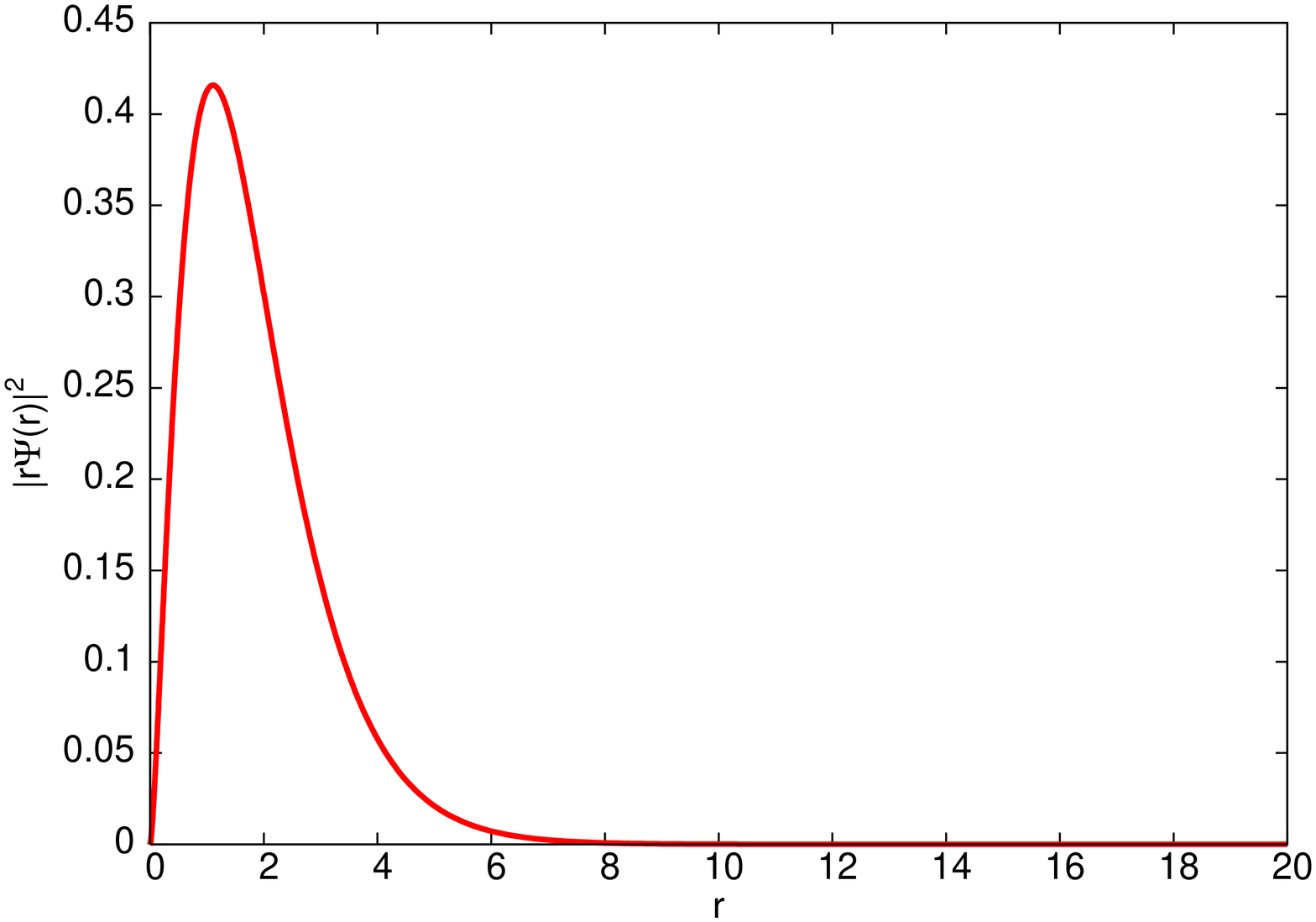}
\includegraphics[angle=270,scale=0.28]{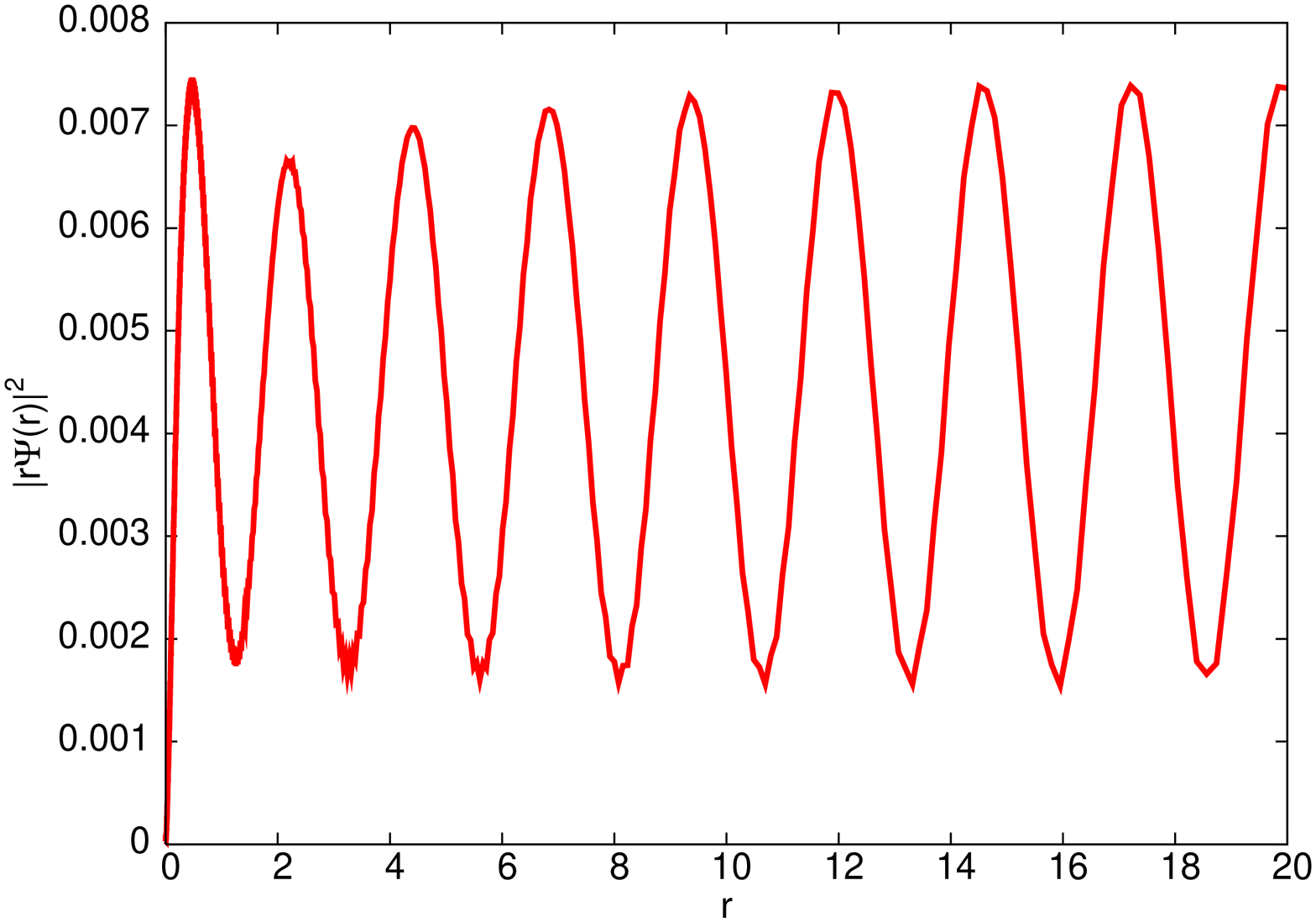}
\caption[$|r\Psi(r)|^2$ for the 1S and an $E=1.5$m$_{\rm e}$c$^2$ state of $U^{91+}$]{\spacing{1} \label{bdcontwf} The left plot is the radial probability density ($|r\Psi(r)|^2$) of the uranium ground state, 1S$_{\frac{1}{2}}$, as function of the radial coordinate (units $\hbar/($m$_{\rm e}$c$) \approx 386$fm). The right plot is the radial probability density of continuum state with $E=1.5$m$_{\rm e}$c$^2$. The bound state is localized since the probability density is concentrated near the origin (the nucleus) and decays exponentially with increasing radial distance. The continuum state oscillates for all distances since it characterizes a non-local state. }
\end{figure}

When the energy of a bound state is in the negative-energy continuum, becoming a supercritical resonance state, its wavefunction will change its character. A bound state is characterized by a wavefunction which is localized around the nucleus and decays exponentially at larger distances. A continuum state is non-localized and always oscillating. Both types of states are shown in Fig.~\ref{bdcontwf} with the bound state being the ground state of uranium and the continuum state $E=1.5$m$_{\rm e}$c$^2$. 
When a state becomes supercritical it takes on the characteristics of both a bound and a continuum state and is called a supercritical resonance.

\subsection{Resonances}\label{resonancessec}
\begin{figure}[ht]\centering
\includegraphics[scale=0.9]{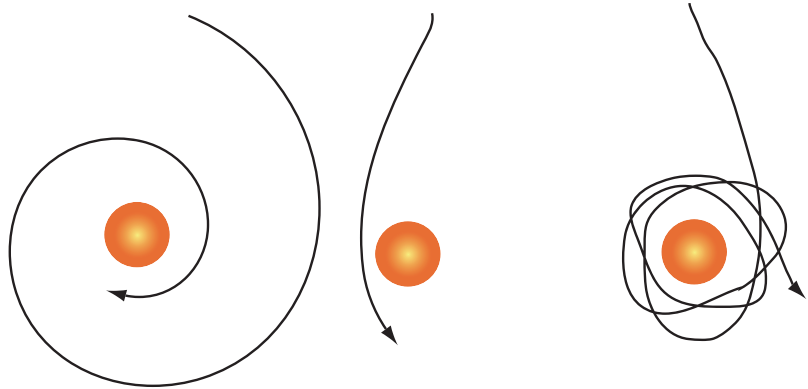}
\caption[Illustration of possible scattering states]{\spacing{1}\label{resdef} The three possibilities of scattering two oppositely charged particles. The left hand illustration represents the projectile becoming bound. The middle illustration shows the projectile being deflected and the third is a mixture of the two and called a resonant state.}\end{figure}

A resonance state in the current context is defined as: a scattering system which has sufficient energy to break up into two or more subsystems, if it is long-lived compared to the collision time \cite{moiseyevrep,reinhardtcs}. Illustrated in Fig.~\ref{resdef} are the three possibilities of the scattering of two oppositely charged bodies. 
The first situation results when the projectile does not have sufficient energy to escape and becomes bound to the target. The second possibility occurs when the projectile is simply scattered by the target, and the third is a mixture of the two. In the last case the projectile becomes bound for a time which is long compared to the collision time and then scatters away. The lifetime of this quasi-bound state is defined by the amount of time the system will stay bound before having a 63\% probability ($1-e^{-1}$) of decaying into two separate parts (projectile and target). 

\begin{figure}[ht]\centering
\includegraphics[width=10cm]{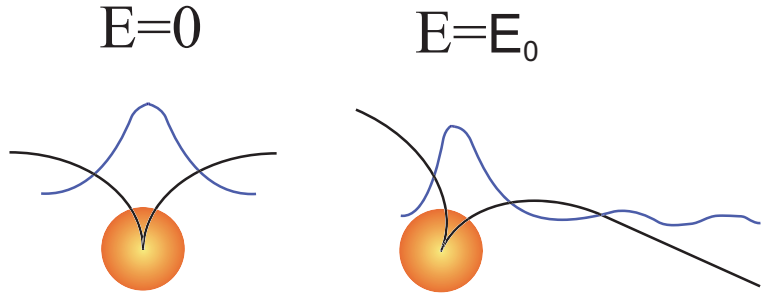}
\caption[Illustration of an atomic resonance]{\spacing{1}\label{hhg} Illustration of the radial Coulomb potential of an unperturbed atom (left) and an atom in the presence of a strong external electric field (right). The potential (along some axis) is the black line and the nucleus is depicted as a solid sphere. The ground state wavefunction is shown in blue with it being a bound state when the electric field is off and a resonance state when the field is on.} \end{figure}

When scattering electrons with different energies from a target with a resonance at some energy, $E_{\rm res}$, a peak will show up in the scattering cross section, $\sigma(E)$ at $E\approx E_{\rm res}$. This peak will have a width, $\Gamma$, which is inversely proportional to the lifetime, $\tau$, of the resonance by $\tau=\hbar/\Gamma$. It will take the shape of a Breit-Wigner distribution (cf. Eq.~\ref{bweqn}). The narrower the peak the longer the projectile electron will stay bound to the target before scattering away. A common example in atomic physics, used in explaining high-harmonic generation, is the resonance formed when the Coulomb potential of an atom is perturbed by a strong electric field (from a laser for instance), illustrated in Fig.~\ref{hhg}.
Electrons scattered off an atom with the external electric field will resonate if their energy is of the order $\Gamma/2$ away from $E_{\rm res}$ or less. This can be thought of as the projectile electron tunneling through the potential barrier to the bound region, remaining in the quasi-bound state for a time and then tunneling back out. Resonance states are therefore different from both a bound or regular continuum state and their time-independent wavefunctions reflect this by being a mixture of the two. A resonance state wavefunction has a bound part which is localized around the potential well (deepest part of the potential), but instead of exponentially decaying the tail oscillates as a continuum state as shown in the right side of the illustration in Fig.~\ref{hhg}. The tail gives some amplitude to the resonance state away from the nucleus where the continuum states have their amplitude. The matrix elements (which determine the connection between different states by $\langle\Psi_1(r)|f(r)|\Psi_2(r)\rangle =\int\Psi_1^*(r) f(r) \Psi_2(r)$ for some coupling $f(r)$ such as an external static electric field) between the resonance state and the continuum states will thus be larger than continuum-bound matrix elements. In this way, the continuum (or scattering) states are connected to the bound part of the resonance state. 

\subsection{Supercritical resonances}\label{spresd}
Supercritical resonance states have the same form for their wavefunctions, but the interpretation differs. In the case of a single supercritical nucleus a ground-state vacancy can decay into a bound electron and a free positron. This takes place despite there being no externally scattered particles. The situation may instead be viewed as the scattering of the negative-energy electrons from the supercritical vacancy state. The behavior is then similar to electrons scattering off an atom in an external electric field. The supercritical resonance has a lifetime and the resonance peak will be at the energy of the supercritical bound state. The interpretation is different, since in the case of supercritical resonances the outgoing state will be a bound-free pair from an initial vacuum state.

The change from bound state to supercritical resonance state results in the exponentially decaying tail being replaced with an oscillating one. The lifetime of the \textit{vacancy} thus becomes finite. When the supercritical resonance state is occupied it is stable. This is due to Pauli-blocking which prevents any tunneling to the (filled) negative-energy continuum. As expected, this atom is stable just as a subcritical atom (an atom which has no supercritical states). The lifetime refers only to an unoccupied supercritical state which then decays by pair creation. This is analogous to the atom in an external electric field since if the resonance state is filled other electrons cannot tunnel in (until the resonance electron tunnels out).

\subsection{Cause of the supercritical resonance}\label{causesc}
An atom perturbed by an external electric field creates a resonance state by modification of the pure Coulomb potential. This creates a finite-size barrier, depicted in Fig.~\ref{hhg}. Three regions result: a bound region near the atom, a region where the potential barrier is high, and an external region where the potential is weak. The region where the potential is high but of finite extent is sometimes called the barrier region, since this is where the electron can tunnel into or out of the resonance state. It is also the region where the wavefunction crosses the border from being a bound state to an oscillating continuum-type state. In most situations in atomic physics where resonances occur it is due to a potential barrier being formed by the mixture of two or more potentials. 

\begin{figure}[H]\centering
\includegraphics[scale=1.2]{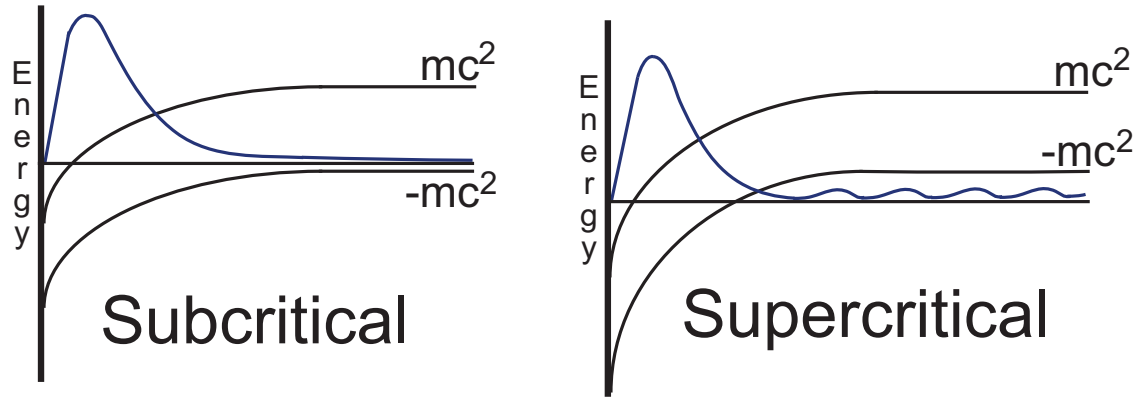}
\caption[Illustration of subcritical and supercritical ground states]{\spacing{1}\label{coulbar} In black the continuum limits, given by $V(r)\pm$m$_{\rm e}$c$^2$, are shown as a function of the radial distance. The probability density, $|r\Psi(r)|^2$, of the ground state for each situation is sketched in blue and displaced vertically using its energy as a baseline. The subcritical ground state has its tail in the gap and thus decays exponentially. The supercritical ground state is more deeply bound and its tail resides in the negative-energy continuum allowing it to oscillate.} \end{figure}

In the case of supercritical resonances no other potential is present except for the modified Coulomb potential (called modified due to the use of a finite-size nucleus). It is not the finite nucleus which is the cause of the potential barrier, since the transition from within the nucleus to outside the nucleus must be continuous. Instead, it is, in fact, the gap between the positive and negative continuum states which acts as the tunneling region and is illustrated in Fig.~\ref{coulbar} for a subcritical and supercritical ground state. 

The illustration shows the ground state vertically displaced by its energy and the energy spectrum is given as a function of the radial distance perturbed by the Coulomb potential. Near the origin the spectrum is highly perturbed but further away less so. In the case of a subcritical potential the tail of the wavefunction is completely contained within the gap between the positive and negative continua. When the potential is supercritical the spectra are further perturbed causing the binding energy of the ground state to increase. This results in the ground state's lobe compressing towards the origin. It also decreases the energy of the state, lowering it in the spectrum. Consequently the tail of the ground state ends up inside the negative-energy continuum. The tail of the ground state can therefore oscillate since it is not in the gap.

The ground state now has the characteristics of a resonance state. The resonance state is, therefore, the result of having two continua separated by a gap and a vector potential, not by the interplay of two separate potentials.

\section{Accessing supercritical resonances}
Stable nuclei with sufficient charge to have a supercritical ground state are unlikely to exist. If they could be created experimentally, their decay time would very likely be much shorter than the decay time of the supercritical ground state which is found to be on the order of 100 zepto-seconds (100$\times10^{-21}$s). Supercritical resonances are predicted by the Dirac equation making them a fundamental prediction of the relativistic theory of the electron. Therefore, other avenues must be sought to verify this prediction. Currently, the most promising approach is to use fully ionized atomic collisions (see section~\ref{prevsearch} for previous attempts).

Atomic collisions with fully ionized heavy atoms yield a supercritical ground state when the nuclei are very close, since the total potential can become supercritical. To visualize the situation it is helpful to use the quasi-molecular picture where the two nuclei are considered to form a slowly varying quasi-molecule with changing internuclear separation. The system then has a single (molecular) spectrum. The potential of the two nuclei is that of two displaced and moving modified Coulomb potentials. At any instant in time, the ground state is then a quasi-molecular ground state (called the 1S$\sigma$ where the additional $\sigma$ states the radial wavefunction is symmetric) which covers both nuclei. While the potential is no longer spherically symmetric or time-independent, it becomes experimentally accessible and yields supercritical resonance states that may be detectable.

\subsection{Dynamical and resonant pair production}\label{dynsppair}
In fully-ionized collisions with a supercritical resonance, two types of pair creation will be present. The first type is due to the spontaneous decay of the supercritical resonance, as has been described in the previous section. These electron-positron pairs are not directly due to the dynamics of the collision. The free positrons (and 1S$\sigma$ electrons) are created due to the instability of the vacuum in the presence of an intense field; within Dirac hole theory the process is represented by the decay of the supercritical resonance state. The process is, therefore, called resonant or spontaneous pair creation. The other type of electron-positron pair creation is what is normally observed in heavy-ion collisions. It is due to the changing potential caused by the moving nuclei. It is referred to as dynamical pair creation. 

\begin{figure}[H]\centering
\includegraphics[scale=0.85]{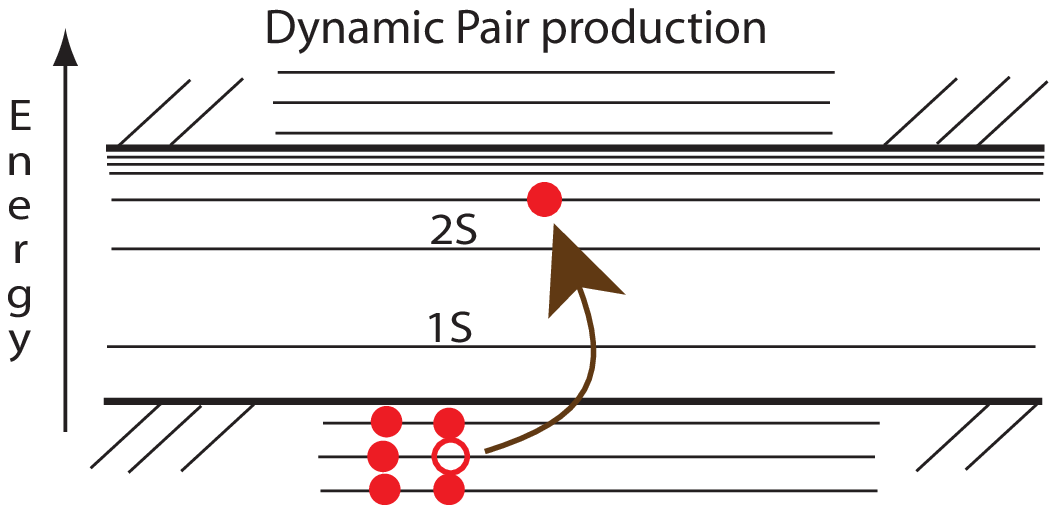}
\includegraphics[scale=0.85]{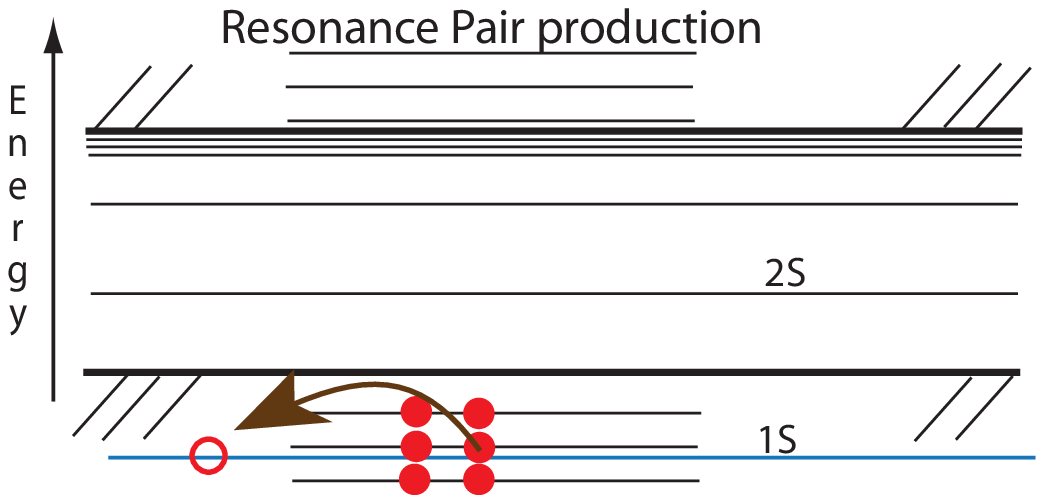}
\caption[Illustration of dynamical and resonant pair creation]{\spacing{1}\label{creats} Illustrative explanation of dynamical pair production (left) and resonant pair production (right). Only a few of the infinite continuum states are shown. The circles represent holes while the filled circles represent electrons. Dynamical pair production occurs when a negative-energy electron absorbs some of the kinetic energy from the motion of the nuclei and is excited to a positive-energy state leaving a vacancy behind. Resonance pair production occurs when the vacant ground state joins the negative-energy continuum which allows a negative-energy electron with $E\approx E_{\rm res}$ to tunnel in and become bound to the nucleus while the vacancy, now in a negative-energy continuum state, escapes.} \end{figure}

Hole theory provides a useful picture of the situations as illustrated in Fig.~\ref{creats}.
The left picture shows dynamical pair creation. In hole theory, this amounts to a negative-energy electron absorbing some of the colliding nuclei's kinetic energy  and jumping into a positive-energy state. The electron leaves a hole in the negative-energy continuum representing a free positron. The electron can jump to a bound state creating a bound-free pair or, with sufficient energy, to a continuum state representing free-free electron-positron pair production. The energy for this transition comes from the kinetic energy of the collision. Since heavy-ion collisions are generally performed at center-of-mass energies well above 200MeV, the 1-2MeV required for this process is small , i.e., it is readily available. In collisions where partially ionized atoms are used the lower bound-free channels require the inner-shell electrons to be excited or else the channel is Pauli-blocked, severely suppressing the channel. 

As the nuclei approach each other, the quasi-molecular ground state of the system is lowered. A negative-energy electron would require less energy to be excited to a bound state. This is why bound-free pair production, or most importantly ground state-free pair production, is the dominant channel. In cases of collisions which involve supercritical resonances, dynamical pair creation is further enhanced. This is again due to the smaller amount of energy required to excite negative-energy states into the (now lower) bound states and into the supercritical ground state. Energetically it may be expected that the negative-energy states closest to the gap would be those most likely to be excited by the collision into a positive-energy state, but this is not the case. The continuum states near the gap are states of low momentum. Thus, they cannot react to the collision or connect to the localized, low-lying, bound states during the collision.

The second type of pair production, illustrated in Fig.~\ref{creats}, has been discussed in the previous section, the decay of a supercritical resonance. During the time in which the system has a supercritical ground state (or possibly other states as well) the bound hole can tunnel out into the negative-energy continuum, or equivalently, a negative-energy electron can tunnel into the bound region of the supercritical state. Clearly, negative-energy states of similar energy to the supercritical resonance are the ones which can tunnel into it, and the resulting positrons will be close (on the order of $\Gamma/2$ away) to the energy of the supercritical resonance. This type of pair creation is exclusively bound-free since the electron is bound to the nuclei once it tunnels in from the negative-energy states.

These two types of pair creation, therefore, have significant overlap since the exclusive resonant pair creation adds to the dominant channel for dynamical pair creation, ground state-free pair production. These two effects interfere with each other making the determination of the single cause (dynamic or resonant) of the electron-positron pair production difficult to discern. An important difference is in the positron spectrum. While the electrons will primarily be in the ground state, since this is the dominant channel for dynamical pair creation and the only channel for resonant pair production, the positrons from resonant pair production will be peaked around the energy of the supercritical resonance. This peak will be smeared, since unlike in the static case the supercritical ground state will change due to the nuclear motion. As the nuclei approach, the state's energy will sink deeper in the negative continuum and then come back up again as the nuclei separate. The resonant pair production will then be a smeared version of all the intermediate supercritical resonances. The smearing is not equal though, since the deeper the resonance the more momentum the negative-energy states have resulting in shorter decay times of the state. The shorter decay time implies that the resonant pair production will be dominated by the deepest supercritical resonance, created at the closest approach of the nuclei. 

The supercritical resonance occurring at closest approach for currently experimentally available fully-ionized atoms has a lifetime of the order of $10^{-19}$s. The interaction time in these collisions at closest approach is of the order of $10^{-21}$s, resulting in a very small signal for the resonance production peak. Maximizing the time the system is supercritical is, therefore, of the utmost importance, if the goal is to observe charged vacuum decay.

\subsection{The trajectory}\label{trajectoryintro}
The main objective of this work is to test the prediction of the Dirac equation with regard to supercritical resonances; consequently we are interested in trajectories which try to maximize the time in which the system is supercritical. To this end a Coulomb trajectory is chosen.

\begin{figure}[H]
\fbox{\includegraphics[scale=0.2]{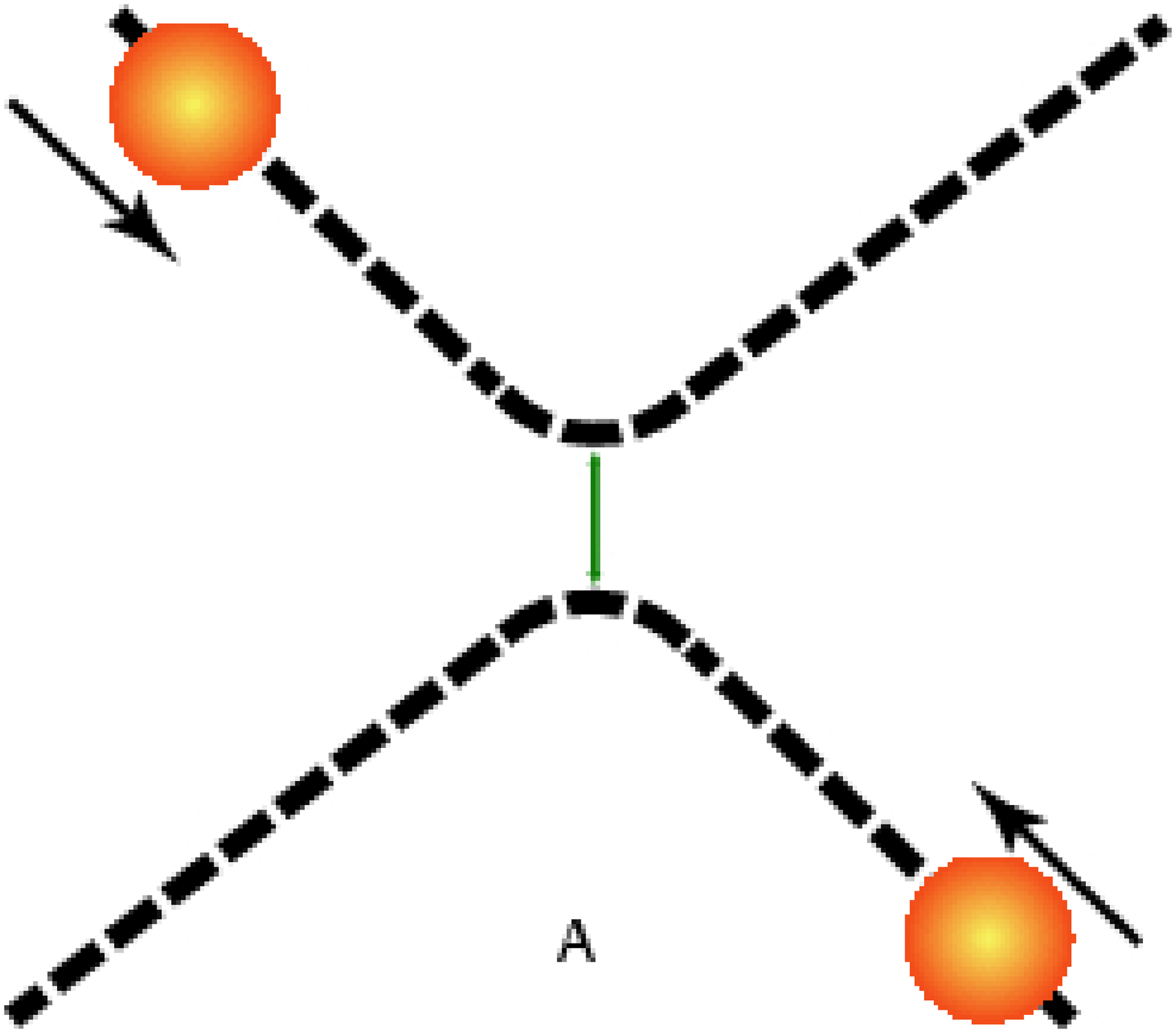}}
\fbox{\includegraphics[scale=0.2]{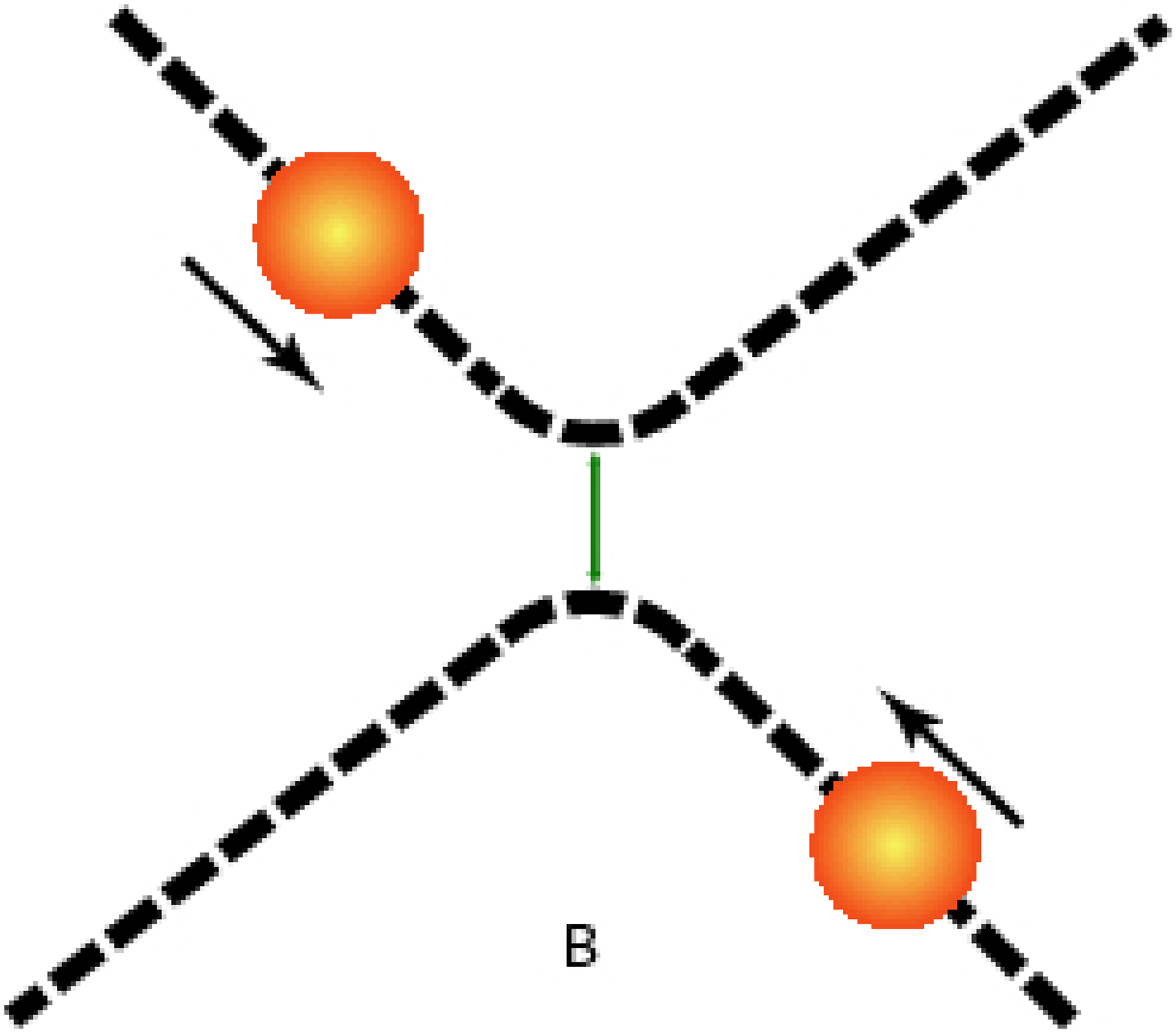}}
\fbox{\includegraphics[scale=0.2]{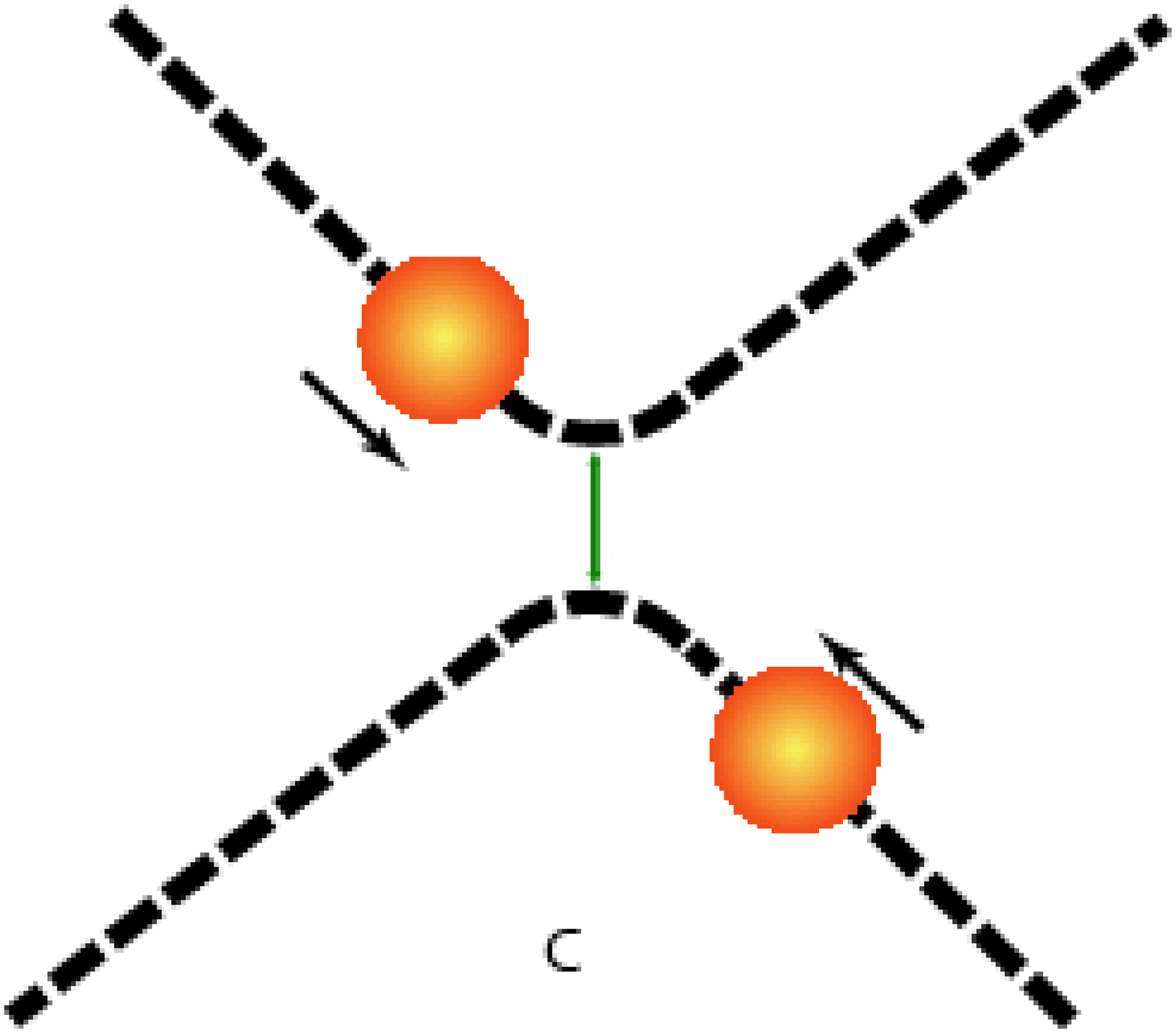}}\\
\fbox{\includegraphics[scale=0.2]{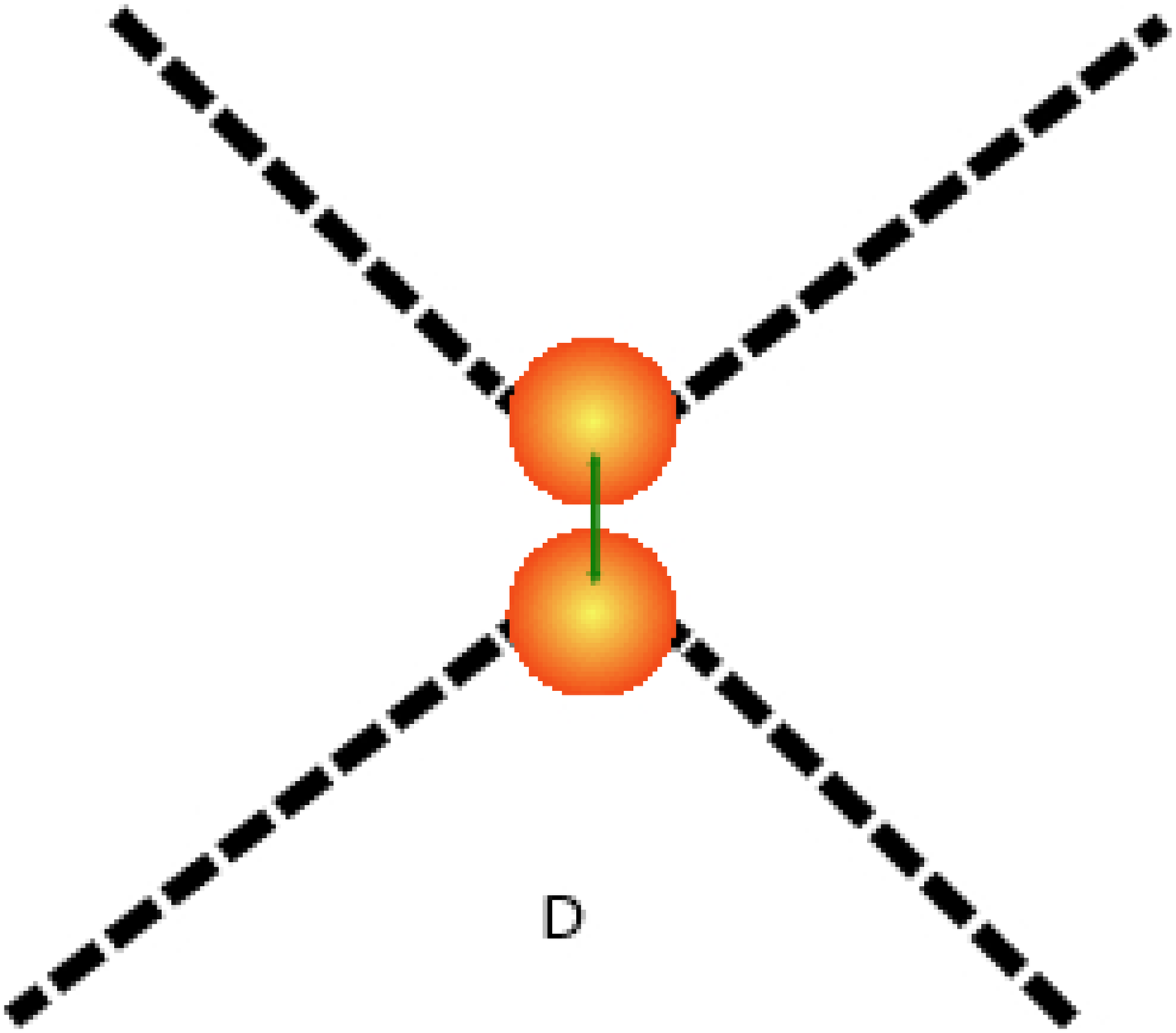}}
\fbox{\includegraphics[scale=0.2]{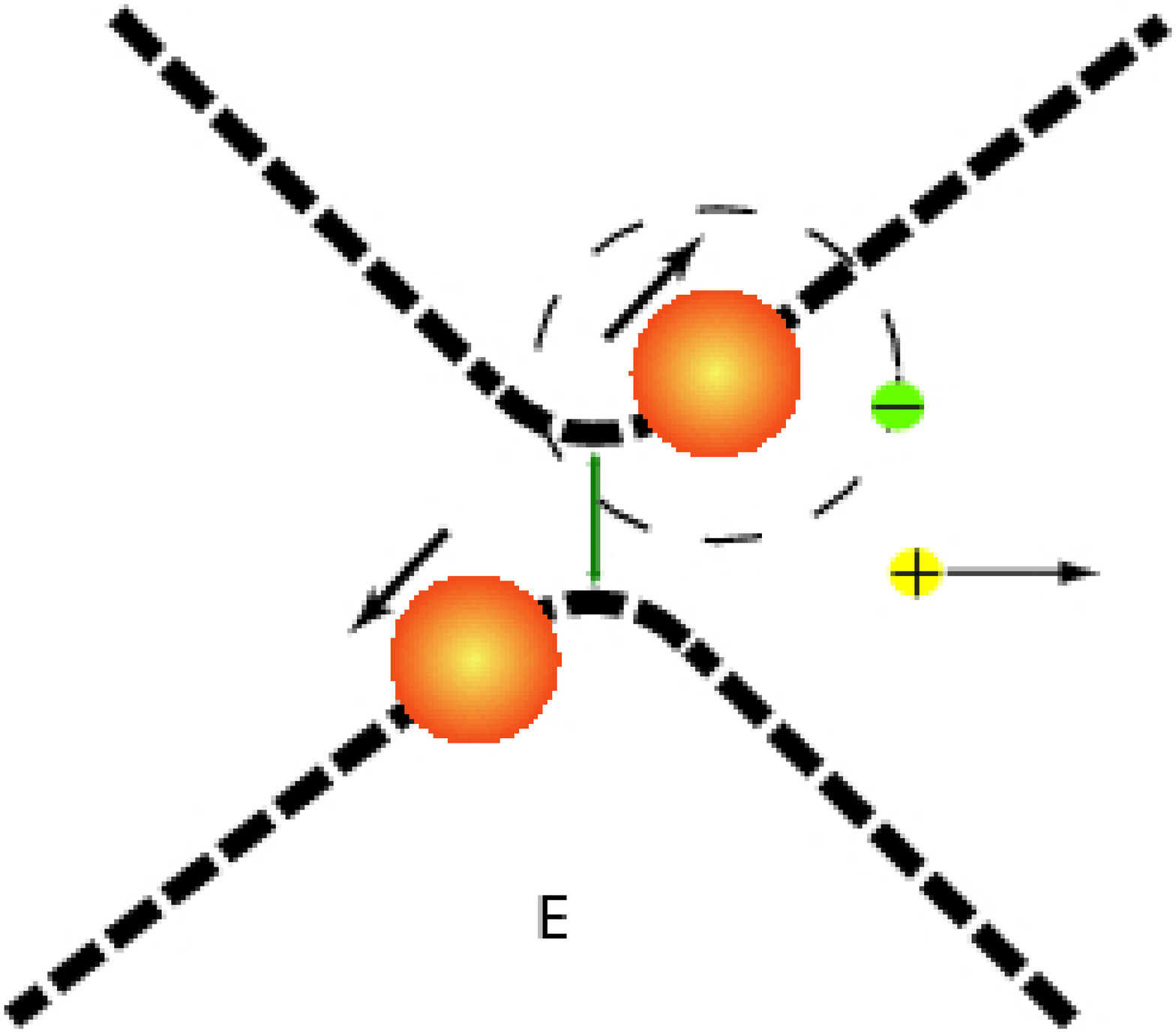}}
\fbox{\includegraphics[scale=0.2]{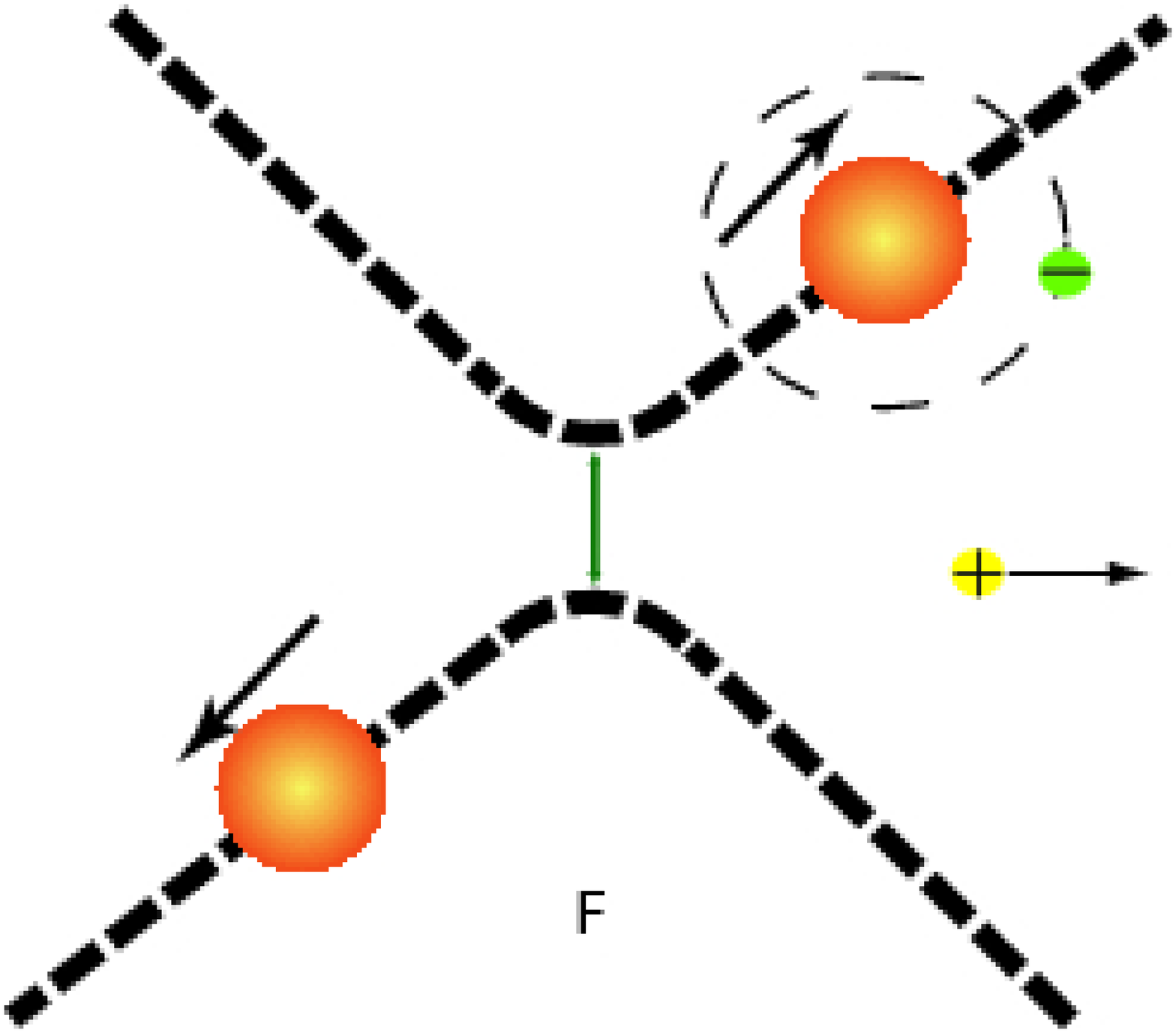}}
\caption[Illustration of a Coulomb trajectory]{\spacing{1}\label{trajil}  An illustration of the Coulomb trajectory for two identical fully-ionized atoms from left to right (using six snapshots). The arrows beside the nuclei represent the velocity showing how the nuclei slow down as they approach due to the Coulomb repulsion. The green arrow shows the distance of closest approach, determined by the collision energy. After reaching closest approach resonant pair creation is depicted with the green electron being bound in the ground state of one of the nuclei and the yellow positron scattering away.}\end{figure}

A Coulomb trajectory for heavy ions involves relatively slow nuclear motion ($v_{\rm max}\approx 0.11$c for bare uranium-uranium) compared with the velocity of a ground state electron. The center-of-mass energy is just enough for the nuclei to slow down as they get closer due to the mutual Coulomb repulsion, and come to a stop at closest approach before being accelerated away. The Coulomb trajectory is obtained by solving the classical force problem due to the repulsion of two nuclei of charges, $Z_1$ and $Z_2$ given by 
\begin{equation}\label{trajeq}
\mu\ddot{r}=-\frac{Z_1 Z_2 \alpha}{r^2}
\end{equation}
in natural units where $\mu$ is the reduced mass of the system. The Coulomb trajectory is illustrated in Fig.~\ref{trajil} where the arrows represent the velocity and show the ions slowing down as they approach each other.
This trajectory offers the advantage that the nuclei are traveling very slowly when the system is supercritical, reducing the dynamical pair production and maximizing the time in which the supercritical resonance has to decay. This differs from a straight-line trajectory where projectiles are of sufficient kinetic energy to have very little deviation due the collision. The Coulomb trajectory in contrast has a much lower center-of-mass energy reducing the amount of dynamical positron production. The ratio of the two pair production channels is therefore optimized since the dynamical production acts effectively as a background to the supercritical signal.

\subsection{The reference frame}
The full Dirac equation is covariant yielding the same results regardless of the reference frame. However, the approximations which are  introduced to solve the collision render the results reference-frame dependent. There are two convenient reference frames for the collision. 

In the target frame, one chooses one of the nuclei, often the heaviest, as the origin. The other nucleus is then treated as a projectile. The projectile's potential is then a perturbation to the modified Coulomb potential of the target. This introduces the time dependence into the system (due to the movement of the projectile). This reference frame has the advantage that the initial states can be taken as simple atomic states since at $t\rightarrow -\infty$ the system is that of a single atom. The final states can also be projected onto the single atomic basis if propagation for $t\rightarrow \infty$ is impractical.

The center-of-mass frame on the other hand treats both nuclei equally, a desirable feature when the nuclei are close and of similar charge. Since the eigenvalues of the states are frame dependent (the eigenvalue is the time component of a four-vector) the center-of-mass frame will have deeper bound states. The supercritical resonance states will also be deeper resulting in larger positron yields. Therefore, care must be taken when choosing the reference frame to ensure meaningful results.

\section{Nuclear sticking}\label{nstick}
Collisions where the nuclei end up within a nuclear radius of the each other result in deformations of the nuclei. This is due to the Coulomb repulsion. There is also the possibility of some effects due to the nuclear force. In deep inelastic heavy-ion collision experiments with lead-uranium, uranium-uranium and uranium-californium collisions, detailed trajectory reconstruction found that the nuclei may have remained in contact for around a zepto-second \cite{PhysRevLett.50.1838}. The causes of this effect are still unclear. Some groups predict that a sticking time of up to 10 zepto-seconds may be achievable \cite{EurPJA.14.191,zagrebaev:031602,Sticking2}. This would give a strong enhancement to the supercritical resonance decay signal due to the increased time for the supercritical resonance to decay (cf. Fig.~\ref{mullerfig}). This would provide strong evidence for resonant pair production as well as long nuclear sticking times.

\section{Modern interpretation of hole theory}
Hole theory has been used in this section to understand the cause of the supercritical resonance and what it is, but a common question arises: what is the validity of hole theory? In the current context hole theory provides a meaningful way to understand the negative-energy states and multi-particle aspects of the theory, but it is only an explanatory tool for the electron-positron field. In the modern view, it has been supplanted by quantum electrodynamics (QED) which deals properly with multiple particles and has only positive-energy states. An example of the failure of hole theory is simple to illustrate using a phenomenon beyond the electron-positron field. Any phenomenon in which the electron or positron numbers are not conserved would suffice. For instance $\beta^+$-decay,
\begin{equation}
p \rightarrow n + e^+ + \nu,
\end{equation}
where a proton, $p$, decays into a neutron, $n$, a positron, $e^+$, and a neutrino, $\nu$. Here a positron was created, a vacancy in hole theory, without an electron being created. This is expected since hole theory is not valid beyond the electromagnetic field interaction and the above process is a weak field decay. 

In the present collision work, the final quantities of interest, discussed in section~\ref{qofint}, are derived from QED as in Ref.~\cite{PhysRevA.45.6296} and in appendix~\ref{appendix2} for a non-orthogonal basis. The quantities necessitate that all initial Dirac wavefunctions be propagated through the collision in order to obtain the desired positron and electron creation spectra in order to search for evidence of a supercritical resonance. Therefore, while hole theory provides some guidance to understanding the supercritical resonance state, it is ultimately QED which directs the search and provides the quantities needed.

\chapter{Time-independent supercritical states}\label{tdscstates}
In single-particle (non-relativistic) quantum mechanics, bound states become resonances when the potential is modified such that the previously bound particle can escape to infinity (e.g., when the linear Stark potential is added to the Coulomb interaction). More generally, a system which has sufficient energy to break up into two or more subsystems is called resonant if it is long lived compared to the collision time \cite{moiseyevrep,reinhardtcs}. Such a state is described by the mean energy position $E_{\mathrm{res}}$ (which is usually shifted from the bound-state eigenenergy), and by the lifetime $\tau$. The latter provides a measure of the decay time of an initially prepared quasi-bound state which undergoes exponential decay. 

In scattering theory, the total cross section resonates  (reaches the unitarity limit) if the particle energy is close to $E_{\mathrm{res}}$ and can be described by a Breit-Wigner distribution (cf. Eq.~\ref{bweqn}) which is characterized by the energy width parameter $\Gamma$. According to the energy-time uncertainty principle, the temporal behavior of a decaying bound state and the cross section shape are related by $\Gamma \tau  \sim \hbar$. A determination of $\Gamma$ and $E_{\rm res}$ and a fit to the Breit-Wigner shape function normally requires a solution of the scattering problem for many energies in the vicinity of $E_{\mathrm{res}}$.

Resonances embedded in the positive continuum are characterized by $E_{\mathrm{R}}=E_{\mathrm{res}}-i \Gamma / 2$. Supercritical resonance energies however are given by $E_{\mathrm{R}}=E_{\mathrm{res}}+i \Gamma / 2$. This is understood by reinterpreting the supercritical resonance for an electron with negative energy as a positive-energy positron resonance propagating backwards in time by CPT symmetry. The time reversal transformation necessitates a positive imaginary part of the energy eigenvalue to ensure that the state decays as time propagates to negative infinity. This follows since the time-dependent part of the wavefunction is given by 
\begin{equation}
e^{-i E_{\rm R}t/(2\hbar)}=e^{-i E_{\rm res}t/(2\hbar)+\Gamma t/(2\hbar)}.
\end{equation} 
This time dependence then ensures that the total probability in the negative-energy hole state decays exponentially with time scale $\hbar/\Gamma$ as $t\rightarrow -\infty$.

We note that some of the work in chapters~\ref{tdscstates} and \ref{CSCAP} was originally published in \cite{ackadcs}.

\section{A supercritical potential}\label{secpot}
We use a physically relevant model of two extended nuclei separated by distances of the order of a few nuclear radii. The nuclear radius is taken to be $R_{\mathrm{n}}=1.2\sqrt[3]{A}$ fm, where $A$ is the number of nucleons \cite{pra10324}. The nuclei are assumed to be displaced homogeneously charged spheres with a separation of $R$ between the charge centers. In the center-of-mass frame the monopole potential for each of the two nuclei with respective charge $Z_i$ and radius $R^{(i)}$ is given (in units of $\hbar =$ c $ =$ m$_{\mathrm{e}}=1,Z=Z_i,R_{\mathrm{n}}=R_{\mathrm{n}}^{(i)} $) by
\begin{equation}\label{potential}
V(r)= \left\{ \begin{array}{cc}
-\frac{Z\alpha}{r} & \mathrm{for}\;r>r_+ \\

-\frac{2Z\alpha}{R_{\mathrm{n}}^3 R}\left[ \frac{(R/2-R_{\mathrm{n}})^3(R/2+3R_{\mathrm{n}})}{16r}-\frac{r_+^2(R/2-2R_{\mathrm{n}})}{4}+\frac{3(R^2/4-R_{\mathrm{n}}^2)r}{8} - \frac{Rr^2}{8} +\frac{r^3}{16}
\right] & \mathrm{for}\; r_- <r<r_+ \\

-\frac{2Z\alpha}{R} & \mathrm{for}\;r <r_- 
\end{array} \right.
\end{equation}
where $r_{\pm} = R/2\pm R_{\mathrm{n}}$. The expression is obtained by using the potential for a homogeneously charged sphere at the origin and displacing it by $R$ along the z-axis and expanding it into Legendre polynomials,
\begin{equation}\label{potexplain}
\sum_{l=0}{V_l P_l(\cos(\theta)) }=\frac{Z\alpha}{2R_{\rm n}}\left( 3-\frac{|r-R|^2}{R_{\rm n}} \right).
\end{equation}
By inverting the sum in Eq.~\ref{potexplain} using the orthogonality relation of Legendre polynomials, $\int_{-1}^{1}P_m(x)P_n(x)dx=-\frac{2}{2n+1}\delta_{n,m}$, Eq.~\ref{potential} is obtained for $V_0(r)$ for each nuclei.

\section{Supercritical states using the mapped Fourier grid method}
The mapped Fourier grid (MFG) method provides approximate resonance parameters directly by two separate means. These methods also illuminate the behavior of the MFG method when a supercritical state is present, which is particularly important for time-dependent heavy-ion collisions in which supercritical intermediate states strongly influence the time evolution of the matter field.

\subsection{The projection method}
Even though it is unlikely that a discretized continuum eigenstate of the Hamiltonian matrix will fall exactly on the mean energy of the resonance, the closer such a state is to the mean energy the more it resembles the resonance state. The bound (small-distance) part of a supercritical 1S$\sigma$ resonant state is similar to a subcritical 1S$\sigma$ state (although with a somewhat compressed lobe). The inner product of a normalized subcritical and supercritical 1S$\sigma$ equals approximately one. Thus, the projection of a subcritical 1S$\sigma$ onto the eigenbasis of a supercritical hamiltonian quantifies how similar the latter states are to the ground state. Therefore, by taking the inner product of each supercritical eigenstate with the subcritical 1S$\sigma$ state one computes a probability density vs energy, $E_{\nu}$, based upon $|\langle \phi(E_{\nu})|\psi_{1S\sigma}\rangle|^2$. This probability distribution displays a Breit-Wigner shape. By fitting to the Breit-Wigner shape (cf. Eq.~\ref{bweqn}), the parameters of the supercritical resonance ($E_{\mathrm{res}},\Gamma$) are obtained. This method does not need to have any background subtracted since it is flat away from $E_{\rm res}$. The result is, however, somewhat sensitive to the range of energies used in the fit and needs enough states to properly cover the distribution.

\subsection{Supercritical resonance states in the density of states}
A further method from which the supercritical resonance parameters can be estimated is based upon the density of states from a diagonalization of the discretized Dirac Hamiltonian. The density of states (in arbitrary units)
\begin{equation}\label{dosfunc}
\rho\left(\frac{E_{\nu+1}+E_{\nu}}{2}\right)  = \frac{1}{E_{\nu+1}-E_{\nu}}
\end{equation}
is calculated from the quasi-continuum eigenvalues, $E_{\nu}$. The fit requires the approximation to the local background density, and, thus, the results are somewhat sensitive to the form of the fit. Using a single rational term for the background with a Breit-Wigner distribution added the density of states can be fitted by 
\begin{equation}\label{fitfunc}
\rho(E_{\nu}) =  \frac{A_1}{\left|E_{\nu}\right|^q}+A_2\frac{\Gamma/2}{(E_{\mathrm{res}}-E_{\nu})^2+\Gamma^2/4} 
\end{equation}
where $A_1$, $A_2$ and $q$ are fit constants.

\subsection{Results obtained using the mapped Fourier grid method}\label{mfgres}
\begin{figure}[!ht]\centering
\includegraphics[angle=270,scale=0.5]{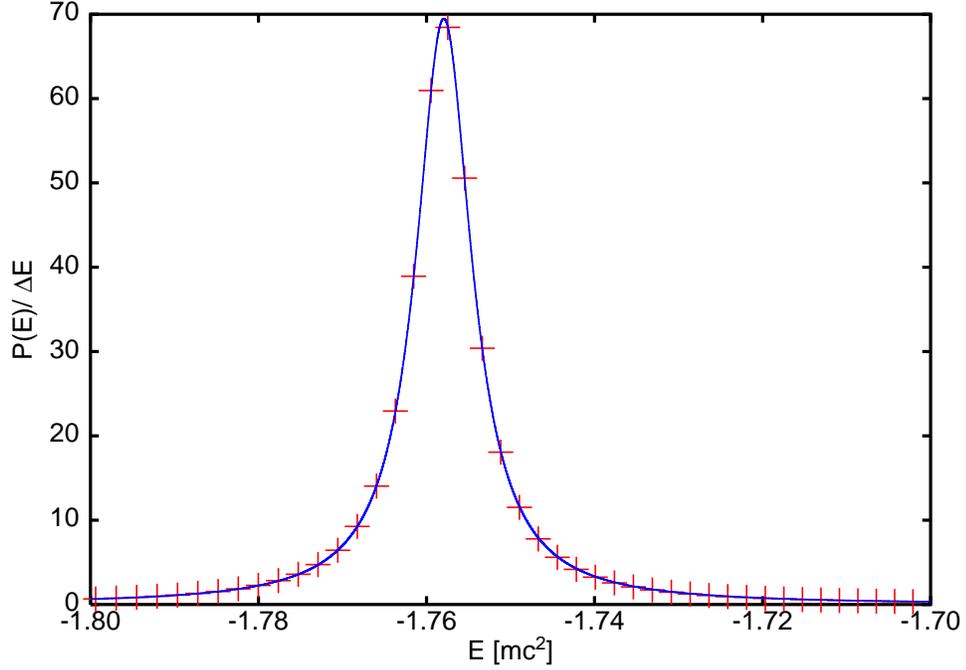}
\caption[Supercritical resonance by the projection method]{\spacing{1}\label{bw} The projections, $2\left|\left\langle \phi(E_\nu) | \psi_{1\mathrm{S}\sigma}\right\rangle \right|^2/(E_{\nu}-E_{\nu -1})$, of a bound $\psi_{1\mathrm{S}\sigma}$ state onto the supercritical U-Cf quasi-continuum states $\phi(E_\nu)$ (calculated for a $N=3900$ and $s=6000$ basis) are shown as crosses (\textcolor{red}{$+$}). The subcritical bound $\psi_{1\mathrm{S}\sigma}$ was obtained for a separation of $R=45$ fm, while the supercritical basis is for $R=20$ fm. The line represents a Breit-Wigner fit to the data with $E_{\mathrm{res}}=-1.75795\,$m$_{\rm e}$c$^2$ and $\Gamma=4.12$ keV for the U-Cf system at $R=20$ fm.}
\end{figure}

The success of the aforementioned methods is demonstrated for the example of the U$^{92+}$-Cf$^{98+}$ quasi-molecular system for an internuclear separation of $R=20$fm where A$_{\rm U}=238$ and A$_{\rm Cf}=251$. The results of $2\left|\left\langle \phi(E_\nu) | \psi_{1\mathrm{S}\sigma}({\rm R}=45 {\rm fm})\right\rangle \right|^2/(E_{\nu}-E_{\nu -1})$ are shown as (red) crosses in Fig.~\ref{bw} (the factor of $2/(E_{\nu}-E_{\nu -1})$ is due to the continuum states being quasi-continuum states and having a width). The $\phi(E_\nu)$ states were obtained by solving the Dirac hamiltonian for the potential in Eq.~\ref{potential} with $R=20$fm. Fitting the results in the range of the resonance to a Breit-Wigner the (blue) line in Fig.~\ref{bw} is obtained with the resonance parameters: $E_{\mathrm{res}}=-1.75795\,$m$_{\rm e}$c$^2$ and $\Gamma=4.12$ keV.

\begin{figure}[!ht]\centering
\includegraphics[angle=270,scale=0.5]{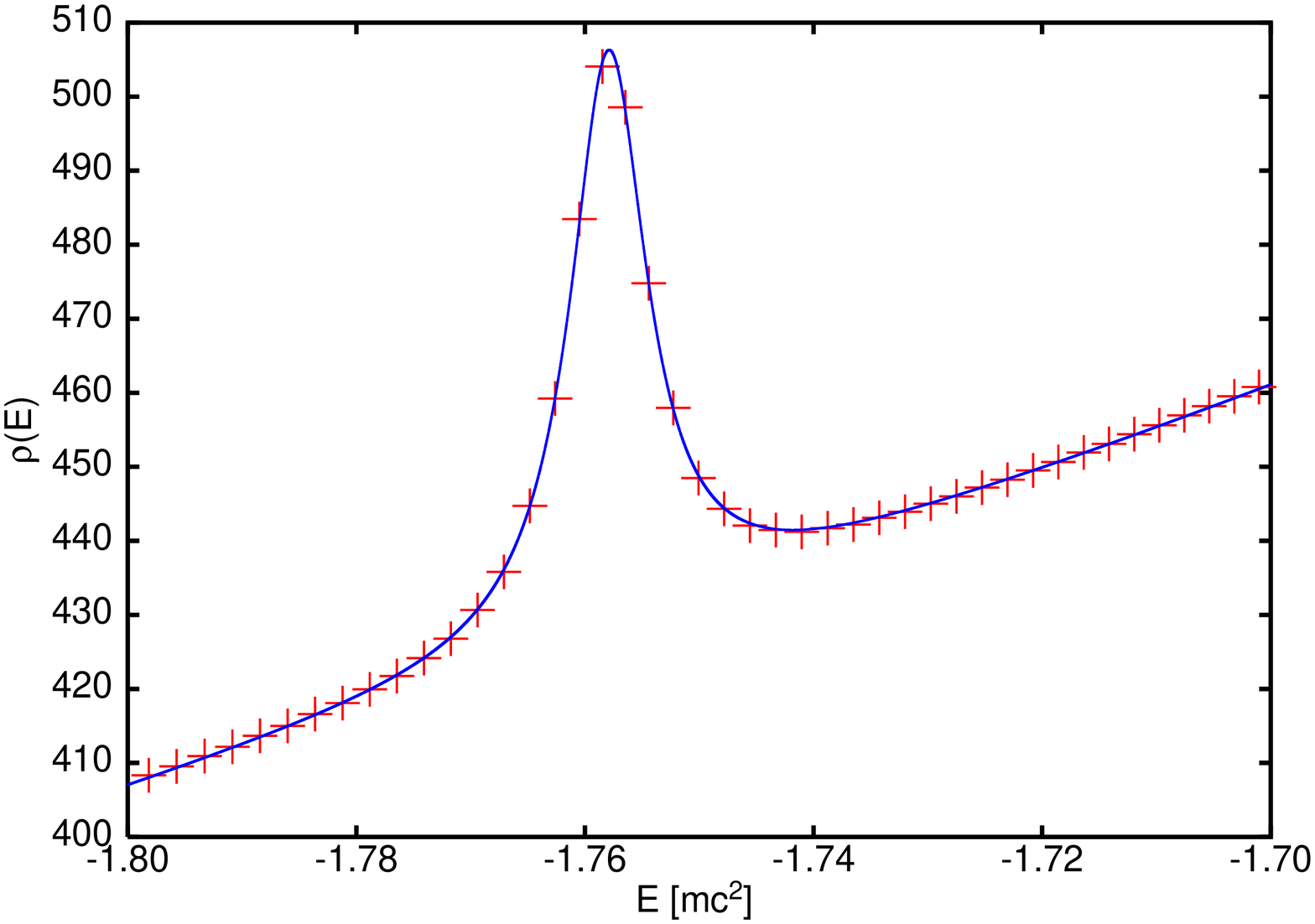}
\caption[Supercritical resonance in the density of states]{\spacing{1}\label{DoS} The density of states Eq.~(\ref{dosfunc}) from a hermitian matrix diagonalization for the U-Cf ($R=20$ fm) Hamiltonian obtained with a basis with $N=3900$ states and $s=6000$, is shown by crosses (\textcolor{red}{$+$}). The line shows a fit with Eq.~(\ref{fitfunc}),
and yields $E_{\mathrm{res}}=-1.75794\, $m$_{\rm e}$c$^2$ and $\Gamma=4.17$ keV.}
\end{figure}

Building the density of states using Eq.~\ref{dosfunc} for the same supercritical system the (red) crosses in Fig.~\ref{DoS} are obtained. Fitting the results with Eq.~\ref{fitfunc} results in the (blue) line. The fit yields $E_{\mathrm{res}}=-1.75794\, $m$_{\rm e}$c$^2$ and $\Gamma=4.17$ keV for the resonance parameters.

These results compare well to those of Ref.~\cite{pra10324} which obtained values of $E_{\rm res}=-1.763$m$_{\rm e}$c$^2$, $\Gamma=4.12$keV using phase-shift analysis and $E_{\rm res}=-1.761$m$_{\rm e}$c$^2$, $\Gamma=4.10$keV using the truncated potential method (described in section~\ref{prevsearch}). The projection method, while sensitive to the energy range used for the fitting, obtains a width closer to these reference values, but a slightly lower $E_{\rm res}$ position. The density-of-states approach requires a fit of the background which is sensitive (especially with regards to the width) to the fit boundaries.

\section{Conclusion}
An important aspect of these methods is that they show how the mapped Fourier grid (MFG) method handles the supercritical resonance state which is important for the time-dependent work in chapter~\ref{tdscchp}. The collision work requires the diagonalization of both subcritical and supercritical matrix representations of the hamiltonian. The above demonstrations show that the supercritical resonance is represented both in the density of states and in the wavefunctions of the eigenstates. Thus, no augmentation or alteration of the MFG method is needed in order to include the supercritical resonance state. 

While the calculations here used a large number of states, which would not be usable by the repeated diagonalization needed for the time-dependent work, it nonetheless shows that the eigenbasis does contain the supercritical resonance state.

\chapter{Analytic continuation methods applied to supercritical states}\label{CSCAP}
Analytic continuation methods take a real hermitian Hamiltonian and transform it into a non-hermitian complex Hamiltonian. A parameter governs this transformation, the analytic continuation parameter (ACP), $\zeta$, and the original, real Hamiltonian is recovered when $\zeta=0$. The goal of these methods is to attenuate the exponentially divergent resonance state, $\phi_{\rm res}(x)$, bringing it into the Hilbert space of the (analytically continued) Hamiltonian. The resonance state is not in the hermitian domain of the real Hamiltonian since it does not satisfy the boundary condition: $\phi_{\rm res}(x) \rightarrow 0$ as $x \rightarrow \infty$ \cite{moiseyevrep}.

In order to obtain the best estimate for resonance position and width, the parameter $\zeta$, should be chosen carefully since the resonance parameters will be $\zeta$-dependent for a finite matrix representation. This is most often accomplished by changing $\zeta$ and looking for stability in the resonance position $E_{\rm res}$ and width $\Gamma$. These parameters, $\{E_{\rm res},\Gamma\}$ can be plotted parametrically as $\zeta$-trajectories:  $(E_{\rm res}(\zeta),\Gamma(\zeta))$. The resonance parameters are determined from a non-hermitian eigenvalue problem with eigenenergy $E_{\mathrm{R}}= E_{\mathrm{res}} \pm i\Gamma /2$ (positive sign for negative-energy resonance states and negative sign for positive-energy resonance states).

Building a matrix representation of the analytically continued Hamiltonian, using the mapped Fourier grid method, at given values of $\zeta$ and diagonalizing the resultant matrix yields a single spectrum of eigenvalues and a single eigenvalue for the resonance. By repeating this procedure for consecutive $\zeta$-values, a set of complex eigenvalues for the resonance is obtained. This set of eigenvalues is then used to find stable values of the resonance parameters as a function of $\zeta$, yielding the best approximation to the resonance parameters (without extrapolation). Some of this work was originally published in \cite{ackadcs}.

\section{Complex scaling}
\begin{figure}[!hb]\centering
\includegraphics[angle=0,scale=0.45]{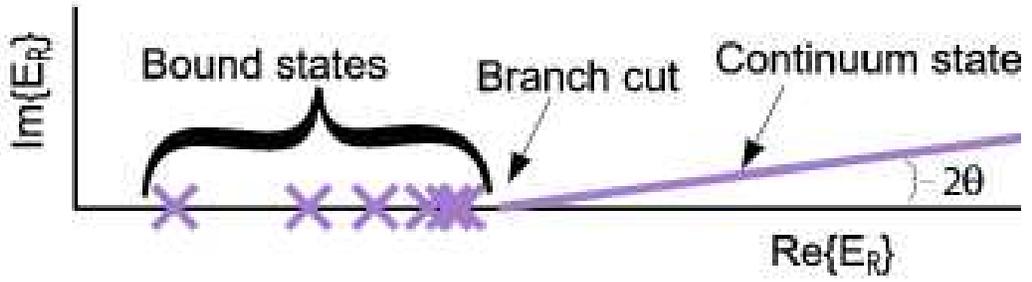}
\caption[Effects of CS on the eigenvalue spectrum]{\spacing{1}\label{csexample} Illustration of the spectrum of a complex scaled non-relativistic hamiltonian where the eigenvalues of the bound states (\textcolor{dark-magenta}{$\times$}) remain on the real axis but the eigenvalues of the continuum states (represented as a continuous line) rotate into the imaginary plane. The pivot point for the rotation is at the branch cut (continuum boundary). The continuum states rotate by $2\theta$ since the non-relativistic momentum is transformed by $p\rightarrow pe^{-i\theta}$ transforming $E=p^2/(2$m$_{\rm e})\rightarrow e^{-2i\theta}p^2/(2$m$_{\rm e})$. }
\end{figure}

Complex scaling introduces an analytic continuation of the Hamiltonian by a transformation of the reaction coordinate,
\begin{equation}\label{CStransformationequation}
r \rightarrow r e^{i \theta}. 
\end{equation}
It has been used extensively in atomic and molecular physics, and has been put on firm mathematical grounds by Reinhardt \cite{reinhardtcs} and Moiseyev \cite{moiseyevrep} and extended to the Dirac equation by \u{S}eba \cite{seba}. Recently it was applied to the case of Stark resonances in the Dirac equation \cite{CSDirac}. Increasing $\theta$ in small steps beginning from zero results in a series of energy spectra where the continuum-state energies are increasingly rotated into the complex energy plane with the branch cut(s) as pivot points shown in Fig.~\ref{csexample} for the positive continuum (in the Dirac case the branch cuts are at $\pm$m$_{\rm e}$c$^2$).
The bound states remain on the real axis and are not affected by the change in $\theta$. As $\theta$ is increased from zero, the continuum states rotate into the imaginary plane by approximately $2\theta$ for non-relativistic states. This can be understood (non-relativistically) by the scaling in the momentum $\hat{p}=-i\frac{d}{dr}\rightarrow \hat{p}e^{-i\theta}$. 

The energy of the (non-relativistic) continuum states is approximately $E=p^2/2$ (in natural units). When the momentum is scaled the energy becomes $E\rightarrow e^{-2i\theta}p^2/2$. In the same fashion, the relativistic energies rotate by $E\rightarrow \pm \sqrt{1+p^2e^{-2i\theta}}$ for small $p$ \cite{CSDirac}. The resonance states (located in the continuum) initially rotate with the neighboring continuum states as $\theta$ is increased from zero, but then stabilize at the resonance energy (i.e., at $E=E_{\rm res}-i \Gamma/2$), and remain relatively stable for a range of $\theta$ values (this is why the method is also referred to as the stabilization method). The closest approximation to the resonance energy occurs at the most stable energy eigenvalue along the $\theta$-trajectory, which is found where $  \left|\frac{dE_{R}}{d\theta} \right|$ is minimized. Complex scaling (rotation) thus is a method to isolate the resonance states from the continuum.

\section{Smooth exterior scaling}\label{sessec}
Smooth exterior scaling (SES) is an extension of the complex scaling (CS) method whose mathematical justification was developed, e.g., by W.P. Reinhardt \cite{reinhardtcs}, and N. Moiseyev \cite{moiseyevrep}. In this thesis SES is extended to the relativistic Dirac equation for supercritical resonances. SES relies on the same justification as CS, but uses a general path in the complex plane that is continuous;  when using non-continuous paths one refers to the method as exterior complex scaling (ECS) \cite{moiseyevrep}. A simple path is obtained by rotating the reaction coordinate into the complex plane about some finite position, $r_{\mathrm{s}}$, instead of the origin. The transformation then has the form,
\begin{equation}\label{ses}
r \rightarrow \left\{
\begin{array}{ccc}
r & & \mathrm{for}\;\; r<r_{\mathrm{s}} \\
\left(r-r_{\mathrm{s}}\right)e^{i\theta} + r_{\mathrm{s}} & & \mathrm{for}\;\; r_{\mathrm{s}} \leq r.
\end{array}\right.
\end{equation}
It offers the advantage of turning on the scaling at a distance $r_{\mathrm{s}}$ which can be chosen appropriately for a given potential shape. It is natural to choose $r_{\mathrm{s}}$ such that the ``bound'' part of the resonance state is not affected directly by the complex scaling (cf. section~\ref{spresd}).
 
This additional freedom introduces some complications not found in CS: unlike CS, SES does not always have a minimum in the $\left| \frac{dE_R}{d\theta} \right|$ curve as a function of $\theta$, which makes it difficult to determine a stabilized $\theta$ value, $\theta_{\mathrm{opt}}$, for $E_R(\theta)$. An approximation can always be made by finding the cusp of the trajectory in $\{E_{\mathrm{res}}, \Gamma\}$ space \cite{Doolencusp}, although this does not allow for a very precise determination of $\theta_{\mathrm{opt}}$ compared with a minimum in $\left| \frac{dE_{R}}{d\theta}\right|$. Alternatively, the parameters can be determined from either the $\frac{dE_{\mathrm{res}}}{d\theta}$ or $\frac{d\Gamma}{d\theta}$ curves. This is because in practice it is usually found that at least one of them will have a minimum, with the best results for each parameter coming from its own derivative minimum, e.g. the minimum of $\frac{d\Gamma}{d\theta}$ will determine the optimal value of $\theta$ for $\Gamma(\theta)$ but not for $E_{\rm res}(\theta)$ \cite{moiseyevcs}. Optimal results are obtained when the three derivatives have the same value of $\theta_{\mathrm{opt}}$, yielding the same values for $E_{\mathrm{res}}$ and $\Gamma$. 

\section{Complex absorbing potentials}
The method of adding a complex absorbing potential (CAP) to the Hamiltonian has been used extensively in atomic and molecular physics \cite{ingrCAP,santraSO,santragf,sahoo}, and was put on firm mathematical grounds by Riss and Meyer \cite{RissCAP}. In this thesis, it is extended to the relativistic Dirac equation for supercritical resonances. The true Hamiltonian is augmented by an imaginary potential, which makes the resulting Hamiltonian operator non-hermitian. In the case of the Dirac Hamiltonian a CAP is added as a scalar,
\begin{equation}
\hat{H}_{\mathrm{CAP}} = \hat{H}-i\eta \hat{\beta}{W}(r),
\end{equation}
where $\eta$ is a small non-negative parameter determining the strength of the CAP and $\hat{\beta}$ is the standard Dirac matrix \cite{greiner}. The function ${W}(r)$ determines the shape of the absorbing potential and can be tailored for the problem at hand. Currently, the most common use of a CAP is as a stabilization method: one solves the system on an equally spaced mesh of values of $\eta$ and takes the minimum of $\left|\eta \frac{dE_{R}}{d\eta}\right|_{\eta=\eta_{\mathrm{opt}}}$ as the closest approximation to the true resonance parameters \cite{RissCAP}.

In the literature, one finds investigations of polynomial CAPs of the form
\begin{equation}\label{cap}
W(r)=\Theta(r-r_{\mathrm{c}}) \left(r-r_{\mathrm{c}}\right)^n
\end{equation}
where $\Theta$ is the Heaviside function and $n$ is an integer. The CAP parameter $r_{\mathrm{c}}$ allows for the turn-on of the absorbing potential outside of the ``bound'' part of the wavefunction \cite{santra}.

Although CAP and SES appear as separate methods, they are not independent. It has been shown that a CAP can be transformed into CS \cite{santrawhycap} and that SES is related to CAP \cite{SES2cap}.

\section{Results using analytic continuation methods}
Calculations were carried out using the potential in Eq.~\ref{potential} for a U$^{92+}$-Cf$^{98+}$ quasi-molecular system for an internuclear separation of $R=20$fm where the atomic mass numbers A$_{\rm U}=238$ and A$_{\rm Cf}=251$. The results from CS, CAP and SES methods are compared here. For the present problem there is no barrier in the potential. However, as explained in section~\ref{causesc}, one can find the region where the supercritical resonance state displays tunneling behavior by looking at the intersection of $E_{\mathrm{res}}$ with the curve $V(r)-$m$_{\rm e}$c$^2$. For the present case, this occurs at $r_V \approx 2$ (in units of $\hbar/$m$_{\rm e}$c).

Due to the aforementioned difficulty (cf. section~\ref{sessec}) in obtaining results using stabilization, smooth exterior scaling will not be compared to CAP or CS in the following sections. Its usefulness will be addressed in chapter~\ref{padechp}.

\subsection{Linear and quadratic CAPs}
\begin{figure}[H]\centering
\includegraphics[angle=270,width=12cm]{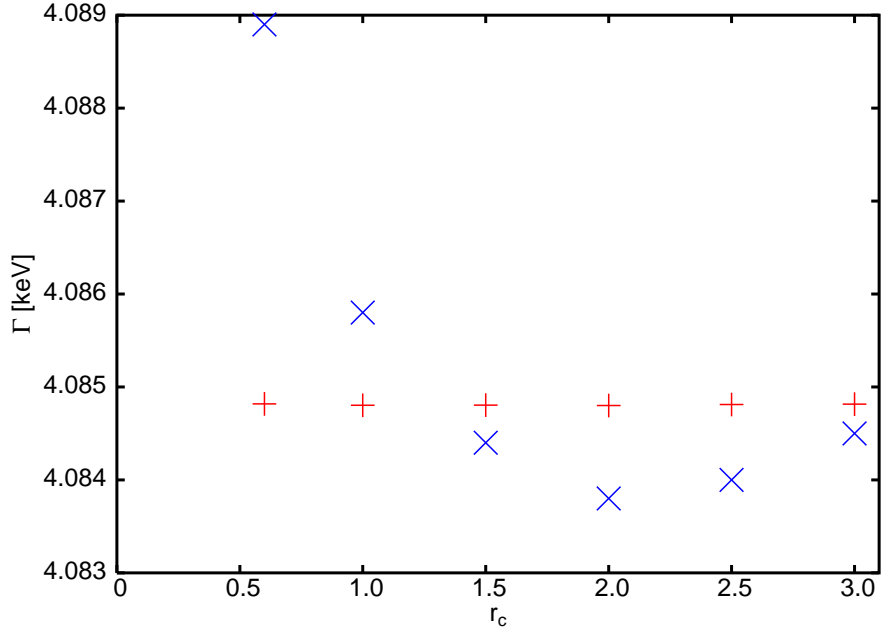}
\caption[Comparison of linear and quadratic CAPs]{\spacing{1}\label{capwidth} The width $\Gamma$(1S$\sigma$) as a function of $r_{\mathrm{c}}$ (in units of $\hbar/$m$_{\rm e}$c) for a linear (\textcolor{blue}{$\times$}) and a quadratic (\textcolor{red}{$+$}) CAP as obtained in a basis specified by $N=2000$ and $s=6000$ for the U-Cf ($R=20$ fm) system.}
\end{figure}

When using a linear absorber ($n=1$ in Eq.~\ref{cap}) it was found that the resonance parameters ($E_{\mathrm{res}},\Gamma$) do depend on the choice of $r_{\mathrm{c}}$, but are least sensitive when $r_{\mathrm{c}} \simeq r_V$. In fact, a minimum is observed for the width parameter, $\Gamma$, at $r_{\mathrm{c}}=r_V$ as seen in Fig.~\ref{capwidth}. For a quadratic CAP ($n=2$), there is minimal variation in the extracted resonance parameters on the same scale.

A few conclusions can be drawn from these results. The linear CAP, through the sensitivity of the result to the value of $r_{\mathrm{c}}$, is providing support to the notion of an effective potential barrier in the calculation at $r \approx r_{\mathrm{c}}$. The quadratic CAP calculation is deemed to be less sensitive to the value of $r_{\mathrm{c}}$ due to the gradual turn-on of the absorber with continuous derivative. It leads to more accurate results and will be used in the subsequent comparisons. Both CAP calculations do provide, however, consistent results for the complex resonance energy position (not shown).

\subsection{Quadratic CAP and CS}
\begin{figure}[!ht]\centering
\includegraphics[angle=270,width=10cm]{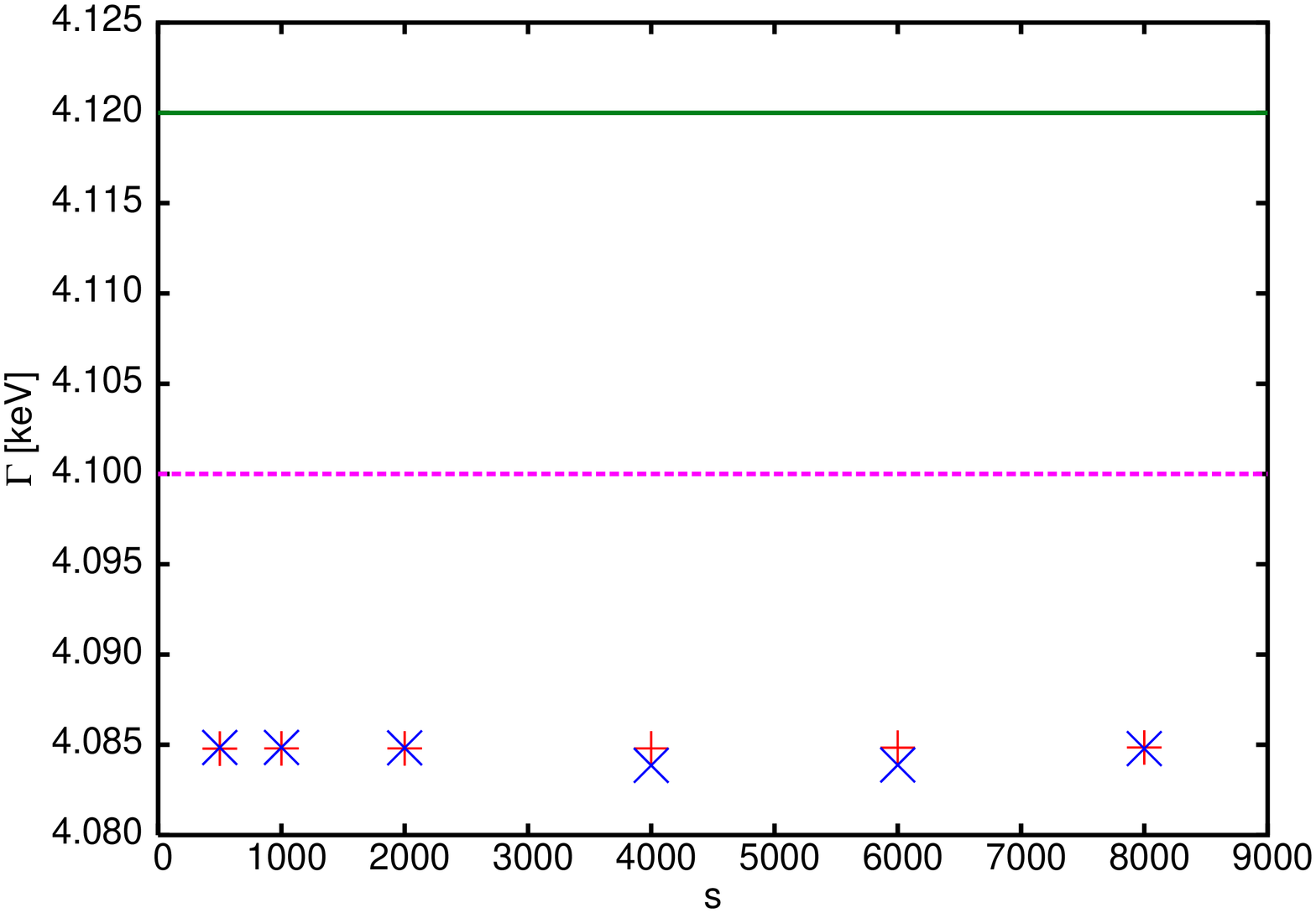}
\includegraphics[angle=270,width=10cm]{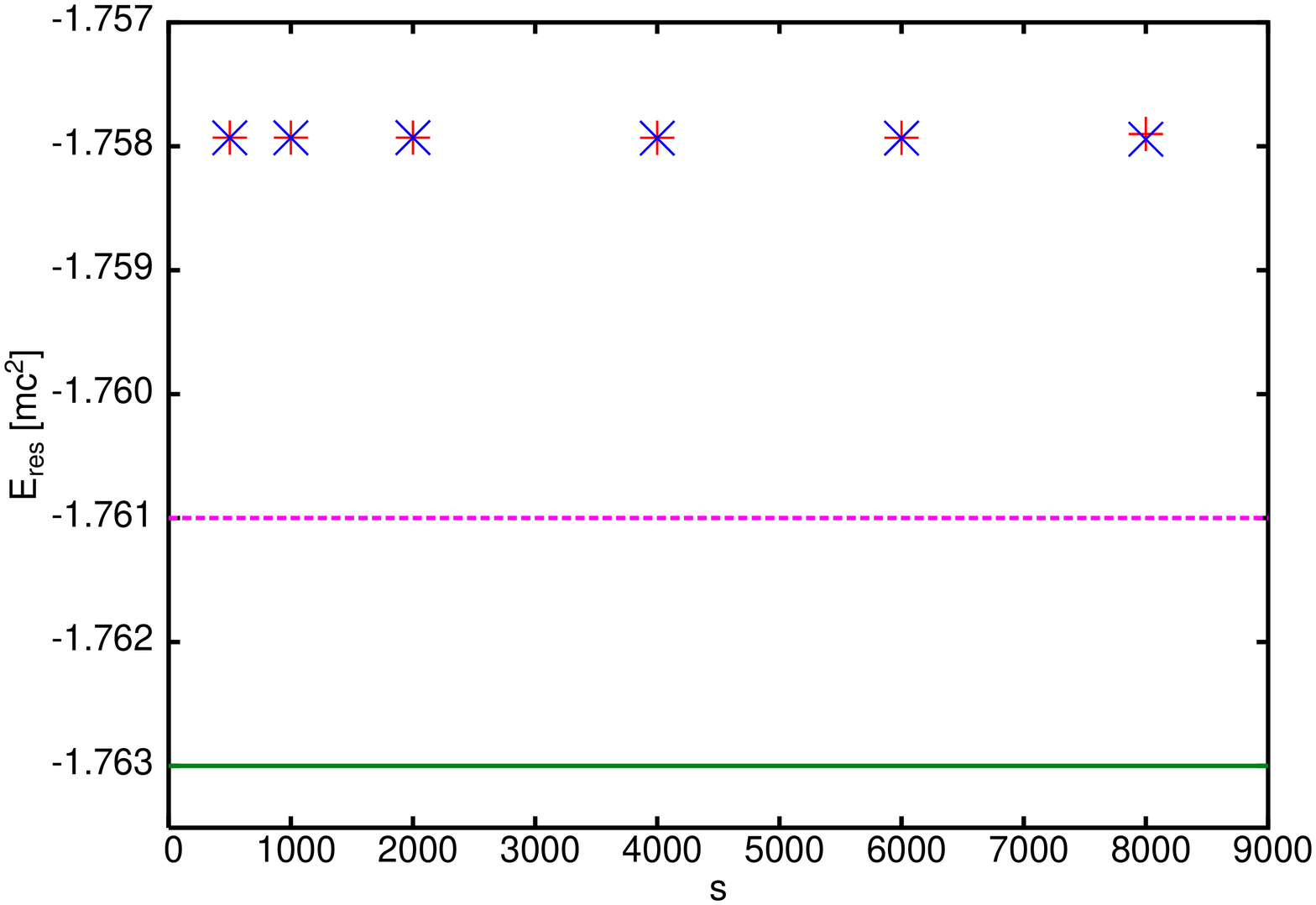}
\caption[Comparison of CS and CAP results as a function of $s$]{\spacing{1}\label{csres}The width $\Gamma$(1S$\sigma$) and mean energy $E_{\mathrm{res}}$(1S$\sigma$) for the U-Cf ($R=20$ fm) system as a function of the mapping parameter $s$ (cf. Eq.~\ref{mapping}) with $N=2500$ points for the CS method (\textcolor{blue}{$\times$}) and $N=2000$ points for the CAP method (\textcolor{red}{$+$}). The lines indicate values from Ref.~\cite{pra10324}, the phase-shift result is given as a solid line, and the truncated potential method result as a dashed line.}
\end{figure}

In Fig.~\ref{csres}, results for the different analytic continuation methods are compared for the U$^{92+}$-Cf$^{98+}$ system at $R=20$ fm. The resonance energy $E_{\mathrm{res}}$ and the width $\Gamma$ are shown as a function of the mapping parameter $s$ (cf. Eq.~\ref{mapping}), for a fixed mesh size of $N=2500$ collocation points for the CS method and $N=2000$ points for the quadratic CAP method. The linear CAP results would be indistinguishable from the $n=2$ results on the scale of the figures.

It was found that the CS and CAP ($n=2$) methods return consistent results within a parameter range of $500 < s < 8000$ with small relative fluctuations. Both $E_{\mathrm{res}}$ and $\Gamma$ are determined with a relative accuracy of better than $10^{-4}$. For values of the mapping parameter $s$ outside of the ideal range the uncertainties increase, particularly so for the width.

Also shown in Fig.~\ref{csres} are two results for the same system from Ref.~\cite{pra10324}. They deviate among themselves on the same scale, approximately, as from the present values derived from the CS and CAP methods. Not shown are the results from the Breit-Wigner fits based on the diagonalization of the hermitean Hamiltonian matrix in the Fourier grid method (cf. section~\ref{mfgres}). We note, however, that $E_{\mathrm{res}}$ is predicted in close agreement with the CAP or CS data, while $\Gamma$ is over-estimated. For the projection method it happens to agree with the result from a phase-shift analysis performed in Ref.~\cite{pra10324}.

\begin{figure}[!hb]\centering
\includegraphics[angle=270,width=12cm]{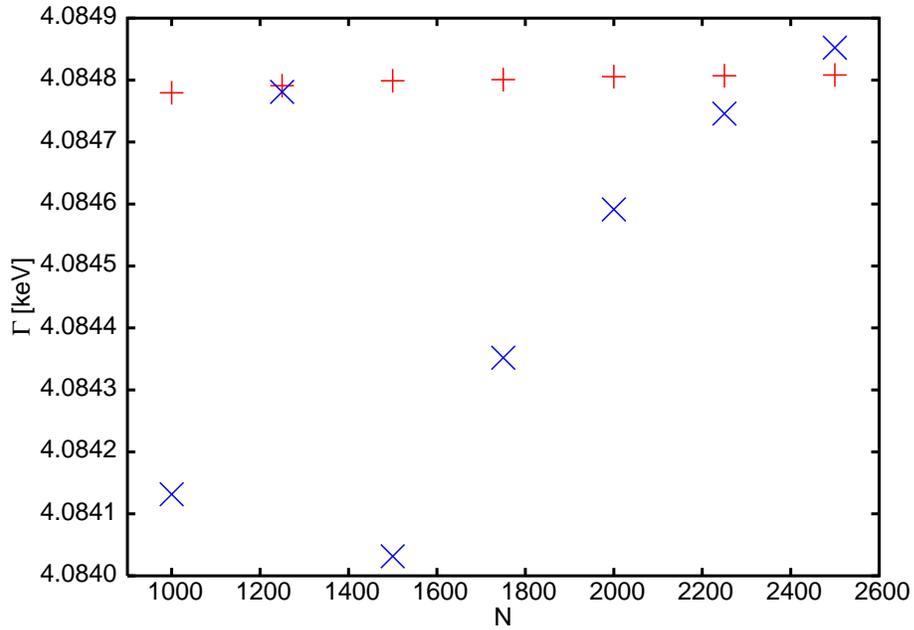}
\caption[CS and CAP stability results]{\spacing{1}\label{Neffect} The width $\Gamma(1{\mathrm{S}}\sigma)$ as a function of mesh size $N$ using the CS (\textcolor{blue}{$\times$}), and quadratic CAP (\textcolor{red}{$+$}) methods with mapping parameter $s=1000$ for the U-Cf ($R=20$ fm) system. }
\end{figure}

In order to demonstrate the convergence properties of the present CAP and CS results, Fig.~\ref{Neffect} shows the dependence of $\Gamma$ on the grid size $N$. A fixed value of the mapping parameter ($s=1000$) is used. 
The data indicate a systematic variation with $N$ for the CS data at large $N$. For $N=2500$, it appears as if the width is established to four significant digits. The CS results fluctuate (on this fine scale) for $N<1750$, before systematically approaching the CAP value. The CS result at $N=1250$ agrees with the CAP result only by chance. The CAP results, on the other hand, display better convergence at $N=1000$ already, with a deviation of less than $20$ meV from the $N=2000$ result. The stability of $E_{\mathrm{res}}$, for different grid sizes, is on the order of $10^{-8}\, $m$_{\rm e}$c$^2$ for both methods meaning it is stable for all $N$ values shown.

\subsection{Other configurations}\label{otherconf}
\begin{figure}[H]\centering
\includegraphics[angle=270,width=12cm]{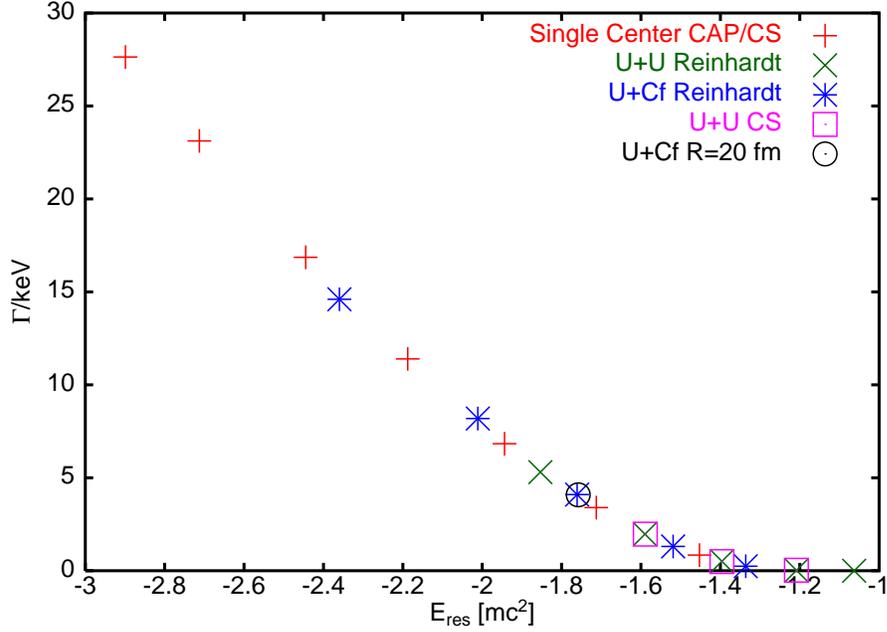}
\caption[Results for different system's ground state resonance parameters]{\spacing{1} \label{EvG} The width $\Gamma(1{\rm S}\sigma)$ as a function of $E_{\rm res}(1{\mathrm{S}}\sigma)$ calculated by different methods and using different supercritical potentials. The U-U and U-Cf data from Ref.~\cite{pra10324} were obtained for the distances $R=$ 16$^*$,16,20,25,30 fm, while the single-center CAP/CS data from the present work were obtained for $Z=$ 195,193,190,187,184,181,178. The leftmost U-U and U-Cf points marked by $R=16^*$ fm were calculated with point-like nuclei.}
\end{figure}

These results are generalizable from the test case of a U-Cf quasi-molecule at $R=20$ fm to other situations. The relationship between the resonance position and width has been explored. In Fig.~\ref{EvG}, $\Gamma$ is shown as a function of $E_{\rm res}$ over a substantial range of system parameters. On the scale of the graph, differences between the present data and previous literature results cannot be noticed, i.e., they follow an almost universal curve. A number of the data points were obtained from different quasi-molecular systems at various internuclear separations with both point and extended nuclear models (in monopole approximation). Data from single-center calculations with one extended nucleus also follow the same relationship. 

\section{Conclusion}
The quadratic CAP and CS are found to give results in close agreement for a wide range of parameters. All methods used gave very stable results as compared with the quoted results from Ref.~\cite{pra10324} and the direct methods of chapter~\ref{tdscstates}. The CAP method has been extended to the Dirac equation for the first time, and a quadratic CAP was found to be superior in determining the supercritical resonance parameters compared with CS.

\clearpage
\chapter{Augmented analytic continuation methods}\label{padechp}
Analytic continuation makes use of a parameter $\zeta$ to turn the original Hamiltonian into a non-hermitian operator. This raises two problems: {\it (i)} one has to optimize $\zeta$ to find the best approximation to the complex resonance eigenenergy; {\it (ii)} one worries about the effect of the unphysical $\zeta$ parameter on the final result. The optimization in step {\it (i)} is performed by stabilizing some measure (e.g., $\left|\frac{dE_{R}}{d\zeta} \right|$) that depends on the eigenenergy as a function of $\zeta$ (cf. chapter~\ref{CSCAP}). 

Recent work based upon the method of adding a complex absorbing potential (CAP) has demonstrated how to obtain more accurate results by extrapolating the complex $E_R(\zeta)$-trajectory to $\zeta=0$ \cite{lebrevePade}. This idea is extended to the relativistic Dirac equation in this chapter using the complementary methods of smooth exterior scaling (SES) and CAP. The stability and parameter independence of the results is demonstrated. The extrapolation technique allows one to obtain highly accurate results for smaller basis size, $N$; thus enabling the extension of the calculations beyond the monopole approximation to the two-center potential. Some of this work was originally published in \cite{ackad:022503} and \cite{eackadconf2007}.

\section{Pad\'e approximant and extrapolation}\label{padeexp}
The goal of the analytic continuation methods is to make the resonance wavefunction $\psi_{\mathrm{res}}$ a bounded function by choosing an analytic continuation parameter $\zeta > \zeta_{\mathrm{crit}}$. Although $\psi_{\mathrm{res}}$ is an eigenfunction of the physical, hermitian, Hamiltonian, it is not in the Hilbert space, since it is exponentially divergent. For a sufficiently large critical value of $\zeta=\zeta_{\mathrm{crit}}$ (called $\theta_{\mathrm{crit}}$ for CS, and $\eta_{\mathrm{crit}}$ for CAP), $\psi_{\mathrm{res}}$ becomes a bounded function and is, therefore, in the Hilbert space of the physical Hamiltonian \cite{moiseyevrep}. Taking the $\lim_{\zeta\rightarrow 0} E_{R}(\zeta)$ always yields a real eigenvalue corresponding to $\hat{H}(\zeta=0)\Psi=E_R(\zeta=0)\Psi$ since $\hat{H}(\zeta=0)$ is hermitian when acting on bounded functions.

The authors of Ref.~\cite{lebrevePade} proposed the following: instead of directly taking the $\zeta=0$ limit, an extrapolation of a part of the trajectory, $E_R(\zeta)$, namely for $\zeta > \zeta_{\mathrm{crit}}$, is used to obtain the complex-valued $E_{\mathrm{Pad\acute{e}}}(\zeta=0)$. The points used for the extrapolation are computed eigenvalues restricted to a region where $\psi_{\mathrm{res}}$ remains in the Hilbert space
(as represented by the finite basis). For the extrapolation the Pad\'e approximant is given by,
\begin{equation}\label{pade}
E_{\mathrm{Pad\acute{e}}}(\zeta)= \frac{\sum_{i=0}^{N_1}p_i \zeta^i}{1+\sum_{j=1}^{N_1+1}q_j \zeta^j}
\end{equation}
where $p_i$ and $q_j$ are complex coefficients \cite{SchlessingerPade}. It is useful to define $N_p=2(N_1+1)$ as the number of points used in the approximant. The extrapolated value of $E_{\mathrm{Pad\acute{e}}}(\zeta=0)$ is given by $p_0$, and follows from a given set of $\zeta$-trajectory points,
\begin{equation}\label{epoints}
\epsilon_i=\left\{ \varepsilon_j=E_R(\zeta_{i}+(j-1)\Delta\zeta),\;j=1...N_{\mathrm{p}} \right\}.
\end{equation}
The smallest $\zeta$-value in the set of points used for the extrapolation is labeled $\zeta_{i}$ and must satisfy $\zeta_i > \zeta_{\mathrm{crit}}$. 

At first glance it would appear that new parameters have been introduced ($N_{\rm p}$, $\zeta_i$) to replace $\zeta$. However, it turns out that the extrapolation is very insensitive to $N_{\rm p}$ for $4\lsim N_{\rm p} \lsim12$. While the extrapolation is also not very sensitive to the initial data point of the trajectory sample $\zeta_i$, we can optimize $\zeta_i$ and the starting parameter ($r_c$ or $r_s$) together and remove the effect of these two parameters on the final result. This is done by finding $p_0$ as a function of $\zeta_i$, i.e. for different starting values of the extrapolation set $\epsilon_i$. The distance between all the $p_0$'s (at different $\zeta$ starting values) for different $r_s$ or $r_c$'s are compared and a minimum in phase-space, $\{E_{\rm res},\Gamma\}$, is found. Averaging over the results for the two resonance parameters gives the best value and the error. This method allows for any number of $r_s$ or $r_c$'s to be compared and even works when mixing SES and CAP results. 

\section{Smooth exterior scaling results using extrapolation}
\begin{figure}[!ht]\centering
\includegraphics[angle=270,scale=0.5]{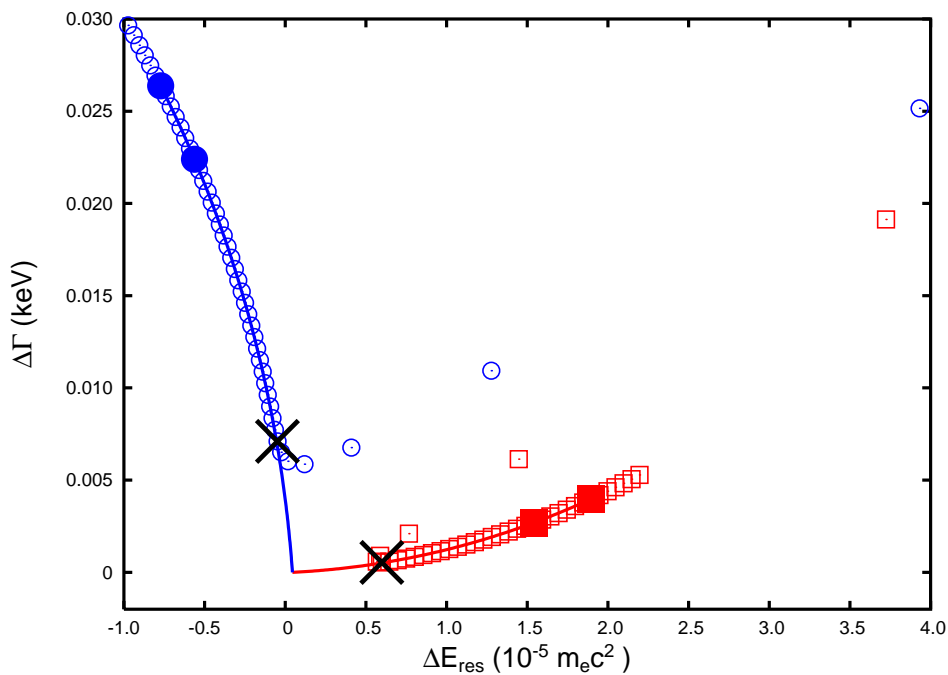}
\caption[Trajectory results for SES]{\spacing{1}\label{trajs}  The $\Gamma(1{\mathrm S}\sigma)$ resonance $\theta$-trajectory from $N=250$, $r_{\mathrm{s}}=2$ (\textcolor{red} {$\boxdot$}) and $r_{\mathrm{s}}=3$ (\textcolor{blue}{$\odot$}) displayed as the difference from a reference calculation (cf. text) for the U-Cf system at $R=20$ fm in monopole approximation. The $\times$'s mark the resonance energy obtained by minimizing $\left|\frac{dE_R}{d\theta}\right|$. The $N_{\mathrm{p}}=8$ data points used for the extrapolations are bracketed by the solid symbols, and a spacing of $\Delta\theta=0.01$ was used.}
\end{figure}

In SES (as in the other methods) the energy $E_R(\theta)$ forms a trajectory which depends on the calculation parameters ($N,s,r_{\mathrm{s}}$). Figure~\ref{trajs} shows the $\theta$-trajectory for $N=250$ (number of collocation points), $s=400$ (mapping scaling parameter), $\Delta\theta=0.01$ (choice of spacing for the trajectory) and $r_{\mathrm{s}}=2$ (exterior scaling radius) and $r_{\mathrm{s}}=3$ respectively as a sequence of squares and circles respectively.
The values are displayed as the deviation from a reference value given by $E_{\mathrm{res}}=-1.757930695(8)$m$_{\mathrm{e}}$c$^2$ and $\Gamma=4.084798(12)$keV, which was obtained by a large-size CAP calculation using $N=3000$, $s=400$ and $r_{\mathrm{c}}=2$. The uncertainties were determined from calculations with different basis parameters, $s$.

In Fig.~\ref{trajs}, the $E_{R}(\theta)$ results for small $\theta$ values start in the upper right hand and abruptly change direction after a few points. This happens in the vicinity of the reference value given by the point (0,0). As $\theta$ increases beyond the cusp in the trajectory, $E_R(\theta)$ changes more slowly, and the $\theta$-trajectory points become denser. Due to the finite number of collocation points $N$, the actual value of $\theta_{\mathrm{crit}}$ (the minimum $\theta$ value such that $\psi_{\mathrm{res}}$ is bounded and properly represented within the finite basis) is higher than theoretically predicted for the $N\rightarrow \infty$ representation \cite{moiseyevrep}, and is close to the cusp-like structure in the $\theta$-trajectory. 

Optimal, directly calculated, approximate values for the resonance energy are chosen by finding the minimum of $\left| \frac{dE_{R}}{d\theta}\right|_{\theta=\theta_{\mathrm{opt}}}$ \cite{moiseyevrep,reinhardtcs}, and are displayed as black crosses for both choices of $r_{\rm s}$. The solid lines represent $N_{\mathrm{p}}=8$ Pad\'e extrapolation curves (cf. Eq.~\ref{pade}). The values of $\varepsilon_1$ and $\varepsilon_{N_{\mathrm{p}}}$ from Eq.~\ref{epoints}, i.e., the bracketing points used to determine the Pad\'e approximant, are indicated by filled symbols. The values of $\theta_1=0.28$ for the $r_{\mathrm{s}}=2$ curve and $\theta_1=0.36$ for the $r_{\mathrm{s}}=3$ curve were chosen, because their $E_{\mathrm{Pad\acute{e}}}(\zeta=0)$ points were found to be closest to each other. Using other starting values for the Pad\'e curves would result in indistinguishable extrapolation curves on the scale of Fig.~\ref{trajs} (as long as the starting point would not be too close to the cusp in the $\theta$-trajectory). From the Pad\'e curves one obtains the best approximation to the resonance parameters by selecting the end points ($E_{\mathrm{Pad\acute{e}}}(\zeta=0)$), located close to the the origin (0,0), i.e., close to the $N=3000$ reference calculation result.

Although very different trajectories from those shown in Fig.~\ref{trajs} were obtained for different $r_{\mathrm{s}}$ values, the Pad\'e curves still extrapolated to nearly the same $\zeta=0$ complex resonance energy ($E_{\rm res}+i\Gamma/2$), as long as $r_{\mathrm{s}}$ was chosen outside the ``bound'' part of the wavefunction and not too large, i.e. $1.5 \lsim r_{\mathrm{s}}\lsim 6$. Larger values of $r_{\rm s}$ pose difficulties for the finite-$N$ computation as it cannot span the large-$r$ region properly. 

Simple complex scaling (CS) is equivalent to SES with $r_{\mathrm{s}}=0$ and, thus, falls outside of the range of acceptable $r_{\mathrm{s}}$ values. It yields results with a much larger deviation from the reference result, even when Pad\'e extrapolation is applied.

\begin{figure}[!ht]\centering
\includegraphics[angle=270,scale=0.5]{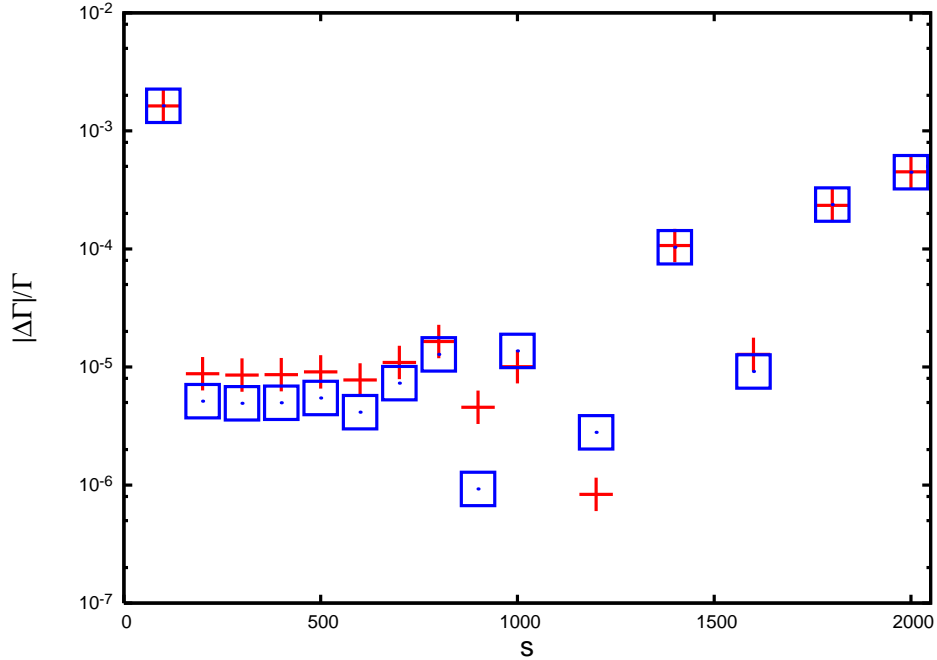}
\caption[SES relative error as a function of $s$]{\spacing{1}\label{NS}  Magnitude of relative error for Pad\'e extrapolated widths $\Delta \Gamma$(1S$\sigma$) for $N=500$ as a function of the scaling parameter $s$, for the U-Cf system at $R=20$fm in monopole approximation. The data were obtained using two reference calculations from the stabilized CAP method with $N=3000$ and $r_{\mathrm{c}}=2$; red plus (\textcolor{red}{$+$}) symbols: $s=400$; blue squares (\textcolor{blue}{$\boxdot$}): $s=750$ (cf. text). }
\end{figure}

The scaling parameter, $s$, introduced by the mapping of the radial coordinate in the mapped Fourier grid method in Eq.~\ref{mapping}, plays a dominant role as far as numerical accuracy is concerned. In Fig.~\ref{NS} the magnitude of the relative error in the width $\Gamma$(1S$\sigma$) is shown as a function of $s$ for SES calculations with $N=500$, while using two reference CAP calculations. The latter were obtained from the stabilization method with a large basis ($N=3000$) and yielded values of $E_{\mathrm{res}}=-1.757930695$, $\Gamma=4.084798$ for scaling parameter $s=400$, and $E_{\mathrm{res}}=-1.757930702$, $\Gamma=4.084808$ for $s=750$ respectively.
Each SES data point was obtained by finding the closest intersection of the $r_{\mathrm{s}}=2,3,4$ Pad\'e curves from all starting $\theta$'s (for $\theta > \theta_{\mathrm{crit}}$), and taking the average of these three best values. The use of different $r_{\mathrm{s}}$ values within the acceptable range would cause changes too small to see on this plot. The resonance position, $E_{\mathrm{res}}$, is much more stable with respect to $s$ and therefore not shown.

The technique of combining information from different $r_{\mathrm{s}}$ calculations in order to select the ideal subset of $\theta$-points helps with the following problem. For each calculation with a fixed value of $r_{\mathrm{s}}$ we looked at extrapolated values for the resonance width as a function of the start value $\theta_{i}$, and observed the deviation from the average value. It was found that numerical noise present in the data was minimal for some range of $\theta_{i}$ values (different for each calculation). This noise was investigated as a function of extrapolation order $N_{\mathrm{p}}$. Combining information from calculations with different values of $r_{\mathrm{s}}$ allows for the elimination of two sources of random uncertainties associated with finite-precision input to a sensitive extrapolation calculation. 

The results demonstrate that the extrapolated ($\theta=0$) SES $N=500$ calculations \emph{exceed} the precision of the stabilized (non-extrapolated) $N=3000$ CAP calculations. Table \ref{sestable} gives the resonance parameter results ($E_{\mathrm{res}},\Gamma$) for two different extrapolations, namely $N_{\mathrm{p}}=4,8$ for basis size $N=500$, and mapping parameter $300 \le s \le 700$, which corresponds to a subset of the data shown in Fig.~\ref{NS}. The standard deviation of the mean, 
\begin{equation}\label{std}
\sigma=\sum_i^{N_i}{\frac{\sqrt{(x_i-\bar{x})^2}}{N_i}},
\end{equation}
where $\bar{x}$ is the average $x$-value, is given separately for resonance position and width. The results are encouraging since extrapolation provides a  marked improvement over the stabilized CS method discussed in chapter~\ref{CSCAP}.  
\begin{table}[H]
	\centering
\begin{tabular}{|c|c|c|c|c|}
\hline 
 $N_{\mathrm{p}}$ & $ E_{\mathrm{res}}$& $\sigma({E_{\mathrm{res}}})$ & $\Gamma$ (keV) & $\sigma({\Gamma})$ \\
\hline\hline
8 &-1.75793073 & 1.7$\times10^{-7}$ &	4.0848085 &	1.4$\times10^{-6}$ \\
\hline
4 & -1.75793072 & 1.6$\times10^{-7}$ & 4.0848031 &	1.4$\times10^{-6}$ \\
\hline
\end{tabular}
	\caption[Extrapolated SES results]{\spacing{1}\label{sestable}Averaged resonance position and width from the extrapolated SES method with basis size $N=500$ using extrapolation orders $N_{\mathrm{p}}=8$ and $N_{\mathrm{p}}=4$. The values were averaged over the stable range $300\le s\le 700$. The standard deviation of the mean is given by $\sigma$ separately for position and width.}
\end{table}

\subsection{Pad\'e extrapolation of CAP trajectories}\label{capcs}
Previously it was shown that a quadratic complex absorbing potential ($n=2$ in Eq.~\ref{cap}) gives better results than simple complex scaling (CS) when used as a stabilization method as done in chapter~\ref{CSCAP}. Further investigation has found that the performance of SES represents an improvement over that of CS (provided a minimum can be found), and that it is competitive with CAP. 

\begin{figure}[!ht]\centering
\includegraphics[angle=270,scale=0.5]{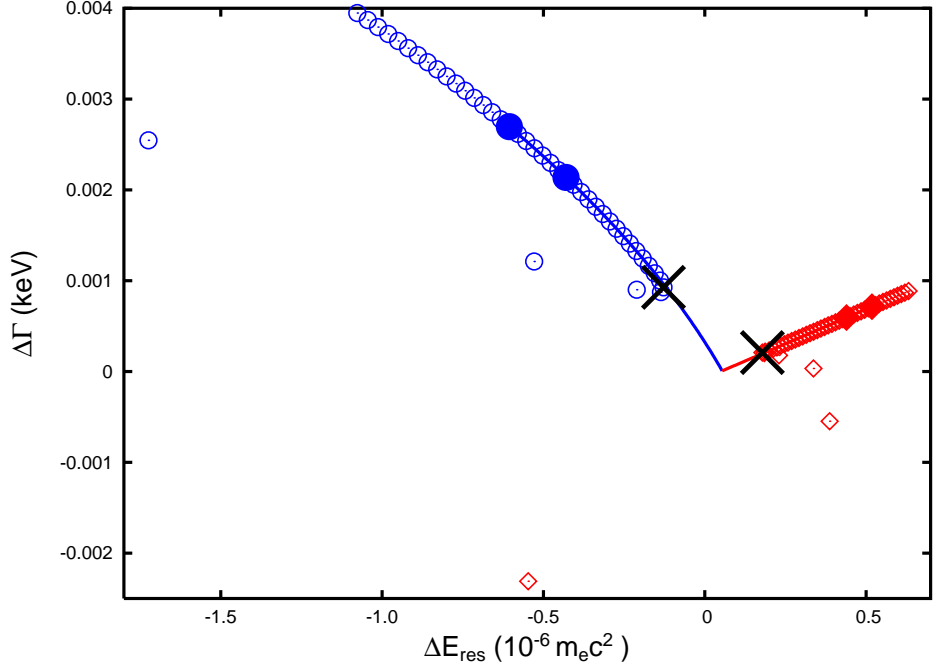}
\caption[SES and CAP trajectories]{\spacing{1} \label{trajs2} SES $\theta$-trajectory with $N=500$, $s=600$, $r_{\mathrm{s}}=3$ (blue circles), and CAP $\eta$-trajectory for $N=500$, $s=600$, $r_{\mathrm{c}}=3$ (red diamonds) displayed as the difference from a reference calculation for the 1S$\sigma$ state in the U-Cf system at $R=20$ fm in monopole approximation. The $\times$'s mark the approximate values obtained from the stabilization method. The reference value was obtained by Pad\'e extrapolation of CAP calculations with $N=3000$, $N_{\mathrm{p}}=4$ for $r_{\mathrm{c}}=2,3,4$ yielding $E_{\mathrm{res}}=-1.7579307062(5)$m$_{\mathrm{e}}$c$^2$ and $\Gamma=4.08481303(8)$keV.}
\end{figure}

Even for the same calculation parameters ($N,s,r_{\mathrm{c}}$), the $\eta$-trajectory for CAP is found to be different from the $\theta$-trajectory in SES. Figure~\ref{trajs2} shows a trajectory pair for identical calculation parameters, namely $N=500$, $s=600$, and $r_{\mathrm{s}}=r_{\mathrm{c}}=3$ respectively.
Even though the trajectories are rather different, the Pad\'e curves extrapolate to rather close values ($E_{\mathrm{Pad\acute{e}}}(\eta=0)$ and $E_{\mathrm{Pad\acute{e}}}(\theta=0)$), i.e. the extrapolated resonance parameters for both methods are very close. When looking at plots of the $N=500$ $\eta=0$ extrapolation results for CAP as a function of $s$, as was done in Fig.~\ref{NS} for SES, almost identical results are observed. In table~\ref{captable}, the average and standard deviation of the mean, $\sigma$ (cf. Eq.~\ref{std}), for the CAP results for $N=500$, $300 \le s \le 700$ are given for two different Pad\'e point numbers, $N_{\mathrm{p}}=4,8$. 

\begin{table}[H]
	\centering
\begin{tabular}{|c|c|c|c|c|}
\hline 
$N_{\mathrm{p}}$ & $E_{\mathrm{res}}$& $\sigma_{E_{\mathrm{res}}}$ & $\Gamma$ (keV) & $\sigma_{\Gamma}$ \\
\hline\hline
8 &-1.75793072 & 1.6$\times10^{-7}$ &	4.0848141&	3.1$\times10^{-6}$ \\
\hline
4 & -1.75793072 & 1.7$\times10^{-7}$ & 4.0848122 &	5.4$\times10^{-6}$ \\
\hline
\end{tabular}
	\caption[Extrapolated CAP results]{\spacing{1}\label{captable}Same as in Table~\ref{sestable}, but for the extrapolated CAP method with basis size $N=500$.}
\end{table}

\subsection{Comparison of CAP and SES results}\label{padeexpl}
Both methods (CAP and SES) show similar behavior with respect to different aspects of the calculation. Both are relatively insensitive to the number of Pad\'e points, $N_{\mathrm{p}}$, used in the extrapolation. It was found that the results were unchanged, to the reported precision, for $4\leq N_{\mathrm{p}} \leq 12$. The effects of changing the distance between the analytic continuation parameter ($\Delta\theta$ for SES and $\Delta\eta$ for CAP) were similarly found to be small compared with $s$. 

It was found that one could choose optimal $\zeta_i$ start values for the Pad\'e approximation by minimizing the deviation between $E_R(\zeta=0)$ from different $r_{\mathrm{s}}$ or $r_{\mathrm{c}}$ calculations. This is simple to implement, and works well for different values of $N_{\mathrm{p}}$. In this way, one obtains results that are independent of the starting point for the rotation ($r_{\mathrm{s}}$ in SES) or the imaginary potential ($r_{\mathrm{c}}$ in CAP). The results are, therefore, very stable with respect to the analytic continuation parameters and depend only on the parameters from the mapped Fourier grid method. As shown in Fig.~\ref{NS} for SES (a very similar graph was obtained for CAP), there is a range of $s$ for which results are stable. For larger basis size $N$ the stable $s$-range increases making a judicious choice of $s$ less important. 

We have averaged the results from both the SES and CAP methods over the stable $s$ region, for a basis size of $N=500$ in tables \ref{sestable} \& \ref{captable}. The standard deviation for the width $\sigma(\Gamma)$ within either method is below $10^{-5}$, while the results differ at this level. It is, therefore, of interest to determine the reliability of the error estimate which is based upon basis parameter variations using larger-$N$ extrapolated calculations. Comparison of the width results with such an estimate based upon $N=3000$ CAP and SES calculations indicates that SES and CAP converge to the same value, closest to the CAP value given in table \ref{captable}.

Concerning the most appropriate order for the extrapolations it is worth noting that the CAP Pad\'e trajectories are rather straight in comparison with the ones for SES, and $N_{\mathrm{p}}=4$ might be more appropriate in this case. Nevertheless, no systematic improvement was found when going to the lower-order approximation (which might be deemed more stable with respect to numerical noise in the trajectory points). 

\clearpage
\section{Beyond the monopole approximation}\label{3chpade}
The ability to compute the resonance parameters to high precision with moderate basis size (e.g., $N=500$) allows for the inclusion of higher multipoles which are present in the two-center interaction. While the effect on the resonance position is expected to be small, the sensitivity of the width to computational details (cf. the different results discussed in chapter~\ref{CSCAP}) indicates that some broadening of the resonance may occur when the two-center nature of the potential is included.

To account for two-center potential effects the wavefunction is expanded using spinor spherical harmonics, $\chi_{\kappa,\mu}$, 
\begin{equation}
\Psi_{\mu}(r,\theta,\phi)=\sum_{\kappa}{ \left(
\begin{array}{c}
G_{\kappa}(r)\chi_{\kappa,\mu}(\theta,\phi) \\
iF_{\kappa}(r)\chi_{-\kappa,\mu}(\theta,\phi)
\end{array}
\right)} \quad ,
\end{equation}
which are labeled by the relativistic angular quantum number $\kappa$ (analogous to $l$ in non-relativistic quantum mechanics) and the magnetic quantum number $\mu$ \cite{greiner}. The Dirac equation for the scaled radial functions, $f(r)=rF(r)$ and $g(r)=rG(r)$, then becomes ($\hbar=$ c $=$ m$_{\rm e}=1$),
\begin{eqnarray}
\label{syseqnG}
\frac{df_{\kappa}}{dr} - \frac{\kappa }{r}f_{\kappa} & = & - \left(E -1 \right)g_{\kappa} + \sum_{\bar{\kappa}=\pm1}^{\pm\infty}{\langle \chi_{\kappa,\mu} \left| V(r,R) \right| \chi_{\bar{\kappa},\mu} \rangle }g_{\bar{\kappa}} \quad , \\
\label{syseqnG2}
\frac{dg_{\kappa}}{dr} + \frac{\kappa}{r}g_{\kappa} & = & \left(E + 1 \right) f_{\kappa} - \sum_{\bar{\kappa}=\pm1}^{\pm\infty}{\langle \chi_{-\kappa,\mu} \left| V(r,R) \right| \chi_{-\bar{\kappa},\mu} \rangle }f_{\bar{\kappa}} \quad ,
\end{eqnarray}
where $V(r,R)$ is the potential for two uniformly charged spheres displaced along the $z$-axis, which is expanded into Legendre polynomials according to $V(r,R)=\sum^{\infty}_{l=0}{V_l(r,R) P_l(\cos{\theta})}$ \cite{greiner}. The monopole term, $V_0$, is given explicitly in Eq.~\ref{potential}, and for the present work we include the coupling terms required for the $\kappa=\pm1,-2$ channels (i.e., the $V_1(r,R)\langle \chi_{\pm\kappa,\mu} \left| P_1 \right| \chi_{\pm\bar{\kappa},\mu} \rangle$ dipole and $V_2(r,R)\langle \chi_{\pm\kappa,\mu} \left| P_2 \right| \chi_{\pm\bar{\kappa},\mu} \rangle$ quadrupole terms). In general, the $\kappa=1,-2$ channels ($P_{1/2}$ and $P_{3/2}$ respectively) have the strongest coupling to the $\kappa=-1$ (S-states) and are therefore expected to have the largest impact on the supercritical ground state.

\begin{table}[!ht]
	\centering
\begin{tabular}{|c||r@{.}l|r@{.}l||r@{.}l|c||c|}
\hline
\multirow{2}{*}{$R$(fm)} & \multicolumn{4}{|c||}{Single-channel ($\kappa=-1$)} & \multicolumn{3}{|c||}{Three-channel ($\kappa=-1,1,2$)} & $\frac{|\Gamma_1-\Gamma_3|}{\Gamma_1}$ \\ 
\cline{2-8}
& \multicolumn{2}{|c|}{$E_{\mathrm{res}}$ (mc$^2$)}& \multicolumn{2}{|c||}{$\Gamma_1$ (keV)} & \multicolumn{2}{|c|}{$E_{\mathrm{res}}$ (mc$^2$)}& $\Gamma_3$ (keV) &  ($\times10^{-3}$)\\
\hline\hline
16 & -2&00635363(3) &  8&148233(4) & -2&00646180(3) & 8.150153(3) & 0.236\\
18 & -1&87487669(4) &  5&881605(1) & -1&87502343(4) & 5.883977(2) &	0.403 \\
20 & -1&75793073(4) &  4&0848148(4)& -1&75811259(4) & 4.087394(1) & 0.631\\
22 & -1&65393272(3) &  2&708987(1) & -1&65414448(3) & 2.711524(2) & 0.937\\
24 & -1&5612122(1)  &  1&6966694(3)& -1&56144826(1) & 1.698947(1) & 1.34\\
26 & -1&47820830(4) &  0&9877052(4)& -1&47846358(4) & 0.989571(1) & 1.89\\
28 & -1&40354329(5) &  0&5221621(4)& -1&40381341(4) & 0.523544(1) & 2.65\\
30 & -1&33603643(6) &  0&2420749(2)& -1&33631779(6) & 0.242976(1) & 3.72\\
32 & -1&27469286(5) & 0&0931429(1)& -1&27498253(5) &  0.093639(5) & 5.32\\
34 & -1&21868219(5) & 0&0271161(1)& -1&21897787(4) &  0.027322(2) & 7.59\\
36 & -1&16731491(6) & 0&0050374(1)& -1&16761472(7) &  0.005091(2)& 10.7\\
38 & -1&12001745(6) & 0&0004170(3)& -1&12031983(6) &  0.000439(4) & 52.5 \\
\hline
\end{tabular}
	\caption[Single- and three-channel U-Cf resonance parameters]{\spacing{1}\label{3captable}Averaged $1{\mathrm S}\sigma$ resonance position and width from the extrapolated CAP method with basis size $N=500$, $N_{\mathrm{p}}=4$, for single-channel (monopole) and three-channel calculations as a function of separation, $R$, in the U-Cf system. The values were averaged over the stable range $300\le s\le 700$ using $r_{\mathrm{c}}=2,3,4$. The digit in parentheses represents the standard deviation from the average and the last column gives the relative difference of the width obtained from three-channel versus single-channel calculations.}
\end{table}

Table \ref{3captable} gives the $1{\mathrm S}{\sigma}$ resonance parameters for different internuclear separations for both the one- and three-channel CAP calculations in the uranium-californium system at an internuclear separation of $R=20$fm. The CAP calculations were performed using a basis size of $N=500$ per channel with $300\le s\le 700$. The CAP parameters of $r_{\mathrm{c}}=2,3,4$ were used to obtain the most stable $\eta$-range for extrapolation, which was carried out with order $N_{\mathrm{p}}=4$. Values in parentheses indicate the standard deviation of the mean (cf. Eq.~\ref{std}). It is clear that the accuracy of the calculations - as indicated by the deviation - is much higher than what is required to measure the effect of the the P-state channels on the $1{\mathrm S}{\sigma}$ resonance. 

The final column contains the relative difference of the width between the three-channel results and the monopole approximation. The correction due to the dipole and quadrupole potentials is seen to increase with internuclear separation $R$. This trend can be expected, as the overall interaction becomes less spherically symmetric in this limit. The effect is small, however, as the strongest contribution towards the binding energy (and thus the supercriticality) is provided by the monopole part. The dipole interaction contributions to the resonance parameters are a result of the relatively small charge asymmetry in the U$^{92+}$-Cf$^{98+}$ potential. 

By further including couplings to the $D$-states ($\kappa=2,-3$ or $D_{3/2}$ and $D_{5/2}$ respectively), which includes an $S-D$ quadrupole coupling, significant corrections to the supercritical resonance parameters are obtained. Table~\ref{5captable} gives the results of the five-channel calculation compared to the single-channel data. While coupling up to $P_5(\cos(\theta))$ are included, it is the quadrupole interaction between the $S$ and $D$ states which causes the change in the resonance parameters. Since the dipole interaction is small for the present system, the next-order effect after the monopole interaction is the quadupole coupling. The large changes to the resonance parameters are due to the inclusion of five-channels resulting in the $S$-channel coupling to the $D$-channel by the quadupole coupling (in three-channel the quadupole only coupled $\kappa=-2$ with $\kappa=1$).
\begin{table}[H]
	\centering
\begin{tabular}{|c||r@{.}l|r@{.}l||r@{.}l|c||c|}
\hline
\multirow{2}{*}{$R$(fm)} & \multicolumn{4}{|c||}{Single-channel ($\kappa=-1$)} & \multicolumn{3}{|c||}{Five-channel ($\kappa=\pm1,\pm2,-3$)} & $\frac{|\Gamma_1-\Gamma_3|}{\Gamma_1}$ \\ 
\cline{2-8}
& \multicolumn{2}{|c|}{$E_{\mathrm{res}}$ (mc$^2$)}& \multicolumn{2}{|c||}{$\Gamma_1$ (keV)} & \multicolumn{2}{|c|}{$E_{\mathrm{res}}$ (mc$^2$)}& $\Gamma_3$ (keV) &  ($\times10^{-3}$)\\
\hline\hline
16 & -2&00635363(3) &  8&148233(4) & -2&03582804(3) & 8.681022(8) & 65.38\\
20 & -1&75793073(4) &  4&0848148(4)& -1&79673776(4) & 4.650882(4) & 138.6\\
24 & -1&5612122(1)  &  1&6966694(3)& -1&60546338(2) & 2.150177(9) & 267.3\\
28 & -1&40354329(5) &  0&5221621(4)& -1&45044923(6) & 0.794256(9) & 521.1\\
32 & -1&27469286(5) & 0&0931429(1)&  -1&32257286(8) & 0.201075(4) & 1158.8\\
\hline
\end{tabular}
	\caption[Single- and five-channel U-Cf resonance parameters]{\spacing{1}\label{5captable}Averaged $1{\mathrm S}\sigma$ resonance position and width from the extrapolated CAP method with basis size $N=500$, $N_{\mathrm{p}}=4$, for single-channel (monopole) and five-channel calculations as a function of separation, $R$, in the U-Cf system. The values were averaged over the stable range $300\le s\le 700$ using $r_{\mathrm{c}}=2,3,4$. The value in parenthesis represents the standard deviation from the average and the last column gives the relative difference of the width for the five- and single-channel calculations.}
\end{table}

\section{Conclusion}
We can summarize the findings from this chapter as follows: the generalization of the extrapolation method from CAP to SES, and the combination of data from both methods allowed us to obtain highly accurate data for relatively small basis size $N$ (utilizing results for an optimum range of the mapped Fourier grid scaling parameter $s$, as well as analytic continuation method parameters $r_c,r_s$). With this breakthrough in computational efficiency we were then able to explore the sensitivity of the resonance parameters to two-center interaction effects. For larger internuclear separations (near the onset of supercriticality) these effects are appreciable. The augmented methods are also applicable to all resonance phenomena to which analytic continuation methods are applied.

\clearpage
\chapter{Time-dependent supercritical states}\label{tdscchp}
While static supercritical resonance states due to a single nucleus are unlikely to exist (or be sufficiently long-lived if such a nucleus was created by fusion), it is possible to gain access to this phenomenon using heavy-ion collisions near the Coulomb barrier. In these collisions, the atoms are taken to be fully-ionized for reasons mentioned in chapter~\ref{physics} and for the practical reason of theoretical simplicity due to the lack of electron-electron interactions (Coulomb or spin-orbit). 

A classical treatment of the nuclear motion by a trajectory $R(t)$ generates an explicitly time-dependent Dirac Hamiltonian for the electron motion which does not conserve energy.  Energy can be taken from the nuclear motion to excite electrons. In the Dirac many-particle interpretation it also means electron-positron pairs can be created (cf. section~\ref{dynsppair}). The potential was previously given in Eq.~\ref{potential} for each nucleus. The potential is modified by the substitution $R\rightarrow R(t)$, since the nuclei are no longer static. 

Direct numerical integration of the Dirac equation over time as a partial differential equation in 3+1 dimensions is practically impossible with present computers, therefore another approach is followed. The method proposed in this thesis relies on the fact that the mapped Fourier grid method generates a complete set of eigenstates (spanning the subspace) for the time-independent Hamiltonian in an efficient manner. All equations are in natural units ($\hbar=$ m$_{\rm e}=$ c $=1$).

\section{Time-dependent Dirac equation with particle creation}
The supercritical resonance, which is the principle focus of this work, primarily influences the pair creation spectra for electrons and positrons. These spectra, and the total amount of pair creation, are, thus, the desired quantities. In contrast to Schr\"odinger quantum mechanics where one has unitary time evolution of individual orbitals, the Dirac theory is to be considered as a multi-particle theory. This is true even if the initial condition involves a single occupied (or unoccupied) state. This follows from treating the Dirac equation as part of a quantized field theory. It requires that the propagator is applied to all initial states which will then allow for the calculation of the quantities of interest, i.e. the observables. This enables the calculation of the electron and positron creation spectra which can then be compared with experimental data.

\subsection{Quantities of interest}\label{qofint}
The signature for the existence of a supercritical resonance state is to be found in the positron and electron energy spectra. This is due to the decay mode of the supercritical resonance state resulting in an electron in the ground state of one of the receding ions and a free positron with an energy close to the supercritical resonance energy (cf. section~\ref{scgs}). 

The probabilities of electron creation ($n_i$ with $i>F$) and positron creation ($\bar{n}_i$ with $i<F$) in a state $i$ are given by
\begin{eqnarray}\label{creat1}
\langle n_i \rangle &=& \sum_{\nu<F}{|\langle \chi^{(+)}_i|\phi_\nu(t_f) \rangle |^2} \\ \label{creat2}
\langle \bar{n}_i\rangle &=& \sum_{\nu>F}{|\langle \chi^{(-)}_i|\phi_\nu(t_f) \rangle |^2}.
\end{eqnarray}
Here $F$ is the Fermi energy (cf. section~\ref{holetheory}), $|\chi_i^{(+)}\rangle$ and $|\chi_i^{(-)}\rangle$ are positive- and negative-eigenstates of the single-center Hamiltonian respectively. The propagated state, $|\phi_\nu(t)\rangle$, corresponds to $|\chi_\nu\rangle$ at $t \rightarrow -\infty$ (as explained in appendix~\ref{appendix2} following the derivation of Ref.~\cite{PhysRevA.45.6296}). Thus, Eq.~\ref{creat1} implies that the population of electron states (bound or free) is obtained from evolving initial states $|\chi_\nu\rangle \rightarrow |\phi_\nu(t)\rangle$ for  $\nu<F$, i.e., for negative-energy states to times $t\geq t_f$ (when the collision is over), and projecting onto the electron states. Similarly, Eq.~\ref{creat2} implies that the positron population at energy $E\approx $m$_{\rm e}$c$^2+p_i^2/(2$m$_{\rm e})$ is obtained by propagating the positive-energy initial states and projecting onto the positron continuum states.

For the positron spectra, the ground state ($|\chi^{(+)}_{\rm 1S}\rangle$) provides the dominant contribution of all initial positive-energy states, since it has the strongest interaction with the negative energy continuum (cf. section~\ref{dynsppair}). The propagated ground state ($|\phi_{\rm 1S}(t)\rangle=U(t,-\infty)|\chi^{(+)}_{\rm 1S}\rangle$, where $U(t_2,t_1)$ is the propagator) thus yields a first approximation to the complete positron spectrum (which has the contributions from all positive-energy states). This lowest-order spectrum is given by
\begin{equation}\label{1scr}
\langle \bar{n}_i \rangle \approx |\langle \chi^{(-)}_i|\phi_{\rm 1S}(t) \rangle |^2.
\end{equation}
The initial 1S state ($|\chi^{(+)}_{\rm 1S}\rangle$) is propagated through the collision. The resultant state ($|\phi_{\rm1S}\rangle$) is projected onto the single-center positron states ($|\chi^{(-)}_i\rangle$). One can then add the contributions from other states in the same manner. 

The first excited state can be expected to play a significant role in dynamical pair production (cf. section~\ref{dynsppair}). It also plays some role in the spontaneous pair production despite remaining subcritical. This is caused by the fact that the time-varying potential couples the first excited state strongly to the ground state even for intermediate nuclear distances. 

Thus, a more complete positron spectrum is obtained by adding the the contribution of the initial first excited state ($|\chi^{(+)}_{\rm 2S}\rangle$) to the spectrum from Eq.~\ref{1scr}. Similarly, the other positive-energy states may be added, one by one. This allows for an additional check on the calculation since the contribution from higher states should diminish.

It is not possible to propagate \textit{all} states above the Fermi level, and the problem is compounded by the difficulty to propagate free states, since, in principle, the spectrum is continuous. To complete the electron spectrum (Eq.~\ref{creat1}) is particularly difficult, since all the states to be propagated are continuum states ($|\chi^{(-)}\rangle$). As explained in chapter~\ref{physics}, the low-lying continuum states ($|E|\approx$ m$_{\rm e}$c$^2$) are unlikely to play a role since their wavenumbers are relatively large, i.e., their coupling due to energetic collision dynamics is weak (small matrix elements). Furthermore, one should expect that states of very high energy, $|E| \gg $ m$_{\rm e}$c$^2$, should not play a significant role. These states represent scattering energies too high to couple into the changing potential on the basis of absorbing energy from the nuclear motion. Therefore, only some range of intermediate-energy scattering states should contribute substantially.

The propagation of continuum states is possible within the mapped Fourier grid method. This is due to the continua being discretized, with adjustable coverage. The density of states is determined by both the scale parameter $s$, and the basis size $N$. The discretization, by a finite number of states, means that one deals with quasi-continuum states. 

These states do not represent a single energy, but a range of energies, as they are localized in a finite volume of space. The energy resolution is limited by the fact that quasi-continuum (wavepacket) states have a finite energy width. The advantage of a discretized $E>$ m$_{\rm e}$c$^2$ and  $E<-$m$_{\rm e}$c$^2$ continua is provided by the ability to propagate the quasi-continuum states as initial states in exactly the same way as for the bound states. One has the freedom to provide adequate coverage for the necessary energy ranges in Eqs.~\ref{creat1} and \ref{creat2}. 

A weakness of the mapped Fourier grid method is that the resultant basis is not perfectly orthogonal (due to numerical noise), which necessitates changes to the expressions for the expansion coefficients (given by $\langle \chi_i|\phi_\nu(t_f)\rangle$ for an orthogonal basis). However, this does not result in any change to Eqs.~\ref{creat1} and \ref{creat2}, since they are derived for the expansion coefficients (cf. appendix~\ref{appendix2}). 

\subsection{Approximating the propagator}\label{tdmfg}
The time evolution of the state $\phi_{\nu}(t)$ during the collision is carried out by the propagator, 
\begin{equation}\label{propeq}
U(t,t_0)=T\left\{ \exp\left(-i \int_{t_0}^t dt' H(t')\right) \right\}
\end{equation}
(where $T$ is the time-ordering symbol), which acts on a state at time $t_0$ and generates the state at $t$. The propagation evolves the state according the the time-dependent collision hamiltonian, $H(t)$. The direct application of the propagator is very difficult. It is, therefore, approximated resulting in a two-step process whereby the state is propagated over a short time segment after projection onto a quasi-stationary eigenbasis. This is achieved by approximating the propagator integral, given by
\begin{equation}\label{Uop}
\int_{t_0}^t dt' (\vec{\mbox{\boldmath $\alpha$}}\cdot \vec{{\bf p}}+\hat{\beta} + V_{\mathrm{2\;center}}(r,R(t')) ),
\end{equation}
for small time intervals as a sum of small, rectangular areas. 

The approximation proceeds as follows. Given an initial state of the system, $| \Psi(\mathbf{r},t_0) \rangle$, the state at a later time is given by,
\begin{equation}\label{expH}
| \Psi(\mathbf{r},t)\rangle=U(t,t_0) | \Psi(\mathbf{r},t_0) \rangle=T\left\{ \exp\left(-i \int_{t_0}^t dt' H(t')\right) \right\}| \Psi(\mathbf{r},t_0) \rangle.
\end{equation}
To approximate Eq.~\ref{expH}, small time steps may be used so that $H(t)$ is assumed constant and may be taken out of the integrand. The propagator for each time step then becomes 
\begin{equation}\label{sprop}
U(t_1,t_0) = \exp\left(-i H(t_1)(t_1 - t_0) \right).
\end{equation}

\begin{figure}[H]\centering
\includegraphics[angle=0,scale=0.93]{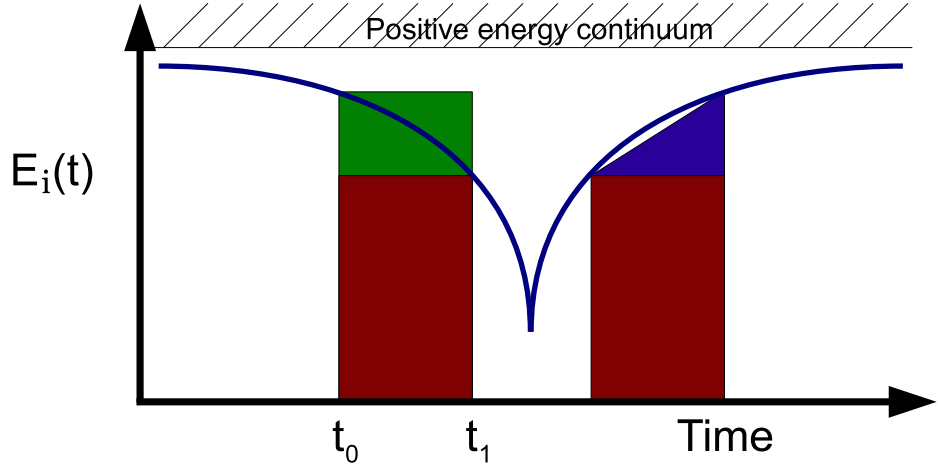}
\caption[Illustration of the propagator approximation]{\spacing{1}\label{Eoft} Illustration of the propagator approximation in an eigenbasis. The (blue) curve shows the eigenenergy of an initial bound state as a function of time. The total left rectangle (red plus green) shows the left-point approximation to the integral in Eq.~\ref{Uop} while the bottom (red) rectangle shows the right-point approximation. The rectangle with a triangle on the top shows the trapezoid approximation actually used in the current work. Translation along the horizontal (time) axis is executed by propagating the state; while movement along the vertical ($E_i(t)$) axis is the result of projecting the state onto the $H(t+\Delta t)$ basis.}
\end{figure}

Constructing a basis for $H(t_1)$ allows for simple time propagation. This is carried out by the projection of the initial state, $| \Psi(\mathbf{r},t_0 ) \rangle$, onto the basis defined by $H(t_1)|\phi_m\rangle=E_m(t_1) |\phi_m\rangle$. Then the propagator is applied to the $\phi_m$-basis. The resultant state is given by
\begin{equation}\label{projprop}
 \Psi(\mathbf{r},t_1 ) \rangle =U(t_1,t_0)| \Psi(\mathbf{r},t_0 ) \rangle= \sum_{\mathrm{m}} a_{\mathrm{m}} \exp\left(-i E_{\mathrm{m}}(t_1 - t_0) \right) |\phi_{\mathrm{m}} \rangle.
\end{equation}
This can be visualized as approximating Eq.~\ref{Uop} by summing a number of small rectangular areas. In an eigenbasis, $H(t)\rightarrow E_i(t)$ for the $i$th eigenstate. Figure~\ref{Eoft} illustrates an $E(t)$ curve for the $i$th initial bound state. 
As the nuclei approach each other, the eigenenergy of the $i$th state decreases and reaches a minimum when the nuclei are at closest approach. The total rectangle starting at $t_0$ (green plus red) shows the left-point approximation used to approximate the integral of Eq.~\ref{Uop}. The right-point approximation (used in Eqs.~\ref{sprop} and \ref{projprop}) is shown as the lower (red) rectangle. $E_i(t)$ is symmetric resulting in both approximations being equal.

The propagation of the state is visualized in Fig.~\ref{Eoft} as a horizontal displacement. The projection of the state onto the $H(t+\Delta t)$ basis is visualized as a vertical displacement. Using the right-point approximation thus amounts to propagating first and then projecting, while the left-point approximation is the reverse. By summing all these rectangles an approximation to Eq.~\ref{Uop} is obtained. 

In the present work, a better approximation to the $U(t_1,t_0)$ operator is employed by taking the average eigenvalue of the previous and current time for Eq.~\ref{Uop} giving
\begin{equation}\label{lprop}
U(t_1,t_0) = \exp\left(-i \frac{H(t_1)-H(t_0)}{2}(t_1 - t_0) \right).
\end{equation}
This approximation to a definite integral is known as the trapezoid rule and has an error of $O(\Delta t^3)$. It is illustrated in Fig~\ref{Eoft} as the triangle plus the rectangle. This approximation allows for the use of a coarser time-mesh to save computational time and does not require much overhead. 

\subsection{Difficulties associated with a non-orthogonal basis}
In appendix~\ref{mfgde} the problem of an non-symmetric representation of the Hamiltonian is discussed in terms of the time-independent problem. Time-dependent problems also require special attention due to the fact that the representation is non-symmetric, since this leads to a set of eigenvectors that can be non-orthogonal due to the numerical noise, but still span the $N$-dimensional subspace (all eigenvalues are unique and real).

In order to accomplish the time propagation of section~\ref{tdmfg} the wavefunctions must be projected onto the new set of states, as shown in Eq.~\ref{projprop}. The set of states spans the Hilbert space, i.e., no norm should be lost due to the change of representation. However, the expansion coefficients are not simply given by $\langle \phi_i|\Psi\rangle$, the inner product of the wavefunction and the $i$th basis state. For a state 
\begin{equation}\label{psia}
\left| \Psi \rangle\right. = \sum_n{a_n \left| \phi_n \rangle\right.} 
\end{equation}
the inner product coefficient is given by 
\begin{equation}\label{cm}
c_m = \langle \phi_m \left|\right. \Psi\rangle.
\end{equation}
Inserting \ref{psia} into \ref{cm} the expansion coefficient, $a_n$, is related to the inner product coefficient according to
\begin{eqnarray}\label{expcoefdef}
c_m & = & \sum_n{a_n \left( \langle \phi_m \left|\right. \phi_n\rangle \right)} \\
\Rightarrow a_n & = & \sum_m{c_m \left( \langle \phi_m \left|\right. \phi_n\rangle \right)^{-1}}.
\end{eqnarray}
It is clear that the non-orthogonality must be taken into account using the inverse of the inner-product matrix. This adds some overhead to the computation, since the inner-product matrix must be computed and inverted, but this process is significantly faster than diagonalization. Therefore, the inversion does not affect the overall order of the computational problem ($O(N^3)$).

\section{Bare uranium-uranium collisions}
There are many different collision systems which become supercritical at closest approach and could be used to demonstrate the existence of supercritical resonance states. The shorter the decay time the larger the positron yield from the supercritical resonance, making it easier to detect. Therefore, the heaviest system which is accessible to experiments is the most desirable. Over the next decade the GSI-FAIR collaboration is planning to perform bare uranium-uranium merged-beam collisions. Thus, we concentrate on this collision system in the present thesis. The computational method can be easily adapted to other bare-nuclei collision systems.

A calculation of the complete collision dynamics is far beyond the currently available computer resources, and thus other approximations are made to the system in addition to the propagator approximation discussed in section~\ref{tdmfg}. The current work is limited to zero-impact-parameter trajectories. This allows one to employ cylindrical symmetry for the Dirac field, eliminating the need for magnetic subchannels in the matrix representation. For zero impact parameter the coupling matrix elements arise from a multipole expansion of the potential (cf. section~\ref{3chpade}). 

In the present work we restrict the potential to the monopole term. This decouples all angular symmetries and we can solve the dominant S-channel ($\kappa=-1$) in isolation. These approximations break the relativistic covariance of the Dirac equation forcing one to make a judicious choice of reference frame.

\subsection{Choosing a reference frame}
While all states need to be propagated, the choice of reference frame has a dramatic effect on the results. The center-of-mass frame, which appears as an obvious choice, is not very well suited for the mapped Fourier grid method in the present work. This is caused by the representation of continuum states as a quasi-continuum with a non-uniform coordinate mesh (cf. appendix~\ref{appendix1}). The low-lying continuum states, which become bound states as the nuclei approach each other, represent electrons with low wavenumber moving too slowly to react efficiently to the (relatively) fast collision.  

Therefore, in order for the calculation to be carried out in the center-of-mass frame the upper continuum must be well sampled and the states should be properly represented. This is due to the restriction to the monopole approximation of the two-center potential which does not localize states in the vicinity of the nuclei for large internuclear separations. The monopole approximation, in fact, is only useful at short distances, since the initial and final states are defined for the hamiltonian of a single nucleus.

In contrast to the center-of-mass frame, the target frame of one of the nuclei allows for the treatment of the second nucleus as a perturbation. The successful solution of the hydrogenic problem (cf. appendix~\ref{appendix1} and Ref.~\cite{me1}) gives us confidence that the initial states are represented properly. Thus, the initial states are taken as eigenstates of hydrogen-like uranium. By starting the calculation at a sufficiently large internuclear separation, $R_{\rm initial}$, the basis change from the initial hydrogen-like uranium states to the perturbed hydrogen-like uranium will not induce any transitions.

There remains a subtle problem of the proper frame to use during supercriticality. In section~\ref{otherconf} the relationship between the resonance energy and width is demonstrated. This is important since the eigenvalue is frame dependent: as a consequence the resonance width (i.e., the positron yield) depends on the choice of frame. The center-of-mass frame gives deeper resonance energies for the same configuration and is more appropriate as the second nucleus is too close to be considered a perturber. It is also the frame in which the monopole approximation to the interaction is defined properly. Therefore, we make a change of basis at an internuclear separation of $R_{\rm c.b.}=0.1$ (38.6fm) both before and after closest approach. This must take place before supercriticality sets in, which yields an optimal choice right before the system becomes supercritical in either frame ($R\approx 36$fm). 

\subsection{The trajectory}
To demonstrate the  supercritical resonance effect the trajectory should maximize the time the system is supercritical, as discussed in section~\ref{trajectoryintro}. Therefore, a Coulomb trajectory of moderate energy (740MeV total center-of-mass energy) is chosen. The mutual repulsion of the nuclei slows them down (Coulomb barrier) resulting in very slow nuclei near closest approach allowing for the system to experience supercriticality.

The trajectory is generated by solving Eq.~\ref{trajeq} numerically using constant time steps within three distinct regions. When the nuclei are far away the time step, $\Delta t$, is increased by an integer factor. This decreases the number of steps needed in the far region and is justified by the slow change of the eigenstates at large distances. By examining the eigenvalue of the 1S$\sigma$ as a function of the internuclear separation, it was found that for uranium-uranium collisions at 740MeV this region occurs around $R>8$ (3089fm). The opposite is true when the nuclei are close. The time step, $\Delta t$, used is reduced for $R<2$ (772fm). The division of the trajectory into these three regions allows for an increased time resolution near closest approach, and saves time-intensive steps (diagonalization and projection) in the far region.

\subsection{Time scales}
At $E_{\rm cm}=740$MeV, the uranium nuclei will reach about 16.5fm (center to center) at closest approach. The supercritical resonance at closest approach will be the deepest and have the shortest life-time (cf. section~\ref{otherconf}). This resonance has a life-time of 392 zepto-seconds ($10^{-21}$s). The collision system will be supercritical for 2.3 zepto-seconds, a time which is shorter by two orders of magnitude.

Even though supercriticality lasts only for a short time, the calculation must span a much longer time frame. The collision system takes 250 zepto-seconds to reach closest approach (when started from far enough away to not induce any transitions due to changing frames from single center to the target frame). After supercriticality, the collision must be propagated for at least another 2 atto-seconds ($10^{-18}$s). This is because there are still transitions taking place. 

Given that the fastest-decaying supercritical resonance has a lifetime much longer than the collision time, one is interested in considering nuclear sticking, where the nuclei remain fixed at closest approach (cf. section~\ref{nstick}).

The sticking times used should be realistic, i.e., experimentally obtainable in deep inelastic collisions. Such times can be on the order of a zepto-second (a few zepto-second sticking has been detected \cite{PhysRevLett.50.1838} and theory predicts slightly more with fully ionized atoms \cite{EurPJA.14.191,zagrebaev:031602,Sticking2}). A sticking time of a few zepto-seconds is still only a few percent of the resonance decay time which would, therefore, seem to escape detection. One would, therefore, not expect to see a sharp supercritical resonance peak superimposed on top of the broad dynamical positron spectrum (cf. section~\ref{dynsppair}). 

The supercritical resonance creation spectrum, for sticking times ($T$) much less than its life-time, may be conjectured as a Breit-Wigner shape (cf. Eq.~\ref{bweqn}) with $\Gamma\approx \hbar/T$. The height of this peak should also be related to the sticking time, $T$, since longer sticking would yield more supercritical resonance decay. Dynamical pair production on the other hand, which is unchanged during sticking, will interfere with the resonant pair production. This interference may provide a suitable method of detection.

\subsection{Results for a fully-ionized uranium-uranium collision}\label{t0ressec}
The collision of fully-ionized uranium nuclei following a Coulomb trajectory at a center-of-mass energy of 740MeV was computed. The parameter space of the work is large, but each parameter has been minimized separately reducing numerical uncertainty to below 0.5\%. The center-of-mass energy of 740MeV corresponds to a closest-approach distance of 16.5fm. This results in a deep resonance with $E_{\rm res}=-1.56$m$_{\rm e}$c$^2$ and $\Gamma=1.68$keV calculated using the technique of chapter~\ref{padechp}. The nuclei are estimated to have a radius of $R_{\rm n}=7.44$fm (using $R_{\rm n}=1.2\times A^{1/3}$ with $A=$238 for uranium). The nuclei are then 1.6fm away from touching. All results are in natural units ($\hbar=$ m$_{\rm e}=$ c $=1$) resulting in an energy of $E=\pm 1$ corresponding to $E=\pm$m$_{\rm e}$c$^2$.

\begin{figure}[!ht]\centering
\includegraphics[angle=270,scale=0.65]{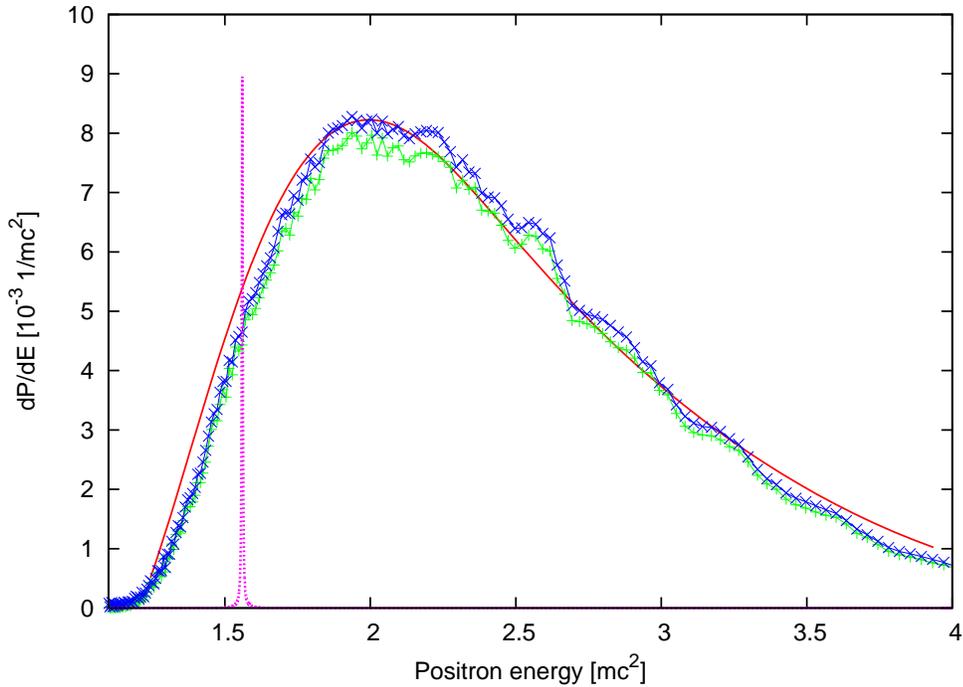}
\caption[Positron spectrum from fully ionized U-U collision]{\spacing{1}\label{T0res} The positron spectrum for fully ionized U-U collisions at $E_{\rm cm}=740$MeV. The \textcolor{green}{$+$}'s are from an initial 1S state only (cf. Eq.~\ref{1scr}) and the \textcolor{blue}{$\times$}'s is from having propagated all bound states. The (red) curve approximates the results from M\"uller \textit{et al.} \cite{PhysRevA.37.1449}. The (magenta) dotted curve is a (scaled) Breit-Wigner shape with the parameters of the supercritical resonance at closest approach ($R=16.5$fm). }
\end{figure}

Figure~\ref{T0res} display the differential probability, $dP/dE$, for the positron spectrum using the parameters: $N=1024$, $s=700$, $R_0=25$ (9.65pm), $R_f=200$ (77.2pm) and $\Delta t=1.88$ (2.42$\times 10^{-21}$s). In the far region ($R >8$) we used $4\times\Delta t$ and in the near region ($R<2$) we used $\Delta t/100$. The (green) curve with \textcolor{green}{$+$}'s is the spectrum obtained by propagating an initial 1S state using Eq.~\ref{1scr}. Each of the \textcolor{green}{+}'s indicate a positron quasi-continuum state ($|\chi_i^{(-)}\rangle$) used for calculating $\langle \bar{n}_i\rangle$ from Eq.~\ref{1scr}. The (blue) curve with the \textcolor{blue}{$\times$}'s is also the positron spectrum, but following propagation of all bound states (Eq.~\ref{creat2} with the sum terminating at the onset of the positive continuum). The (red) curve approximates the results from M\"uller \textit{et al.} \cite{PhysRevA.37.1449} shown in Fig.~\ref{mullerfig}.
Propagating all bound states adds less than six percent to $\Delta P/\Delta E$ at any energy. The addition of the excited bound states does not alter the spectrum significantly. As was expected, the positron states near the $|E|\approx$m$_{\rm e}$c$^2$ boundary do not participate. 

The results obtained are in close agreement with M\"uller \textit{et al.}'s close-coupling calculation \cite{PhysRevA.37.1449}. The only notable difference is on the lower energy side of the peak. The peak position is in close agreement for both calculations. 

The (magenta) dotted curve is a plot of the supercritical resonance structure (a Breit-Wigner using Eq.~\ref{bweqn} but scaled to the peak of the positron spectrum) at closest approach ($R=16.5$fm). The dynamically calculated positron peak is very broad compared to the supercritical resonance and is not centered at its peak. There is no indication of the supercritical resonance in this spectrum. One may adopt the view that the positron spectrum is too heavily dominated by dynamical positron production for the resonance signal to be observed.

While the energy density near the supercritical resonance position is high, it would not necessarily be enough to show the sharp resonance peak. This may lead one to speculate that the peak is simply unresolved. The results from chapter~\ref{tdscstates} address this. Despite it being unlikely that an eigenstate will fall at the resonance energy ($\chi_{E=E_{\rm res}}^{(-)}$), the surrounding eigenstates together represent the supercritical resonance state. This is due to the continuum states of the mapped Fourier grid method being quasi-continuum states with a finite energy width. Thus, if a sharp resonance peak was to be found in the spectrum the neighboring eigenstates would have shown it.

\subsection{Enhancement due to nuclear sticking}
Calculations for uranium-uranium collisions with nuclear sticking have also been performed. The same parameters as  in section~\ref{t0ressec} have been used. The only modification is that the nuclei are held fixed at closest approach for a time, $T$, given in zepto-seconds ($10^{-21}$s). 

As long as the nuclei stick together propagation occurs with a frozen $H(t_i)$ only. No projection is done since the basis is unchanged. This results in static occupation probabilities until the sticking ceases. The basis must, therefore, be large enough to allow for the created negative-energy wavepacket to propagate to larger radii without distortion. With a basis size of $N=1024$ this limits the computationally feasible sticking times to less than $T\approx150$ zepto-seconds for uranium-uranium at $E_{\rm cm}=740$MeV.

\begin{figure}[!ht]\centering
\includegraphics[angle=270,scale=0.55]{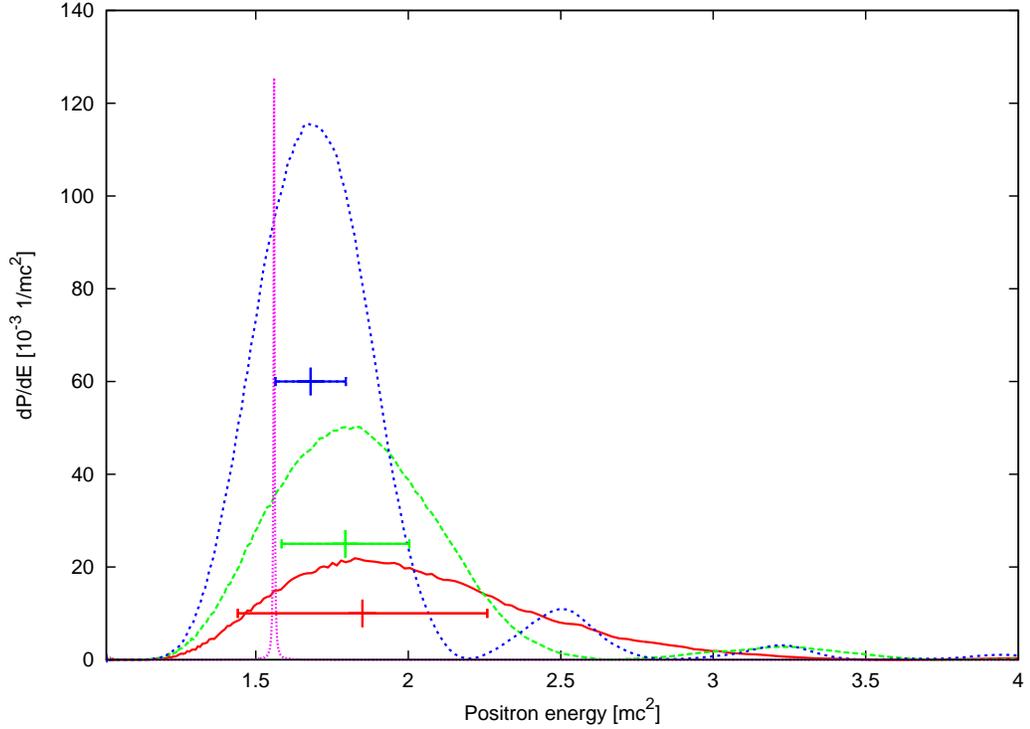}
\caption[Positron spectrum from U-U collision with sticking]{\spacing{1}\label{Tsres} The positron spectrum for fully ionized U-U collisions at $E_{\rm cm}=740$MeV for trajectories with nuclear sticking times of $T=2,5,10$ zepto-seconds ($10^{-21}$s) ordered from lowest to highest main peak. Each curve was calculated using all the bound states. The (magenta) dotted curve is a (scaled) Breit-Wigner shape with the parameters of the supercritical resonance at closest approach ($R=16.5$fm). The points with horizontal error bars of the same color represent the conjectured width ($\hbar/(T+T_0)$ with $T_0=1.15\times10^{-21}$s) of the resonant peak.}
\end{figure}

Figure~\ref{Tsres} displays the positron spectrum, $\langle\bar{n}_i\rangle$, for trajectories with nuclear sticking times of $T=2,5,10$ ordered from lowest to highest peak. 
All the bound states were used as initial states. The (red) curve, representing a sticking time of $T=2$, has its peak shifted to the left as compared to the $T=0$ spectrum by approximately 0.15m$_{\rm e}$c$^2$ (76.6keV). Longer sticking times result in higher and narrower peaks. The (green) long-dashed curve, representing a sticking time of $T=5$ has a secondary peak at $E=3.5$m$_{\rm e}$c$^2$. The (blue) curve, representing the longest sticking time predicted by other theory groups \cite{EurPJA.14.191,zagrebaev:031602,Sticking2}, $T=10$, shows an even higher and narrower peak with a width of 0.45m$_{\rm e}$c$^2$ (229keV). It was analyzed at $R_f=290$ (instead of $R_f=200$) to reduce the numerical noise. 

This curve also has two visible secondary peaks at $E=2.50$m$_{\rm e}$c$^2$ and $E=3.23$m$_{\rm e}$c$^2$. The second peak is higher than the primary peak of the $T=0$ calculation shown in figure~\ref{T0res}. The results shown are in close agreement with those of Ref.~\cite{PhysRevA.37.1449} shown in Fig.~\ref{mullerfig}.

At half of the respective peak's height, a width estimate of the resonance decay spectrum $\hbar/(T+T_0)$ is shown in the same color as the curve. Here $T_0$ is half the time the system is supercritical without sticking, chosen to estimate the supercritical decay without sticking. The comparison shows that the width of the dominant peak structures are not explained perfectly. There can be various reasons for this. The estimate ignores the interplay between dynamical and spontaneous positron production. Furthermore, the calculation includes only the contributions from the propagated initial bound states. The inclusion of initial discretized continuum states with energy $E_n$ has been attempted, but with limited success. These calculations do not converge reasonably with increasing $E_n$. This failure is attributed to problems with poor orthogonality of these states (both among themselves, and to a lesser degree with the bound states). Thus, the present calculation does not include contributions from direct free positron-electron pair production.

The (magenta) dotted curve represents the (scaled) resonance. As the sticking time increases the supercritical resonance decay will play a more dominant role. Indeed, we see in Fig.~\ref{Tsres} that the peak is becoming narrower and more centered near the supercritical resonance energy as the sticking increases. As the width of the resonant pair production spectra (estimated at $\Gamma\approx \hbar/T$) decreases with longer sticking times the interference peaks also shift to lower energy. This is due to the increased dominance of the spectrum by the resonant pair production. At longer sticking times ($20<T \ll \hbar/\Gamma $) very little interference will remain as the dynamical pair production should be dwarfed by the resonant pair production. Currently we know of no way to separate out the resonant and dynamical contributions.

One can interpret the results as follows: naively one would think that the spontaneous decay results in positrons emitted at the resonance energy $E_{\rm res}$ within the natural width $\Gamma$. This would be expected for sticking times of the order of $T \approx\hbar/\Gamma$. However, realistic sticking times are substantially shorter and result in a structure comparable to the dynamical positron spectrum. We speculate that the complicated spectra in Fig.~\ref{Tsres} are a result of the two pair-creation processes (resonant and dynamical) interfering in a complicated way.

\section{Conclusion}
The collision of fully ionized uranium-uranium has been solved in the monopole approximation. The results obtained are consistent with previously published results \cite{PhysRevA.37.1449}. All bound states (obtained in the diagonalization of the single-center hamiltonian) have been used as initial states and the continuum was well covered.

In collisions with no nuclear sticking, there is no evidence of the supercritical resonance decay. The positron spectrum is shown to be a wide peak away from the resonance peak (at closest approach).

The enhancement due to nuclear sticking can be significant due the the narrowing and shifting it causes in the main positron peak. This effect is large and also results in secondary peaks for a sticking time of $T\geq5$. Thus, supercritical resonance states may be detectable by their interference with dynamical pair production.

\appendix
\chapter[The mapped Fourier grid method]{The mapped Fourier grid method \footnote{Some of the following was originally published in Ref.~\cite{me1}.}}\label{appendix1}
The mapped Fourier grid method has been proposed for the non-relativistic Schr\"odinger problem and has been applied successfully in the context of atomic and chemical physics \cite{mfg,symmse,newest}. We have extended the method to the Dirac equation, and have tested it in Ref.~\cite{me1} for the exactly solvable problem of hydrogenic states. We show in this appendix how it yields a matrix representation for the Dirac-Coulomb problem. 
\section[The mapped Fourier grid method for the Dirac equation]{The mapped Fourier grid method for the Dirac equation}\label{mfgde}

The coupled radial Dirac equations for the upper and lower components $G$ and $F$ for a spherically symmetric potential, $V(r)$, are commonly written as \cite{greiner},\cite{sakurai}
\begin{eqnarray}
-c\hbar \frac{dF}{dr} - \frac{(1-\kappa)\hbar c }{r} F & = & (E - V(r) -mc^2) G, \\
c\hbar \frac{dG}{dr} + \frac{(1+\kappa)\hbar c }{r} G & = & (E - V(r) +mc^2) F,
\end{eqnarray}
where $\kappa$ is a quantum number analogous to the angular momentum quantum number $l$ in non-relativistic quantum mechanics. Using $f(r)=rF(r)$, $g(r)=rG(r)$ and the Coulomb potential for a charge $Z$, in natural units $\hbar = m_e=c=1$, the equations become
\begin{eqnarray}
\label{syseqn}
\frac{df}{dr} - \frac{\kappa }{r}f & = & - \left(E + \frac{Z\alpha}{r} -1 \right) g, \\
\label{syseqn2}
\frac{dg}{dr} + \frac{\kappa }{r}g & = & \left(E + \frac{Z\alpha}{r} +1 \right) f,
\end{eqnarray}
 where $\alpha=e^2/\hbar c$ is the fine-structure constant. It is useful to write Eqs.~(\ref{syseqn}) and (\ref{syseqn2}) in matrix notation,
\begin{equation}
\label{Hr}
\left(
\begin{array}{cc}
-\left( \frac{Z \alpha}{r}-1 \right)  & -\left( \frac{d}{dr}-\frac{\kappa}{r}\right)
\\
\left( \frac{d}{dr}+\frac{\kappa }{r}\right)  & -\left( \frac{Z\alpha }{r} +1\right)
\end{array}\right) \left( \begin{array}{c} g(r) \\  f(r)
\end{array} \right) = E \left( \begin{array}{c} g(r) \\  f(r)
\end{array} \right).
\end{equation}

The mapping for the Schr\"odinger-Coulomb problem from Ref.~\cite{Kosloff96} is first modified by introducing an additional scaling parameter $s$ giving,
\begin{equation}
\label{mappingorig}
r(\theta) = s\theta - A\arctan{ \frac{s \theta}{A}}.
\end{equation}
For the relativistic calculation we improve upon this map adding a second-order singularity at $\theta = \pi$, giving
\begin{equation}
\label{mapping}
r(\theta) = \frac{s\theta - A\arctan{ \frac{s \theta}{A}}}{(\pi-\theta)^2}.
\end{equation}
This choice, with $A=4000$, was motivated in Ref.~\cite{Kosloff96} by looking at the coverage of the classical phase space for an optimal representation of about 20 bound hydrogenic states. The introduction of the singularity at $\theta=\pi$ does not modify the behavior at small $\theta$, and thus the choice of $A=4000$ is also good for the relativistic treatment of hydrogen. The added scale parameter $s$ allows  the tailoring of the coverage in phase space in order to achieve an optimal representation for more bound states or for a combination of bound and continuum parts of the spectrum. The latter is important for time-dependent applications where both excitation and ionization are present. 

The singularity at $\theta=\pi$ permits to map the semi-infinite $r$-range to $\theta \in \left[ 0,\pi \right.)$. This feature allows for incorporation of the correct boundary condition as $r \rightarrow \infty$, which leads to improved orthogonality of the computed eigenstates.

In terms of the independent variable $\theta$, Eq.~(\ref{Hr}) becomes,
\begin{equation}
\label{Htheta}
\left(
\begin{array}{cc}
-\left( \frac{Z\alpha }{r(\theta)}-1\right)  & -\left( \frac{1}{J(\theta)}\frac{d}{d\theta}-\frac{\kappa}{r(\theta)}\right)  \\
\left( \frac{1}{J(\theta)} \frac{d}{d\theta}+\frac{\kappa }{r(\theta)}\right)  & -\left( \frac{Z\alpha }{r(\theta)}+1\right)
\end{array}\right)
\left( \begin{array}{c} g(r(\theta)) \\  f(r(\theta))
\end{array} \right) =
E \left( \begin{array}{c} g(r(\theta)) \\   f(r(\theta))
\end{array} \right),
\end{equation}
where $J(\theta)$ is the Jacobian of the transformation from $r$ to $\theta$ given by
\begin{equation}
\label{J}
J(\theta) = \frac{dr}{d\theta} = \frac{s - \frac{s}{1+(\frac{s \theta}{A})^2}}{(\pi-\theta)^2}+2\frac{s\theta - A\arctan{\frac{s \theta}{A}} }{(\pi-\theta)^3}.
\end{equation}
This Jacobian weakens the singularity in the second derivative at the origin (due to the antisymmetric extension into the $-r$ region) in comparison with the original mapping given by Fattal et al. \cite{Kosloff96}.

A finite sine series truncated at order $N$ is chosen to represent each component of the spinor, i.e.,
\begin{equation}
\label{sinseries}
\left( \begin{array}{c} g(r(\theta)) \\  f(r(\theta))
\end{array} \right)
 =
\left(
\begin{array}{c}
\sum_{m=1}^{N}a_m \sin(m \theta) \\ \sum_{n=1}^{N} b_n \sin(n \theta)
\end{array} \right).
\end{equation}
Eq.~(\ref{sinseries}) when inserted into Eq.~(\ref{Htheta}) yields an analytic expression for the derivative,
\begin{equation}
\label{contderiv}
\frac{d}{d\theta} 
\left(
\begin{array}{c}
\sum_{m=1}^{N}a_m \sin(m \theta) \\  \sum_{n=1}^{N}b_n \sin(n \theta)
\end{array} \right) = 
\left(
\begin{array}{c}
\sum_{m=1}^{N} m a_m \cos(m \theta) \\
\sum_{n=1}^{N} n b_n \cos(n \theta)
\end{array} \right).
\end{equation}
In order to solve the problem numerically the $\theta$ variable is discretized. The number of points used for the $\theta$ mesh is set to $N$, yielding a square ($2N\times 2N$) Hamiltonian matrix. The spinor is now represented at $\theta=\theta_i$ by an eigenvector of the Hamiltonian matrix with the two components combined into a vector of length $2N$. The $\theta_i$ points are evenly spaced by $\Delta \theta = \frac{\pi}{N+1}$.  The furthest point on the mesh is given by $r_{\mathrm{max}} = r(N \Delta \theta)$, which is controlled by the scaling parameters. Note that $r_{\mathrm{max}}$ increases with $s$ in a non-linear fashion. Thus, for given $N$, $s$ determines the number of bound states and pseudo-Rydberg states that are obtained, as well as their respective accuracies. 

The expansion coefficients in Eq.~(\ref{sinseries}) are obtained from projection by the discrete sums:
\begin{eqnarray}
a_k & = & \frac{2}{N+1} \sum_{j=1}^N g(\theta_j) \sin(k\theta_j),
\end{eqnarray}
and
\begin{eqnarray}
b_k & = & \frac{2}{N+1} \sum_{j=1}^N f(\theta_j) \sin(k\theta_j).
\end{eqnarray}
The spinor (\ref{sinseries}) is therefore written explicitly as,
\begin{equation}
\left(
\begin{array}{c}
\frac{2}{N+1} \sum_{j=1}^N \sum_{m=1}^{N} \sin(m\theta_j) \sin(m \theta_i)  g(\theta_j)\\
\frac{2}{N+1} \sum_{j=1}^N \sum_{n=1}^{N} \sin(n \theta_j) \sin(n \theta_i)  f(\theta_j)
\end{array} \right).
\end{equation}
Introduction of the cosine-sine matrix, 
\begin{equation}
D_{ij} = \frac{2}{N+1} \sum_{m=1}^{N} m \cos(m \theta_i) \sin(m\theta_j),
\end{equation}
simplifies the discretized Eq.~(\ref{contderiv}) to
\begin{equation}
\label{disderiv}
\left(
\begin{array}{c}
\frac{2}{N+1} \left(\frac{d}{d\theta} \sum_{j=1}^N \sum_{m=1}^{N} \sin(m\theta_j) \sin(m \theta)  g(\theta_j)
\right)_i\\
\frac{2 }{N+1} \left(  \frac{d}{d\theta} \sum_{j=1}^N \sum_{n=1}^{N} \sin(n \theta_j) \sin(n \theta) 
f(\theta_j)\right)_i
\end{array} \right)
= 
\left(
\begin{array}{c}
 \sum_{j=1}^N D_{ij}  g(\theta_j)\\
 \sum_{j=1}^N D_{ij}  f(\theta_j)
\end{array} \right),
\end{equation}
where the index $i$ refers to the fact that the spinor derivative is evaluated at $r=r_i=r(\theta_i)$.

The discretized Hamiltonian is now written in block-component form as:
\begin{equation}
\label{H}
\left(
\begin{tabular}{cc}
$-\left( \frac{Z\alpha }{r(\theta_i)}-1\right) \delta_{i,j} $ & $-\left(
\frac{D_{ij}}{J(\theta_i)}-\frac{\kappa \delta_{i,j}}{r(\theta_i)}\right) $ \\
$\left( \frac{D_{ij}}{J(\theta_i)} + \frac{\kappa \delta_{i,j} }{r(\theta_i)}\right)  $ & $-\left( \frac{Z\alpha
}{r(\theta_i)} +1\right)\delta_{i,j}$
\end{tabular}\right).
\end{equation}

Due to the derivative terms, the representation of the Hamiltonian in the Fourier-sine basis is non-symmetric. In practice, one should be concerned about the appearance of complex eigenvalues mostly due to numerical errors. Thus, it is desirable to find alternate symmetric representations. In the non-relativistic 2$^{nd}$ order differential equation this can be achieved by re-mapping the eigenfunctions before the Fourier grid method is applied \cite{symmse}. While this approach guarantees real eigenvalues, it is also possible to lose efficiency as the introduced effective potential has a more complicated structure than the original one.


\chapter{Particle spectra from a non-orthogonal basis}\label{appendix2}
The mapped Fourier grid method yields a non-orthogonal complete set of states due to numerical noise. The derivation of the lepton/antilepton number operators is performed while taking into account the non-orthogonality, thereby correcting for it and yielding more accurate results. The derivation follows that of Wells \textit{et al.} \cite{PhysRevA.45.6296}.

\section{Basis states}
In the target frame of a two-body collision, the limit of $t\rightarrow \pm \infty$ results in a single-center potential for the target nucleus. It is, therefore, in this limit that we define the final particle states. The particles in the two-center basis (for finite times) are viewed as quasi-particles since they do not directly correspond to final-state electrons and positrons. In order to differentiate between particle and quasi-particle states, a subcritical QED ground state, $\left| \Phi_0(t)\rangle\right.$, is chosen. It is taken to be the time-independent state of the bare, target nucleus, with the Hamiltonian in natural units ($\hbar=$ c $=$ m$_{\rm e}$ = 1),
\begin{equation}
H_{T}=\vec{\mbox{\boldmath $\alpha$}}\cdot \vec{{\bf p}}+ \hat{\beta}+V(\vec{r}).
\end{equation} 
For the particle basis the states $\left| \chi_i \rangle \right.$ are defined as solutions to the time-independent Dirac equation for a single nucleus
\begin{equation}\label{tistates}
(\vec{\mbox{\boldmath $\alpha$}} \cdot \vec{{\bf p}}+ \hat{\beta} + V_{\mathrm{target}}(r) )\left| \chi_i \rangle \right. = E_i \left| \chi_i \rangle \right..
\end{equation}
The basis corresponding to the quasi-particles $\left| \phi_i(t) \rangle \right.$ are solutions to the time-dependent Dirac equation for the full collision potential, the two-center modified Coulomb potential $V_{\mathrm{2\;center}}(r,R)$, given by 
\begin{equation}\label{tdstates}
(\vec{\mbox{\boldmath $\alpha$}}\cdot \vec{{\bf p}}+\hat{\beta} + V_{\mathrm{2\;center}}(r,R) )\left| \phi_i(t) \rangle \right. = i \frac{\partial }{\partial t}\left| \phi_i(t) \rangle \right..
\end{equation}
Since the states obtained using the mapped Fourier grid method are complete but are not perfectly orthogonal (due to numerical noise \cite{me1}) the inner-product matrices for each are needed. For each basis they are defined as
\begin{eqnarray}
S_{i,j} = \langle \chi_i | \chi_j \rangle, \\
\overline{S}_{i,j} = \langle \phi_i (t) | \phi_j (t)\rangle.
\end{eqnarray}
Lastly, to distinguish between particle and quasi-particle states we use different symbols for their creation and annihilation operators. The operators $b_i^\dagger,a_i^\dagger$ are the creation operators for positrons and electrons respectively operating on the time-independent states of Eq.~\ref{tistates}.  The operators $\beta_i^\dagger,\alpha_i^\dagger$ are the creation operators for quasiparticles and act on the states in Eq.~\ref{tdstates}.  The $i$ index on the quasi-particle operator corresponds to the index for the time-dependent state $| \phi_i (t)\rangle$ when $t\rightarrow -\infty$ which is proportional to $\left| \chi_i \rangle \right.$ making $i$ a good quantum number in this limit. 

\section{Operator relations}
It is the time-evolved vacuum state, $|\Phi_0(t)\rangle$, which is calculated in practice allowing for the direct application of the quasi-particle operators. We are interested in the number of particles (not quasi-particles) making a relation between the two particle-operators types necessary. To do this we use two expansions for the field operator. 

The field equation for the field operator, $| \Psi_i(t) \rangle$, in the two-center collision system is given by
\begin{equation}
(\vec{\mbox{\boldmath $\alpha$}}\cdot \vec{{\bf p}}+\hat{\beta} + V_{\mathrm{2\;center}}(r,R(t)) )\left| \Psi_i(t) \rangle \right. = i \frac{\partial }{\partial t}\left| \Psi_i(t) \rangle \right. .
\end{equation}
Firstly, the field operator is expanded into solutions to the time-dependent basis $\left| \phi_j(t) \rangle \right.$ (with the usual switch for negative-energy states) of Eq.~\ref{tdstates}, giving
\begin{equation}\label{field}
\left| \Psi_i(t) \rangle \right. = \sum_{r<F} \beta_r^\dagger \left| \phi_r(t) \rangle \right. +\sum_{s>F} \alpha_s \left| \phi_s(t) \rangle \right..
\end{equation}
The symbol, $F$ denotes the Fermi level which is placed above the negative-energy states (making positron states denoted by $r<F$) and below the ground state. This definition works only for subcritical systems, but this poses no problem since we are only interested in the final particle content when the nuclei are separated. The state $\left| \phi_r(t) \rangle \right.$ is the $r$th initial state in Eq.~\ref{tdstates}, i.e. the 1S, 2S, etc. Since  $\lim_{t \rightarrow -\infty} \phi_j(t) \rightarrow \chi_j e^{-iE_jt}$, $\{ \chi_i\}$ satisfies the appropriate boundary conditions. Therefore we can also expand $\left| \Psi_i(t)\rangle \right.$ into the static target basis $\{ \chi_i\}$,
\begin{equation}\label{field2}
\left| \Psi_i(t ) \rangle \right. = \sum_{r<F} b_r^\dagger \left| \chi_r \rangle \right.e^{-iE_rt} +\sum_{s>F} a_s \left| \chi_s \rangle e^{-iE_st}\right..
\end{equation}
Equating the two expansions in Eq.~\ref{field} and \ref{field2} we get
\begin{equation}
\sum_{r<F} b_r^\dagger \left| \chi_r \rangle \right.e^{-iE_rt} +\sum_{s>F} a_s \left| \chi_s \rangle e^{-iE_st}\right.= \sum_{\rho<F} \beta_\rho^\dagger \left| \phi_\rho(t) \rangle \right. +\sum_{\sigma>F} \alpha_\sigma \left| \phi_\sigma(t) \rangle \right..
\end{equation}

Multiplying on the left by $\sum_\nu \langle \left.\chi_\nu \right| S_{k,\nu}^{-1}$ yields
\begin{eqnarray}
\sum_{r<F }\sum_{\nu} b_r^\dagger \langle \left.\chi_\nu \right| S_{k,\nu}^{-1}\left| \chi_r \rangle \right.e^{-iE_rt} &+&\sum_{s>F}\sum_{\nu} a_s \langle \left.\chi_\nu \right| S_{k,\nu}^{-1}\left| \chi_s \rangle e^{-iE_st}\right.  \\ 
= \sum_{\rho<F}\sum_{\nu} \beta_\rho^\dagger \langle \left.\chi_\nu \right| S_{k,\nu}^{-1}\left| \phi_\rho(t) \rangle \right. &+&\sum_{\sigma>F}\sum_{\nu} \alpha_\sigma \langle \left.\chi_\nu \right| S_{k,\nu}^{-1} \left| \phi_\sigma(t) \rangle \right.  \\
\Leftrightarrow \sum_{r<F}\sum_{\nu} b_r^\dagger S_{k,\nu}^{-1} \langle \chi_\nu  \left| \chi_r \rangle \right.e^{-iE_rt} &+&\sum_{s>F}\sum_{\nu} a_s S_{k,\nu}^{-1}\langle\chi_\nu \left| \chi_s \rangle e^{-iE_st}\right.  \\ 
= \sum_{\rho<F}\sum_{\nu} \beta_\rho^\dagger S_{k,\nu}^{-1}\langle \chi_\nu \left| \phi_\rho(t) \rangle \right. &+&\sum_{\sigma>F}\sum_{\nu} \alpha_\sigma S_{k,\nu}^{-1}\langle\chi_\nu \left| \phi_\sigma(t) \rangle \right.  \\
\Leftrightarrow \sum_{r<F}\sum_{\nu} b_r^\dagger S_{k,\nu}^{-1} S_{\nu,r} e^{-iE_rt} &+&\sum_{s>F}\sum_{\nu} a_s S_{k,\nu}^{-1}S_{\nu,s} e^{-iE_st}  \\ = \sum_{\rho<F}\sum_{\nu} \beta_\rho^\dagger S_{k,\nu}^{-1}\langle \chi_\nu \left| \phi_\rho(t) \rangle \right. &+&\sum_{\sigma>F}\sum_{\nu} \alpha_\sigma S_{k,\nu}^{-1}\langle\chi_\nu \left| \phi_\sigma(t) \rangle \right.  \\
\Leftrightarrow \sum_{r<F} b_r^\dagger \delta_{k,r} e^{-iE_rt}  &+&\sum_{s>F} a_k \delta_{k,s} e^{-iE_rt}  \\ = \sum_{\rho<F}\sum_{\nu} \beta_\rho^\dagger S_{k,\nu}^{-1}\langle \chi_\nu \left| \phi_\rho(t) \rangle \right. &+&\sum_{\sigma>F}\sum_{\nu} \alpha_\sigma S_{k,\nu}^{-1}\langle\chi_\nu \left| \phi_\sigma(t) \rangle \right.  
\end{eqnarray}

We have now obtained the particle creation and annihilation operators in terms of the quasi-particle operators giving
\begin{eqnarray}
b_k^\dagger   = &\sum_{\rho,\nu} \beta_\rho^\dagger S_{k,\nu}^{-1}\langle \chi_\nu \left| \phi_\rho(t) \rangle \right.e^{iE_kt} & +\sum_{\sigma,\nu} \alpha_\sigma S_{k,\nu}^{-1}\langle\chi_\nu \left| \phi_\sigma(t) \rangle \right.e^{iE_kt}\\
b_k   = &\sum_{\rho,\nu} \beta_\rho S_{k,\nu}^{-1}\langle \chi_\nu \left| \phi_\rho(t) \rangle^* \right.e^{-iE_kt} &+\sum_{\sigma,\nu} \alpha_\sigma^\dagger S_{k,\nu}^{-1}\langle\chi_\nu \left| \phi_\sigma(t) \rangle^* \right.e^{-iE_kt} \\
a_k^\dagger   = &\sum_{\rho,\nu} \beta_\rho S_{k,\nu}^{-1}\langle \chi_\nu \left| \phi_\rho(t) \rangle^* \right.e^{-iE_kt} &+\sum_{\sigma,\nu} \alpha_\sigma^\dagger S_{k,\nu}^{-1}\langle\chi_\nu \left| \phi_\sigma(t) \rangle^* \right.e^{-iE_kt} \\
a_k = &\sum_{\rho,\nu} \beta_\rho^\dagger S_{k,\nu}^{-1}\langle \chi_\nu \left| \phi_\rho(t) \rangle \right.e^{iE_kt} &+\sum_{\sigma,\nu} \alpha_\sigma S_{k,\nu}^{-1}\langle\chi_\nu \left| \phi_\sigma(t) \rangle \right.e^{iE_kt} ,
\end{eqnarray}
where $\sigma>F$, $\rho<F$, $k<F$ for $b_k^\dagger$ and $b_k$ while $k>F$ for $a_k^\dagger$ and $a_k$.

We now simplify the notation by defining the expansion coefficients in a non-orthogonal basis. In such a basis the expansion coefficients are not given by the inner products of the state and the basis states, $\langle \varphi_n |\Psi\rangle$, but are derived as follows,
\begin{eqnarray}
\left| \Psi \rangle\right. &=& \sum_n{a_n \left| \phi_n \rangle\right.} \\
c_m &=& \langle \phi_m \left|\right. \Psi\rangle \\
&=& \sum_n{a_n \left( \langle \phi_m \left|\right. \phi_n\rangle \right)} \\
\Rightarrow a_n & = &  \sum_m{c_m \left( \langle \phi_m \left|\right. \phi_n\rangle \right)^{-1}}
\end{eqnarray}
Therefore, we can rewrite the creation and annihilation relations using 
\begin{equation}
a_{k,\nu}(t)= \sum_n{S_{k,n}^{-1}\langle \chi_n \left|\right. \phi_\nu (t) \rangle} 
\end{equation}
(note that $a$ with a single index is an annihilation operator but with 2 indices it is an expansion coefficient) where $\nu$ is the state which is propagated, for instance the 1S ground state. This simplifies the relation and puts it in terms of the expansion coefficients, $a_{k,\nu}(t)$ with the result
\begin{eqnarray}\label{b1}
b_k^\dagger  = &\sum_{\rho<F} \beta_\rho^\dagger a_{k,\rho} e^{iE_kt} +\sum_{\sigma>F} \alpha_\sigma a_{k,\sigma} e^{iE_kt} & k<F\\ \label{b2}
b_k = &\sum_{\rho<F} \beta_\rho a^*_{k,\rho} e^{-iE_kt} +\sum_{\sigma>F} \alpha_\sigma^\dagger a^*_{k,\sigma}e^{-iE_kt} & k<F \\
\label{b3}
a_k^\dagger   = &\sum_{\rho<F} \beta_\rho a^*_{k,\rho}e^{-iE_kt} +\sum_{\sigma>F} \alpha_\sigma^\dagger a^*_{k,\sigma}e^{-iE_kt} & k>F \\
\label{b4}
a_k = &\sum_{\rho<F} \beta_\rho^\dagger a_{k,\rho} e^{iE_kt} +\sum_{\sigma>F} \alpha_\sigma a_{k,\sigma} e^{iE_kt} & k>F
\end{eqnarray}
It is noteworthy that the phase-factors are with respect to the energy of the $k$th state and are not part of the sums. This implies that they do not contribute when evaluating number operators, e.g. $\langle \Phi_0(t)| b_k^\dagger b_k|\Phi_0(t)\rangle$.

\section{Observables}
We are interested in the number of particles that will be created after the collision. Therefore, the particle number expectation value needs to be evaluated. Using the relations in Eqs.~\ref{b1}, \ref{b2}, \ref{b3} and \ref{b4} we can write the particle number operator in terms of the quasi-particle operators and operate on the time-evolved vacuum state $|\Phi_0(t)\rangle$. 

We work out the positron number as an example. The number of positrons created in the $k$th state due to the initial states $\rho$ and $\sigma$ are given by
\begin{eqnarray*}
\langle \bar{n}_k \rangle &=& \left\langle \Phi_0(t) \left| b_k^\dagger b_k \right|\Phi_0(t) \right\rangle\\
&=& \left\langle \Phi_0(t) \left| \left( \sum_{\rho<F} \beta_\rho^{\dagger} a_{k,\rho} e^{iE_kt} +\sum_{\sigma>F} \alpha_\sigma a_{k,\sigma} e^{iE_kt} \right) \right. \right. \\
& & \left. \left. \times\left(\sum_{\rho'<F} \beta_{\rho'} a^*_{k,\rho'} e^{-iE_kt} +\sum_{\sigma'>F} \alpha_{\sigma'}^\dagger a^*_{k,\sigma'} e^{-iE_kt}\right)\right| \Phi_0(t) \right\rangle 
\end{eqnarray*}
where it is understood that $k<F$.
The only non-zero term is $\alpha \alpha^\dagger$,
\begin{eqnarray}\label{fn}
\langle \bar{n}_k \rangle &=&  \sum_{\sigma,\sigma'>F} a_{k,\sigma} a^*_{k,\sigma'}\left\langle \Phi_0(t) \left|\alpha_\sigma \alpha_{\sigma'}^\dagger \right| \Phi_0(t) \right\rangle \\
&=&  \sum_{\sigma>F} a_{k,\sigma} a^*_{k,\sigma}
\end{eqnarray}
For example, the number of positrons created in state $k$ (at $t\rightarrow \infty$) by propagating the 1S only is
\begin{equation}
\langle \bar{n}_k \rangle =  |a_{k,1S}|^2. 
\end{equation}
These are the same expressions obtained when an orthogonal basis is used. Indeed the electron number also works out to
\begin{equation}\label{fna}
\langle n_k \rangle = \sum_{\sigma<F} a_{k,\sigma} a^*_{k,\sigma}
\end{equation}
where $k>F$. Therefore, we need only be concerned with how the expansion coefficients are formed in order to use the usual results for the particle observables.


\bibliography{thesis}
\bibliographystyle{unsrt}
\end{document}